\documentclass[11pt,draftcls,onecolumn]{IEEEtran}

\newcommand{\vx}{\mathbf x}
\newcommand{\zerob}{\mathbf 0}

\newcommand{\vax}{\hat{\mathbf x}}

\newcommand{\va}{\mathbf a}

\newcommand{\vb}{\mathbf b}

\newcommand{\vr}{\mathbf r}

\newcommand{\vu}{\mathbf u}

\newcommand{\vy}{\mathbf y}

\newcommand{\MI}{\mathbf I}

\newcommand{\MP}{\mathbf P}

\newcommand{\MW}{\mathbf W}

\newcommand{\MPhi}{\mathbf{{\Phi}}}
\newcommand{\vphi}{\boldsymbol{\varphi}}

\usepackage{cite}
\usepackage[pdftex]{graphicx}
\usepackage[cmex10]{amsmath}
\usepackage{amssymb, amsfonts, mathrsfs}
\usepackage[tight,footnotesize, FIGTOPCAP]{subfigure}

\begin{document}
\title{Sparse Vector Distributions and Recovery from Compressed Sensing}

\author{Bob L. Sturm \\ Department of Architecture, Design and Media Technology, Aalborg University Copenhagen, 
Lautrupvang 15, 2750 Ballerup, Denmark}

\maketitle

\begin{abstract}
\noindent It is well known that the performance of sparse vector recovery algorithms from 
compressive measurements can depend on the distribution 
underlying the non-zero elements of a sparse vector.
However, the extent of these effects has yet to be explored,
and formally presented.
In this paper, I empirically investigate this dependence
for seven distributions and fifteen recovery algorithms.
The two morals of this work are: 
1) any judgement of the recovery performance of one algorithm 
over that of another must be prefaced by the conditions
for which this is observed to be true, 
including sparse vector distributions,
and the criterion for exact recovery;
and 2) a recovery algorithm must be selected carefully based on
what distribution one expects to underlie the sensed sparse signal.
\end{abstract}

\section{Introduction}
Several researchers have observed that 
the distribution of the non-zero elements of a sparse vector
can impact its recoverability from its observation, e.g., 
\cite{Tropp2004,Dai2009,Donoho2009,Maleki2010,Qui2010}.
The degrees to which recovery algorithms are affected
by the underlying distribution of a sparse vector
have not been thoroughly investigated, both empirically and analytically,
except perhaps in the case of $\ell_1$-regularization \cite{Donoho2009}
and OMP \cite{Jin2008}.
The worst case scenario appears to be 
sparse vectors distributed Bernoulli equiprobable in $\{-a,a\}$
(or ``constant amplitude random sign'') \cite{Dai2009, Maleki2010}.
Maleki and Donoho \cite{Maleki2010} use these types of vectors to
tune three recovery algorithms.
They also briefly investigate the change in 
recovery performance by their tuned algorithms
for sparse vectors distributed Cauchy, uniform in $[-1,1]$, and Laplacian \cite{Maleki2010};
but this is just to show that vectors distributed
differently than Bernoulli have better recoverability.
Their reasoning goes, if one's sparse vector has a different distribution,
the possibility of recovery from compressive measurements 
will be better than expected.

In this article, I empirically compare the performance
of fifteen algorithms for the recovery of 
compressively sampled sparse vectors 
having elements sampled from seven distributions. 
The algorithms I test can be grouped into four categories: 
greedy iterative descent, thresholding, convex relaxation, and majorization.
The greedy approaches I test are:
orthogonal matching pursuit (OMP) \cite{Pati1993, Tropp2004},
stagewise OMP \cite{Donoho2006},
regularized OMP \cite{Needell2010}, 
and probabilistic OMP (PrOMP) \cite{Divekar2010}.
For the thresholding algorithms,
I test the recommended algorithms produced by Maleki and Donoho \cite{Maleki2010}
--- which includes iterative hard and soft thresholding (IHT, IST) \cite{Blumensath2009}, 
and two-stage thresholding (TST) ---
Approximate Message Passing (AMP) \cite{Donoho2009},
Subspace Pursuit (SP) \cite{Dai2009},
and CoSaMP \cite{Needell2009},
and Algebraic Pursuit with 1-memory (ALPS) \cite{Cevher2011}.
The convex relaxation methods I test inclde
$\ell_1$-minimization (BP) \cite{Chen1998},
iteratively reweighted $\ell_1$-minimization (IRl1) \cite{Candes2008b},
and Gradient Projection for Sparse Reconstruction (GPSR) \cite{Figueiredo2007}.
Finally, the majorization approach is the smoothed $\ell_0$ technique (SL0) \cite{Mohimani2009}.
The seven distributions from which I sample sparse vectors are:
Normal; Laplacian; uniform;
Bernoulli; bimodal Gaussian; bimodal uniform; and bimodal Rayleigh.
I test several pairs of problem sparsities and indeterminacies
for sensing matrices sampled from the uniform spherical ensemble.

These comparisons provide many interesting observations,
many of which are known but yet to be formally stated.
First, I find that there is no one algorithm that
outperforms the others for all distributions I test
when using exact recovery of the full support.
SL0 and BP/AMP however, do appear to be the best in all cases.
Some algorithms are extremely sensitive to the distribution
underlying sparse vectors, and others are not.
Greedy methods and majorization perform better than
$\ell_1$-minimization approaches
for vectors distributed with probability density concentrated at zero.
Recovery of Bernoulli distributed vectors shows the lowest rates.
Thus, choosing a recovery algorithm should be
guided by the expected underlying distribution of the sparse vector.
I also find that one can obtain an inflated recovery performance 
with a seemingly reasonable criterion for exact recovery.
For some distributions, such a criterion may not produce results
that accurately reflect the true performance of an algorithm.

In the next section,
I briefly review compressed sensing and my notation.
Then I describe each of the fifteen algorithms that I test.
Section 3 provides the main menu of results 
from my numerous simulations.
Here I look at the effects on performance of perfect recovery criterion,
and the effects of sparse vector distributions.
I also compare all algorithms for each of the
distributions I test through inspecting 
their phase transitions.
I conclude with a summary, and avenues for future research.

\section{Compressed Sensing and Signal Recovery}
Given a vector of ambient dimension $N$, $\vx \in \mathbb{C}^N$,
we {\em sense} it by $\MPhi : \mathbb{C}^N \to \mathbb{C}^m$,
producing a measurement vector $\vu = \MPhi\vx$.
Several methods have been developed and studied for
recovering $\vx$ given $\vu$ and $\MPhi$,
which are obviously sensitive to
both the size and content of $\MPhi$ relative to $\vx$.
The overdetermined problem ($N \le m$) has been studied in
statistics, linear algebra, frame theory, and others, e.g., \cite{Christensen2003,Mallat2009}.
The underdetermined problem ($N > m$), 
with $\vx$ a {\em sparse} vector,
and $\MPhi$ a random matrix,
has been studied in approximation theory \cite{DeVore1998, Mallat2009},
and more recently compressed sensing \cite{Candes2006,Donoho2006b, Tropp2010}.
From here on, we are working with vectors that have ambient dimensions 
larger than the number of measurements, i.e., $N > m$.
Below, I briefly review the fifteen algorithms I test,
and provide details about their implementations.

\subsection{Notations}
We define the support of a vector $\vx \in \mathbb{C}^N$ 
as the indices of those elements with non-zero values, i.e., 
\begin{equation}
\mathcal{S}(\vx) := \{n \in \Omega : [\vx]_n \ne 0 \}
\end{equation}
where $[\va]_n$ is the $n$th component of the vector,
and $\Omega := \{1, 2, \ldots, N\}$.
A vector $\vx$ is called $s${\em-sparse} 
when at most $s$ of its entries are nonzero, i.e., $|\mathcal{S}(\vx)| \le s$.
As the work of this paper is purely empirical,
I only consider finite-energy complex vectors 
in a Hilbert space of dimension $N < \infty$.
Here, the inner product of two vectors is defined
$\langle \va, \vb \rangle := \vb^*\va$
where $\vb^*$ is the conjugate transpose;
and the $\ell_p$-norm for $1 \le p < \infty$
of any vector in this space is defined
\begin{equation}
|| \va ||_p^p := \sum_{n=1}^N | [\va]_n |^p.
\end{equation} 

In compressed sensing, the sensing matrix $\MPhi = [\vphi_1 | \vphi_2 | \ldots | \vphi_N]$
maps a length-$N$ vector to a lower $m$-dimensional space.
I define $\MPhi_{\Omega_k}$ as a matrix of the $k$ columns of $\MPhi$ 
indexed by the ordered set $\Omega_k \subseteq \Omega$,
i.e., $\MPhi_{\Omega_k} = [\vphi_{\Omega_k(1)} | \vphi_{\Omega_k(2)} | \cdots | \vphi_{\Omega_k(k)}]$
where $\Omega_k(1)$ is the first element of $\Omega_k$.
I notate a set difference as $\Omega \backslash \Omega_k$.
Alternatively, one may speak of a {\em dictionary} of {\em atoms},
$\mathcal{D} = \{\vphi_n \in \mathbb{C}^m \}_{n \in \Omega}$.
For the problem of recovering the sensed vector from its measurements,
one defines its {\em indeterminacy} as $\delta := m/N$,
and its {\em sparsity} as $\rho := s/m$.
The problem indeterminacy
describes the undersampling of a vector in its measurements;
and the problem sparsity describes the proportion of the sensing matrix 
active in the measurements.

\subsection{Recovery by Greedy Pursuit}

Greedy pursuits entail the iterative augmentation of a set of 
atoms selected from the dictionary,
and an updating of the residual.
I initialize all the following greedy methods 
by $\Omega_0 = \emptyset$, and $\vr_0 = \vu$,
and define the stopping criteria to be 
$|| \vr_k ||_2 \le 10^{-5} ||\vu||_2$, or $|\Omega_k| > 2s$,
unless otherwise specified.

\subsubsection{Orthogonal Matching Pursuit (OMP) \cite{Pati1993}}
OMP augments the $k$th index set 
$\Omega_{k+1} = \Omega_k \cup \{n_k\}$ by selecting a new index according to
\begin{equation}
n_k = \arg \min_{n \in \Omega} || \vr_k - \langle \vr_k, \vphi_n \rangle \vphi_n ||_2^2 = \arg \max_{n \in \Omega} |\langle \vr_k, \vphi_n \rangle|
\label{eq:MPselection}
\end{equation}
where the $k$th residual is defined
as the projection of the measurements onto the
subspace orthogonal to that spanned by the 
dictionary elements indexed by $\Omega_k$, i.e.,
\begin{equation}
\vr_k := \MP_k^\perp \vu = (\MI_m - \MPhi_{\Omega_{k}}\MPhi_{\Omega_{k}}^\dagger)\vu
\label{eq:OMPresidual}
\end{equation}
where $\MPhi_{\Omega_{k}}^\dagger := (\MPhi_{\Omega_{k}}^*\MPhi_{\Omega_{k}})^{-1}\MPhi_{\Omega_{k}}^*$,
and $\MI_m$ is the size-$m$ identity matrix.
OMP thereby ensures each residual is orthogonal
to the space spanned by the atoms indexed by $\Omega_k$.
OMP creates the $k$th solution by
\begin{equation}
\vx_k = \arg \min_\vx || \vu - \MPhi\MI_{\Omega_{k}}\vx||_2
\label{eq:OMPsolution}
\end{equation}
where the $N$ square matrix \([\MI_{\Omega_{k}}]_{jj} = 1 \; \forall j \in \Omega_{k}\), and zero elsewhere.
The implementation I use\footnote{{\tt SolveOMP.m} in SparseLab: http://sparselab.stanford.edu/}
involves a QR decomposition to efficiently perform the projection step.

\subsubsection{Probabilistic OMP (PrOMP) \cite{Divekar2010}}
PrOMP augments the $k$th index set $\Omega_{k+1} = \Omega_k \cup \{n_k\}$
by sampling from $n_k \sim P( n \in \Omega^* | \vr_k )$,
where the set $\Omega^*$ denotes the true indices of the non-zero entries of $\vx$,
and the residual is defined in (\ref{eq:OMPresidual}).
Divekar et al. \cite{Divekar2010} estimates this distribution by
\begin{equation}
P( n \in \Omega^* | \vr_k ) = \begin{cases}
(1-p)/l, & | \langle \vr_k, \vphi_n \rangle | \ge f_l \\
p/(N-k-l), & 0 < | \langle \vr_k, \vphi_n \rangle | < f_l \\
0, & | \langle \vr_k, \vphi_n \rangle | = 0
\end{cases}
\end{equation}
where $0 \le p \le 1$, and $f_l$ is the $l$th largest value in 
$\{| \langle \vr_k, \vphi_n \rangle | : n \in \Omega \}$, $1 \le l < N-k$.
In this way, PrOMP produces several candidate solutions (\ref{eq:OMPsolution}),
which can include that found by OMP,
and selects the one that produces the smallest residual $|| \vr_k ||_2$.
In my implementation, I set $p = 0.001$, $l=2$, 
and have PrOMP generate at most $10$ solutions.
I determined these values by experimentation, but not formalized tuning.
I make PrOMP stop generating each solution 
using the same stopping criteria of OMP.


\subsubsection{Regularized OMP (ROMP) \cite{Needell2010}}
ROMP augments the $k$th index set 
\(\Omega_{k+1} = \Omega_k \cup J^*\) 
where \(J^* \subset \Omega\) 
is a set of indices determined in the following way.
First, ROMP defines the set 
\begin{equation}
I := \{i \in \Omega : |\langle \vr_k, \vphi_i \rangle| > 0\}
\end{equation}
where \(\vr_k\) is defined in (\ref{eq:OMPresidual}),
and because of this we know \(I\cap\Omega_k = \emptyset\).
Next, ROMP finds the set of $L$ disjoint sets \(\mathcal{J} = \{J_1, J_2, \ldots, J_L : |J_l| \le s\}\) where
\begin{align}
J_1 & = \{ j \in I : | \langle \vr_k, \vphi_j \rangle | \ge \frac{1}{2} \max_{i\in I} | \langle \vr_k, \vphi_i \rangle | \} \\
J_2 & = \{ j \in I\backslash J_1 : | \langle \vr_k, \vphi_j \rangle | \ge \frac{1}{2} \max_{i\in I\backslash J_1} | \langle \vr_k, \vphi_i \rangle | \} \\
\vdots \nonumber \\
J_L & = \{ j \in I\backslash \cup_{l=1}^L J_l : | \langle \vr_k, \vphi_j \rangle | \ge \frac{1}{2} \max_{i\in I\backslash \cup_{l=1}^L J_l} | \langle \vr_k, \vphi_i \rangle | \}.
\end{align}
From this set, ROMP chooses the best defined by
\begin{equation}
J^* = \arg \max_{J \in \mathcal{J} } || \MPhi_J^* \vr_k||_2. 
\end{equation}
The solution at a given iteration is defined in (\ref{eq:OMPsolution}).
Note that ROMP requires one to specify the solution sparsity desired.

\subsubsection{Stagewise OMP (StOMP) \cite{Donoho2006}}
StOMP augments the $k$th index set 
\(\Omega_{k+1} = \Omega_k \cup J^*\) 
where
\begin{equation}
J^* := \{i \in \Omega : |\langle \vr_k, \vphi_i \rangle| > t_k\sigma_k\}
\end{equation}
where \(t_k\sigma_k\) is a threshold parameter, 
and \(\vr_k\) is defined in (\ref{eq:OMPresidual}).
The solution at a given iteration is defined in (\ref{eq:OMPsolution}).
In their work \cite{Donoho2006}, Donoho et al. define \(\sigma_k = ||\vr_k||_2/\sqrt{m}\), and \(2\le t_k \le 3\), 
motivated from either avoiding false alarms (which requires knowing the sparsity of the solution) 
or missed detection (false discovery rate).
I retain the defaults of the implementation I use\footnote{{\tt SolveStOMP.m} 
in SparseLab: http://sparselab.stanford.edu/},
which means the threshold is set by the false discovery rate with \(t_k = 2\).
Donoho et al. recommend running this procedure only 5-10 times,
but here I ran it up to \(2s\) times which I find does not degrade its performance. 

\subsection{Recovery by Thresholding}
Thresholding techniques entail the iterative refinement of
a solution, and an update of the residual.
I initialize all the following thresholding methods 
by $\vx_0 = \zerob$,
and restrict the number of refinements to $k \le 300$, 
or when $|| \vr_k ||_2 < 10^{-5} ||\vu||_2$.

\subsubsection{Iterative Thresholding \cite{Blumensath2009}}
Iterative thresholding refines the solution $\vx_k$ according to
\begin{equation}
\vx_{k+1} = T\left ( \vx_k + \kappa \MPhi^* \vr_k; \tau_k \right )
\label{eq:thresholding}
\end{equation}
where  \(T(\vy)\) is a thresholding function applied element-wise to $\vy$,
$\tau_k \ge 0$ is a threshold,
\(0 < \kappa < 1\) is a relaxation parameter,
and the residual is $\vr_k = \vu - \MPhi\vx_k$.
For iterative {\em hard} thresholding (IHT), this function is defined
\begin{equation}
T(x; \tau_k) := \begin{cases}
x, & |x| > \tau_k \\
0, & \textrm{else}.
\end{cases}
\end{equation}
For iterative {\em soft} thresholding (IST), this function is defined
\begin{equation}
T(x; \tau_k) = \begin{cases}
\mathrm{sgn}(x) (|x| - \tau_k), & |x| > \tau_k \\
0, & \textrm{else}.
\end{cases}
\label{eq:softthresholding}
\end{equation}
The implementations of IHT and IST I use 
are the ``recommended versions'' by Maleki and Donoho \cite{Maleki2010},
where they set $\kappa = 0.65$ for IHT and $0.6$ for IST,
and adaptively set $\tau_k$ 
according to a false alarm rate,
the problem indeterminacy,
and an estimate of the variance of the residual.\footnote{See http://sparselab.stanford.edu/OptimalTuning/main.htm}
These settings come from extensive empirical tests 
for sparse signals distributed Bernoulli.

\subsubsection{Compressive Sampling MP (CoSaMP) \cite{Needell2009}}
CoSaMP refines the solution $\vx_k$ by two stages of thresholding.
First, given a sparsity $s$ CoSaMP finds the support
\begin{equation}
J := \mathcal{S} \left [T_{2s} ( \mathbf{\Phi}^*\vr_k ; \epsilon) \right ]
\label{eq:CoSaMPT1}
\end{equation}
where \(T_{2s}(\vy)\) nulls all elements of $\vy$ 
except for the $2s$ ones with the largest magnitudes above $\epsilon \ge 0$.
CoSaMP then thresholds again to find the new support
\begin{equation}
\Omega_{k+1} = \mathcal{S}\left [ T_s ( \arg \min_{\vx} || \vu - \MPhi \MI_{\mathcal{S}(\vx)\cup J} \vx ||_2 ; \epsilon) \right ]
\end{equation}
where \(T_s(\vy)\) nulls all elements of $\vy$ 
except for the $s$ ones with the largest magnitudes above $\epsilon \ge 0$.
The new solution $\vx_{k+1}$ is then computed by (\ref{eq:OMPsolution}).
In my implementation of CoSaMP,
I set $\epsilon = 10^{-12}$ to avoid numerical errors;
and in addition to the stopping criterion mentioned above,
I exit the refinements if $||\vr_k|| >||\vr_{k-1}||$,
in which case I choose the previous solution.

\subsubsection{Subspace Pursuit (SP) \cite{Dai2009}}
SP operates in the same manner as CoSaMP,
but instead of retaining the $2s$ largest magnitudes in (\ref{eq:CoSaMPT1}),
it keeps only $s$.
The stopping criteria of my implementation of SP
are the same as for CoSaMP.

\subsubsection{Two-stage Thresholding (TST) \cite{Maleki2010}}
Noting the similarity between the two,
Maleki and Donoho generalize CoSaMP and SP into TST.
Given $\vx_k$, $\vr_k$, and $s$, TST with parameters $(\alpha,\beta)$ finds
\begin{equation}
J := \mathcal{S} \left [T_{\alpha s} ( \vx_k + \kappa \mathbf{\Phi}^*\vr_k ) \right ]
\label{eq:TSTT1}
\end{equation}
where \(0 < \kappa < 1\) is a relaxation parameter,
and \(T_{\alpha s}(\vy)\) nulls all elements of $\vy$ 
except for the $\alpha s$ ones with the largest magnitudes.
TST then thresholds again to find the new support
\begin{equation}
\Omega_{k+1} = \mathcal{S}\left [ T_{\beta s} ( \arg \min_{\vx} || \vu - \MPhi \MI_{\mathcal{S}(\vx)\cup J} \vx ||_2) \right ]
\end{equation}
where \(T_{\beta s}(\vy)\) nulls all elements of $\vy$ 
except for the $\beta s$ ones with the largest magnitudes.
The new solution $\vx_{k+1}$ is then computed by (\ref{eq:OMPsolution}).
Obviously, when $\alpha = 2$ and $\beta = 1$, 
TST becomes similar to CoSaMP;
and to become similar to SP, $\alpha = \beta = 1$.
The implementation of TST I use 
is that recommended by Maleki and Donoho \cite{Maleki2010},\footnote{See http://sparselab.stanford.edu/OptimalTuning/main.htm}
where they choose $\alpha = \beta = 1$, define $\kappa = 0.6$, and estimate the sparsity 
\begin{equation}
s = \lfloor (0.044417\delta^2+ 0.34142\delta +0.14844) m \rfloor.
\end{equation}
These values come from their extensive experiments.
Note that one iteration of SP (or CoSaMP) and one iteration of TST are only equivalent when
\begin{equation}
\mathcal{S} \left [T_{s} ( \vx_k + \kappa \mathbf{\Phi}^*\vr_k ) \right ] = \mathcal{S}(\vx_k) \bigcup \mathcal{S} \left [T_{s} ( \mathbf{\Phi}^*\vr_k ) \right ].
\end{equation}
Since the thresholding is non-linear, TST will not produce the same results as SP (or as CoSaMP).

\subsubsection{Approximate Message Passing \cite{Donoho2009}}
AMP proceeds as iterative thresholding, 
but with a critical difference in how it defines the residual in (\ref{eq:thresholding}).
Given \(\vx_k\), \(\vx_{k-1}\) and $\tau_k$ as the $m$th largest magnitude in \(\vx_{k-1}\), 
AMP with soft thresholding defines the \(k\)-order residual \(\vr_k\)
\begin{equation}
\vr_k = \vu - \mathbf{\Phi}\vx_k + \vr_{k-1} \frac{1}{m} || \vx_{k-1} + \mathbf{\Phi}^* \vr_{k-1} ||_{0,\tau_k}
\end{equation}
where 
$|| \vy ||_{0,\tau} := |\{ n : |[\vy]_n| \ge \tau\}|$.
AMP then refines the solution \(\vx_k\) by soft thresholding (\ref{eq:softthresholding})
\begin{equation}
\vx_{k+1} = T\left ( \vx_k + \mathbf{\Phi}^* \vr_k; \tau_k \right ).
\end{equation}
AMP repeats the procedure above until some stopping condition.
In the implementation I use,\footnote{http://people.epfl.ch/ulugbek.kamilov}
I make AMP stop refinement when $||\vr_k||_2 \le 10^{-5} ||\vu||_2 $ or $k > 300$.

\subsubsection{Algebraic pursuit (ALPS) with 1-memory \cite{Cevher2011}}
ALPS essentially entails accelerated iterative hard thresholding with memory.
Given $\vx_k$ and a desired sparsity $s$,
ALPS with 1-memory refines the solution
\begin{equation}
\vx_{k+1} = T_s(\vb_k) + \mu_k[T_s(\vb_k) - T_s(\vb_{k-1})]
\end{equation}
where $\mu_k \in (0,1]$, and defining the expanded support set
$\mathcal{I}_k := \mathcal{S}(\vx_k) \cup \mathcal{S}(T_s[\MI_{\Omega \backslash \mathcal{S}(\vx_k)}\MPhi^*(\vu - \MPhi\vx_k)])$ 
\begin{equation}
\vb_{k} := \vx_k + \kappa_k \MI_{\mathcal{I}_k} \MPhi^*(\vu - \MPhi\vx_k)
\end{equation}
where the optimal step size is given by
\begin{equation}
\kappa_k = \frac{|| \MI_{\mathcal{I}_k} \MPhi^*(\vu - \MPhi\vx_k)||_2}{|| \MPhi\MI_{\mathcal{I}_k} \MPhi^*(\vu - \MPhi\vx_k)||_2}.
\end{equation}
In the implementation I use,\footnote{http://lions.epfl.ch/ALPS}
ALPS computes the best weight at each step $\mu_k$ according to FISTA \cite{Beck2009}.
Note that for ALPS I specify the correct sparsity, as I do for ROMP, CoSaMP and SP.

\subsection{Recovery by Conxex Relaxation}
The following methods attempt to solve the sparse recovery problem
\begin{equation}
\min | \mathcal{S}(\vx)| \; \textrm{subject to} \; \vu = \MPhi\vx
\label{eq:sparseproblem}
\end{equation}
by replacing the non-convex measure of strict sparsity 
with a relaxed and convex measure.

\subsubsection{$\ell_1$-minimization \cite{Chen1998}} 
The principle of Basis Pursuit (BP) replaces strict sense sparsity
$|\mathcal{S}(x)|$ with the relaxed and convex $\ell_1$ norm.
We can rewrite (\ref{eq:sparseproblem}) in the following two ways
\begin{align}
& \min || \vx ||_1 \; \textrm{subject to} \: \vu = \MPhi \vx \label{eq:l1minimizationeq} \\
& \min \frac{1}{2} ||\vu - \MPhi \vx||_2 + \lambda || \vx ||_1 \label{eq:l1minimizationlag}
\end{align}
both of which can be solved by several methods,
e.g., simplex and interior-point methods \cite{Boyd2004},
as a linear program \cite{Chen1998},
and by gradient methods \cite{Figueiredo2007}.
In my implementation, I solve (\ref{eq:l1minimizationeq}) 
using the CVX toolbox \cite{Grant2011}.

\subsubsection{Gradient Projection for Sparse Reconstruction (GPSR) \cite{Figueiredo2007}}
GPSR provides a computationally light and iterative approach to solving
(\ref{eq:l1minimizationlag}), or at least taking one to the neighborhood of the solution,
by using gradient projection, thresholding, and line search.
The details of the implementation are out of the scope of this paper,
but suffice it to say I am using the ``Basic" implementation
provided by the authors \cite{Figueiredo2007} with their defaults,
and $\lambda = 0.005||\MPhi^*\vu||_\infty$.

\subsubsection{Iteratively Reweighted $\ell_1$-minimization (IRl1) \cite{Candes2008b}}
Given a diagonal $N$ square matrix $\MW_k$,
IRl1 solves
\begin{equation}
\min || \MW_k \vx ||_1 \; \textrm{subject to} \: \vu =  \MPhi \vx.
\label{eq:weightedl1minimization}
\end{equation}
Using the solution, IRl1 constructs a new weighting matrix
\begin{equation}
[\MW_{k+1}]_{nn} = 1/(| [\vx_{k+1}]_n | + \epsilon )
\label{eq:weightupdate}
\end{equation}
where \(\epsilon > 0\) is set for stability.
IRl1 then solves (\ref{eq:weightedl1minimization}) with these new weights,
and continues in this way until some stopping criterion is met.
For initialization, $\MW_0 = \MI$.
In my implementation --- which uses CVX as for BP ---
I make $\epsilon = 0.1$ as done in \cite{Candes2008b},
and limit the number of iterations to 4, or until $|| \vr_k ||_2 \le 10^{-5}||\vu||_2$.

\subsubsection{Smoothed $\ell_0$ (SL0) \cite{Mohimani2009}}
If we define
\begin{equation}
J(\vx; \sigma) := \sum_{i=1}^m \exp \left [ -[\vx]_i^2/ 2\sigma^2 \right ].
\end{equation}
then the strict sense sparsity in (\ref{eq:sparseproblem}) can be expressed
\begin{equation}
|\mathcal{S}(\vx)|  = m - \lim_{\sigma \to 0} J(\vx; \sigma).
\end{equation}
For a decreasing set of variances $\{\sigma , \sigma d, \sigma d^2, \ldots \}$
for $0 < d < 1$, SL0 finds solutions to
\begin{equation}
\vx_k = \arg \min_\vx \frac{1}{2} ||\vu - \MPhi \vx ||_2 - \lambda J(\vx; \sigma_k)
\end{equation}
using steepest descent with soft thresholding,
and reprojection back to the feasible set.
Finally, depending on the last \(\sigma_L\), 
SL0 arrives at a solution that will be quite close to the unique sparsest solution.
In the implementation I use provided by the authors,\footnote{http://ee.sharif.ir/$\sim$SLzero}
$\sigma := 2 || \MPhi^\dagger\vu ||_\infty$
where here $\MPhi^\dagger = \MPhi^*(\MPhi \MPhi^*)^{-1}$.
I set $d = 0.95$, and restrict the last variance to be no larger than $4 \times 10^{-5}$.

\section{Computer Simulations}
My problem suite is nearly the same as that used 
in the empirical work of Maleki and Donoho \cite{Maleki2010},
except here I make $N=400$ instead of $800$,
and test 50 realizations (instead of 100) of each sparse vector
at each pair of problem sparsity and indetereminacy.\footnote{I am currently
running simulations at the original dimensions used by
Maleki and Donoho, but preliminary results do not show much
difference with those presented below.}
I sample each real sensing matrix from the {\em uniform spherical ensemble},
where each column of $\MPhi$ is a point on the $N$-dimensional unit sphere,
and each element of each column of $\MPhi$ is 
independently and identically distributed zero-mean Gaussian.

\begin{figure}[t]
\centering
\subfigure{
\includegraphics[width=0.495\textwidth]{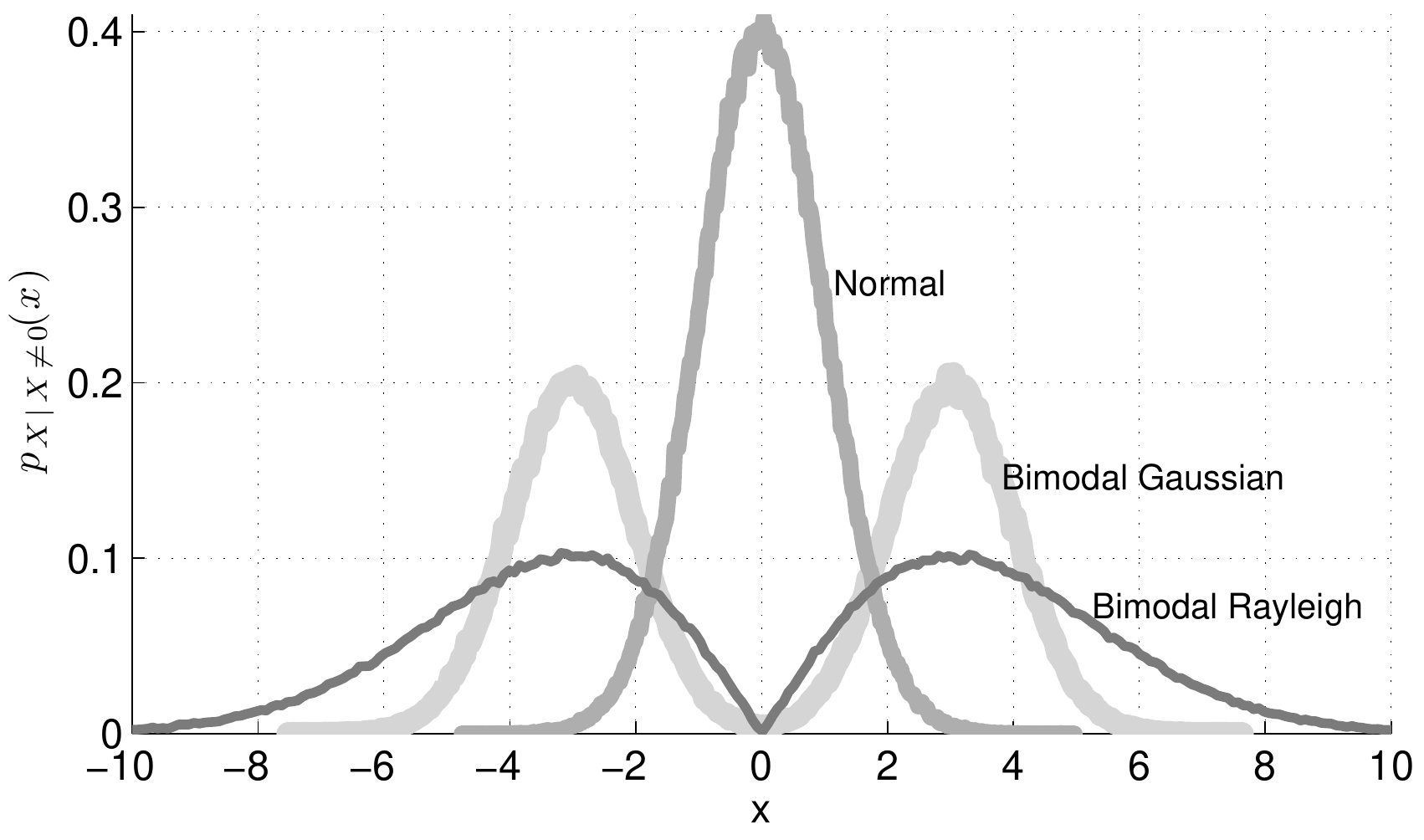}}\hspace{-0.1in}
\subfigure{
\includegraphics[width=0.485\textwidth]{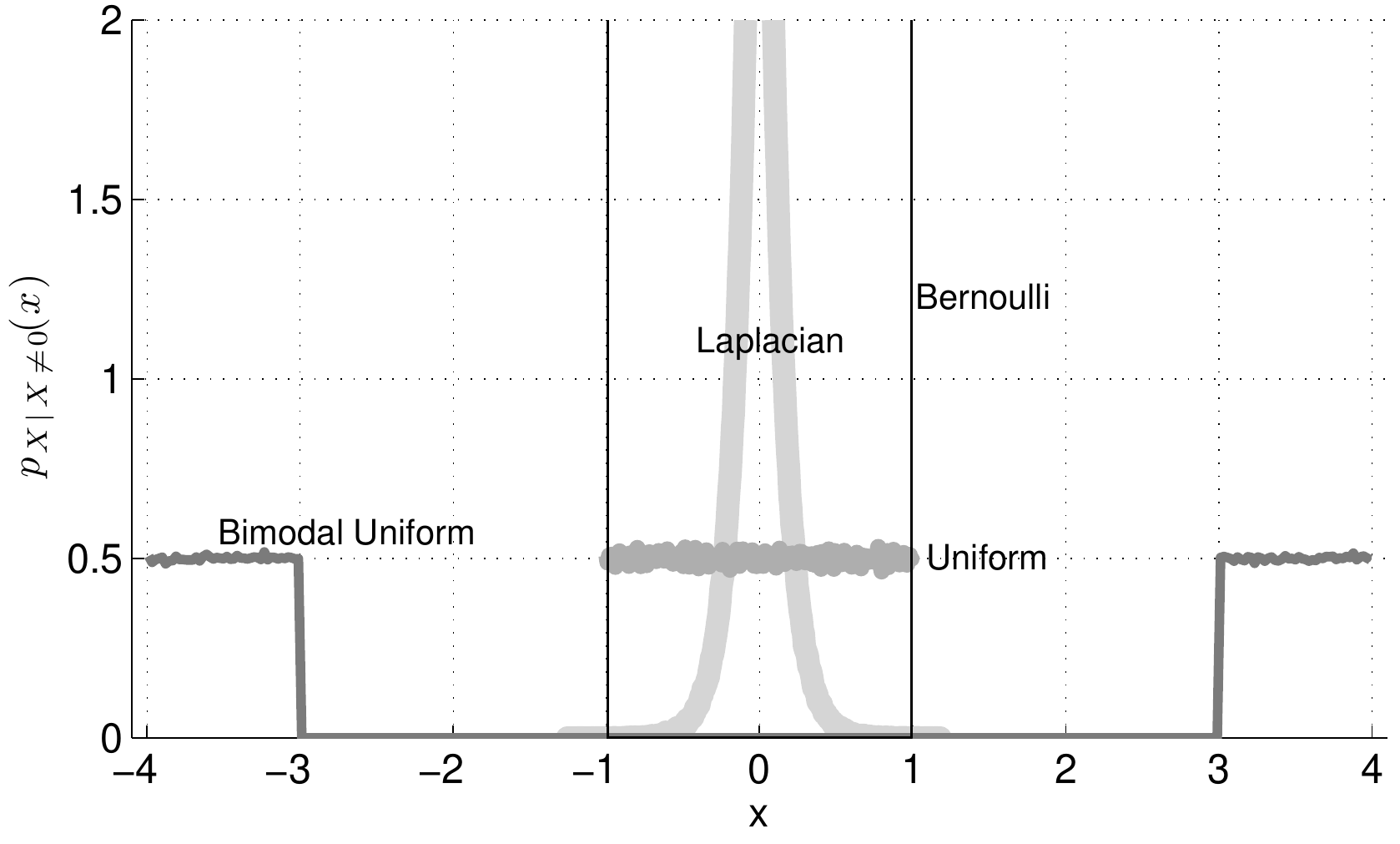}}\\ \vspace{-0.1in}
\caption{Empirical probability distributions of all 
non-zero elements in the sparse signals of my test suite.}
\label{fig:empiricalPDFs}
\end{figure}

I generate each real $s$-sparse vector in the following way.
First, I determine the indices for the non-zero elements by drawing 
$s$ elements at random (without replacement) from $\Omega$.
Then, for the vector elements at these indices,
I independently sample from the same distribution of the following:
\begin{enumerate}
\item Standard Normal (N)
\item Zero-mean Laplacian (L) , with parameter $\lambda = 10$
\item Zero-mean uniform (U) in the support $[-1,1]$
\item Bernoulli (B) equiprobable in $\{-1,1\}$ (``Constant Amplitude Random Sign'' in \cite{Maleki2010})
\item Bimodal Gaussian (BG) as a sum of two unit variance Gaussians with means $\pm 3$
\item Bimodal uniform (BU) as a sum of two uniform distributions in $[-4,-2]$ and $[2, 4]$
\item Bimodal Rayleigh (BR), where the magnitude value is distributed Rayleigh with parameter $\sigma = 3$.
\end{enumerate}
I sample from a distribution until I obtain a value with a magnitude greater than $10^{-10}$.
This ensures that all sparse components have magnitudes 100 times greater
than my specification of numerical precision ($\epsilon < 10^{-12}$) in CoSaMP and SP.
Figure \ref{fig:empiricalPDFs} shows the empirical distributions 
of all the signals I compressively sample.

The problems I test have 30 different linearly-spaced sparsities $\rho \in [0.05, 1]$.
For each of these, I test 16 different linearly-spaced indeterminacies $\delta \in [0.05, 0.5414]$.\footnote{I am testing a
wider range in my current simulations, but in my opinion, 
the most important part is when a signal is compressed to a high degree.}
For each sparsity and indeterminacy pair then,
I find the proportion of 50 vectors sensed by a $\MPhi$ that are recovered 
by the fifteen algorithms I describe above.
Before I test for recovery, I ``debias'' \cite{Figueiredo2007}
each solution in the following way.
I first find the effective support of a solution $\vax$
\begin{equation}
\mathcal{I}_{\vax} := \mathcal{S}(T_l[\vx_k]) : \textrm{rank}\{ \MPhi_{\mathcal{S}(T_l[\vx_k])}\} = l, \min_{l \le m} || \vx_k - T_l(\vx_k) ||_2.
\end{equation}
In other words, I find a skinny submatrix of $\MPhi$ that has the
largest (full column) rank of $l$ associated with the $l$ largest magnitudes in the solution.
Finally, I produce the debiased solution by solving (\ref{eq:OMPsolution})
with $\Omega_k = \mathcal{I}_{\vax}$,
and finally hard thresholding, i.e., 
\begin{equation}
\vax = T( \arg \min_\vx || \vu - \MPhi \MI_{\mathcal{I}_{\vax}} \vx ||_2 ; 10^{-10})
\end{equation}
At each problem indeterminacy,
I interpolate the recovery results over all sparsities to find 
where successful recovery occurs with a probability of 0.5.
This creates an empirical phase transition plot showing the boundary
above which most recoveries fail, and below which most recoveries succeed.

\subsection{Exact Recovery Criteria and Their Effects}
I consider two different criteria for exact recovery.
First, I consider a solution recovered exactly when 
\begin{equation*}
\frac{||\vx - \vax||_2}{||\vx||_2} \le \epsilon_\vx
\tag{$R_{\ell_2}$}
\label{eq:successcriterion1}
\end{equation*}
for some $\epsilon_\vx \ge 0$.
We can relate ($R_{\ell_2}$) to the stopping criterion
of the algorithms above, i.e., when
\begin{equation}
\frac{||\vu - \MPhi \vax ||_2}{||\vu||_2} \le \epsilon_\vu
\label{eq:stoppingcriterion}
\end{equation}
with $\epsilon_\vu \ge 0$,
and $\vax $ is defined by (\ref{eq:OMPsolution}).
We can rewrite this as
\begin{equation}
\frac{|| \MPhi (\vx - \vax) ||_2}{|| \MPhi \vx ||_2} \le \epsilon_\vu
\end{equation}
and bound it considering the frame bounds of the sensing matrix, i.e., 
\begin{equation}
A|| \vx ||_2 \le || \MPhi \vx ||_2 \le B || \vx ||_2 \; \forall \vx.
\end{equation}
with $0< A \le B< \infty$.
Noticing that $|| \MPhi (\vx - \vax) ||_2 \ge A|| \vx - \vax ||_2$,
and $||\MPhi \vx ||_2 \le B || \vx ||_2$,
we can bound the expression from below by
\begin{equation}
\frac{A || \vx - \vax ||_2 }{B || \vx ||_2} \le \frac{|| \MPhi (\vx - \vax) ||_2}{|| \MPhi \vx ||_2} \le \epsilon_\vu
\end{equation}
and thus
\begin{equation}
\frac{|| \vx - \vax ||_2 }{|| \vx ||_2} \le \frac{B}{A} \epsilon_\vu.
\end{equation}
From these we see that given (\ref{eq:stoppingcriterion})
the recovery condition ($R_{\ell_2}$) must be true when
\begin{equation}
\frac{B}{A} \epsilon_\vu \le \epsilon_\vx.
\end{equation}
In my experiments, I specify $\epsilon_\vx = 10^{-2}$ as done in \cite{Maleki2010}, 
and set $\epsilon_\vu = 10^{-5}$ (but Maleki and Donoho set the latter $\epsilon_\vu = 10^{-3}$).

The second recovery condition I use is the recovery of the support,
in which case I consider a solution recovered exactly when 
\begin{equation*}
\mathcal{S}(\vax) = \mathcal{S}(\vx).
\tag{$R_{\mathcal{S}}$}
\label{eq:successcriterion2}
\end{equation*}
When the criterion (\ref{eq:successcriterion2}) is true,
(\ref{eq:successcriterion1}) is necessarily true by virtue of (\ref{eq:OMPsolution})
with $\Omega_k = \mathcal{S}(\vx)$.
We might consider relating this recovery condition
to that in (\ref{eq:successcriterion1}) in the following sense.
Given that only a portion of the true support is recovered,
what are the conditions that criterion (\ref{eq:successcriterion1}) is true?
Consider without loss of generality
that an algorithm has recovered 
the true support except the first $0 < |\mathcal{T}| < |\mathcal{S}(\vx)|$ elements, 
i.e., $\mathcal{S}(\vax) \cup \mathcal{T} = \mathcal{S}(\vx)$,
and $\mathcal{S}(\vax) \cap \mathcal{T} = \emptyset$.
For simplicity I denote $\widehat{\mathcal{S}} = \mathcal{S}(\vax)$
and $\mathcal{S} = \mathcal{S}(\vx)$.
If we assume that the atoms associated with the elements in $\mathcal{T}$
are orthogonal to the rest of the atoms in the support $\widehat{\mathcal{S}}$,
then we can write the solution as
\begin{equation}
\vx = \MPhi_{\widehat{\mathcal{S}}}^\dagger \vu + \MPhi_{\mathcal{T}}^\dagger \vu = \vax +\MPhi_{\mathcal{T}}^\dagger \vu.
\end{equation}
Using the orthogonality assumption,
we can write the left hand side of (\ref{eq:successcriterion1}) as
\begin{equation}
\frac{||\vx - \vax||_2^2}{||\vx||_2^2} = \frac{|| \MPhi_{\mathcal{T}}^\dagger \vu||_2^2}{||\vx||_2^2} = 
\frac{|| \MPhi_{\mathcal{T}}^\dagger \MPhi \vx ||_2^2}{||\vx||_2^2} =
\frac{|| \MPhi_{\mathcal{T}}^\dagger [\MPhi_{\mathcal{T}} | \square] \vx ||_2^2}{||\vx||_2^2} =
\frac{|| [ \MPhi_{\mathcal{T}}^\dagger \MPhi_{\mathcal{T}} | \zerob] \vx ||_2^2}{||\vx||_2^2} =
\frac{||\vx_\mathcal{T} ||_2^2}{||\vx||_2^2}
\end{equation}
as expected.
And thus, we see one way to guarantee condition (\ref{eq:successcriterion1}) is for
$||\vx_\mathcal{T} ||_2 \le  \epsilon_\vx ||\vx||_2$.
If we remove the orthogonality assumption, this becomes 
\begin{equation}
||\vx_{\mathcal{S}}^\perp||_2 := ||\vx - \MPhi_{\mathcal{S}} \MPhi_{\mathcal{S}}^\dagger \vx||_2 \le  \epsilon_\vx ||\vx||_2.
\end{equation}

Now, consider that $|\mathcal{T}| = |\mathcal{S}| - 1$,
i.e., that we have missed all but one of the elements,
but the recovery algorithm has found and precisely estimated 
the largest element with a magnitude $\alpha$.
Assume that all other non-zero elements have magnitudes 
less than or equal to $\beta$.
Thus, $||\vx - \vax||_2^2 \le (|\mathcal{S}|-1)\beta^2$;
and $||\vx||_2^2 \le (|\mathcal{S}|-1)\beta^2 + \alpha^2$.
From the above analysis, we see
\begin{equation}
\frac{||\vx - \vax||_2^2}{||\vx||_2^2} \le \frac{(|\mathcal{S}|-1)\beta^2}{(|\mathcal{S}|-1)\beta^2 + \alpha^2} \le \epsilon_\vx^2
\end{equation}
We can see that the criterion (\ref{eq:successcriterion1}) is still guaranteed as long as
\begin{equation}
\frac{\alpha^2}{\beta^2} \ge \frac{(1-\epsilon_\vx^2)(|\mathcal{S}|-1)}{\epsilon_\vx^2}.
\end{equation}

This analysis shows that we can substantially violate the criterion (\ref{eq:successcriterion2})
and still meet (\ref{eq:successcriterion1}) as long as the distribution of the sparse signal permits it.
For instance, if all the elements of $\vx$ have unit magnitudes, 
then missing $|\mathcal{T}|$ elements of the support produces
$||\vx_{\mathcal{S}}^\perp||_2^2/||\vx||_2^2 = |\mathcal{T}|/|\mathcal{S}|$;
and to satisfy (\ref{eq:successcriterion1}) this 
requires $\sqrt{|\mathcal{T}|/|\mathcal{S}|} \le \epsilon_\vx $.
This is not likely unless $|\mathcal{S}|$ is very large and we miss only a few elements.
If instead our sparse signal is distributed such that it has
only a few extremely large magnitudes and the rest small,
then we can miss much more of the support and still satisfy (\ref{eq:successcriterion1}).

\begin{figure}[tb]
\centering
\subfigure[Normal]{
\includegraphics[width=0.49\textwidth]{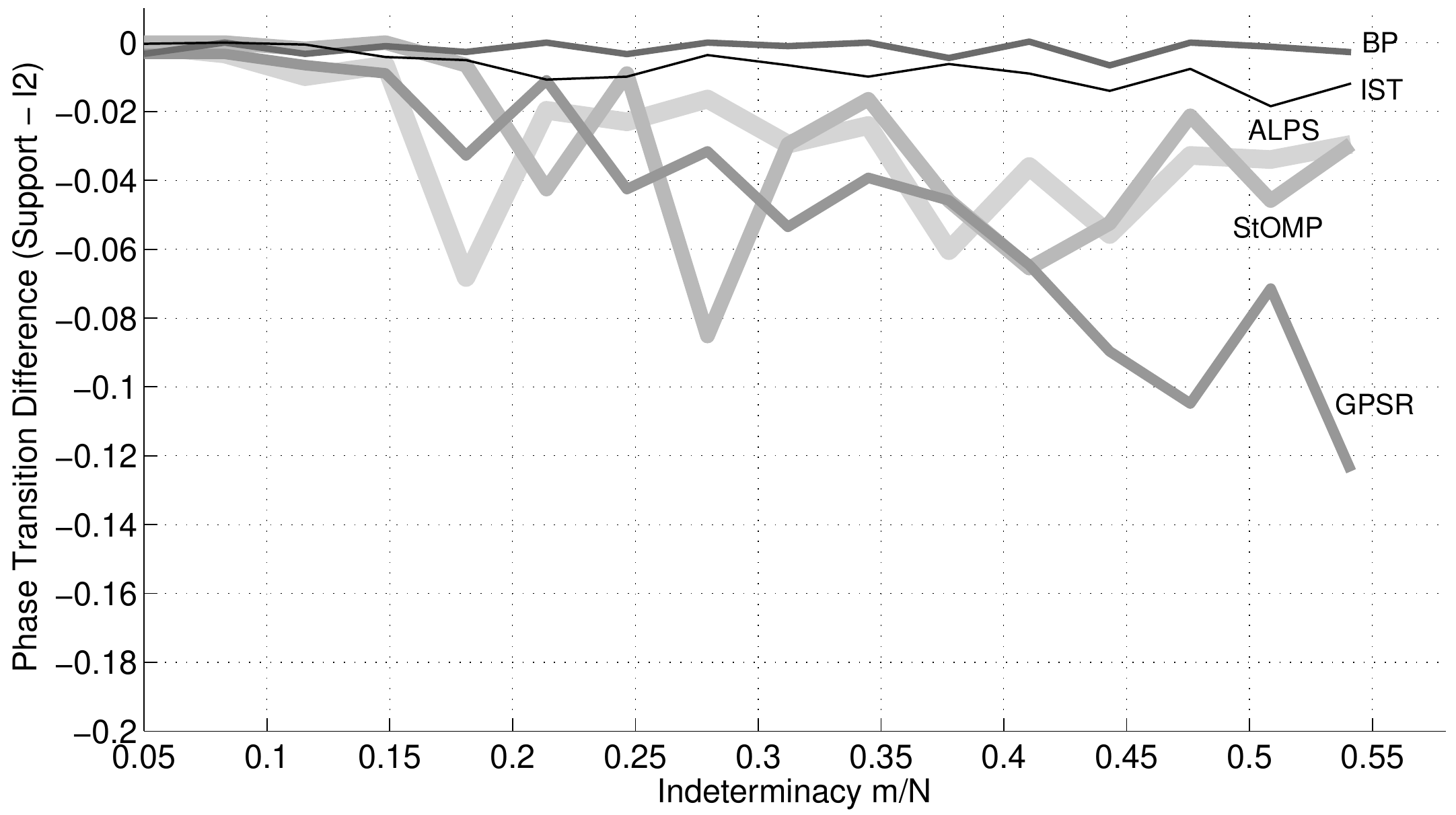}}\hspace{-0.1in}
\subfigure[Laplacian]{
\includegraphics[width=0.49\textwidth]{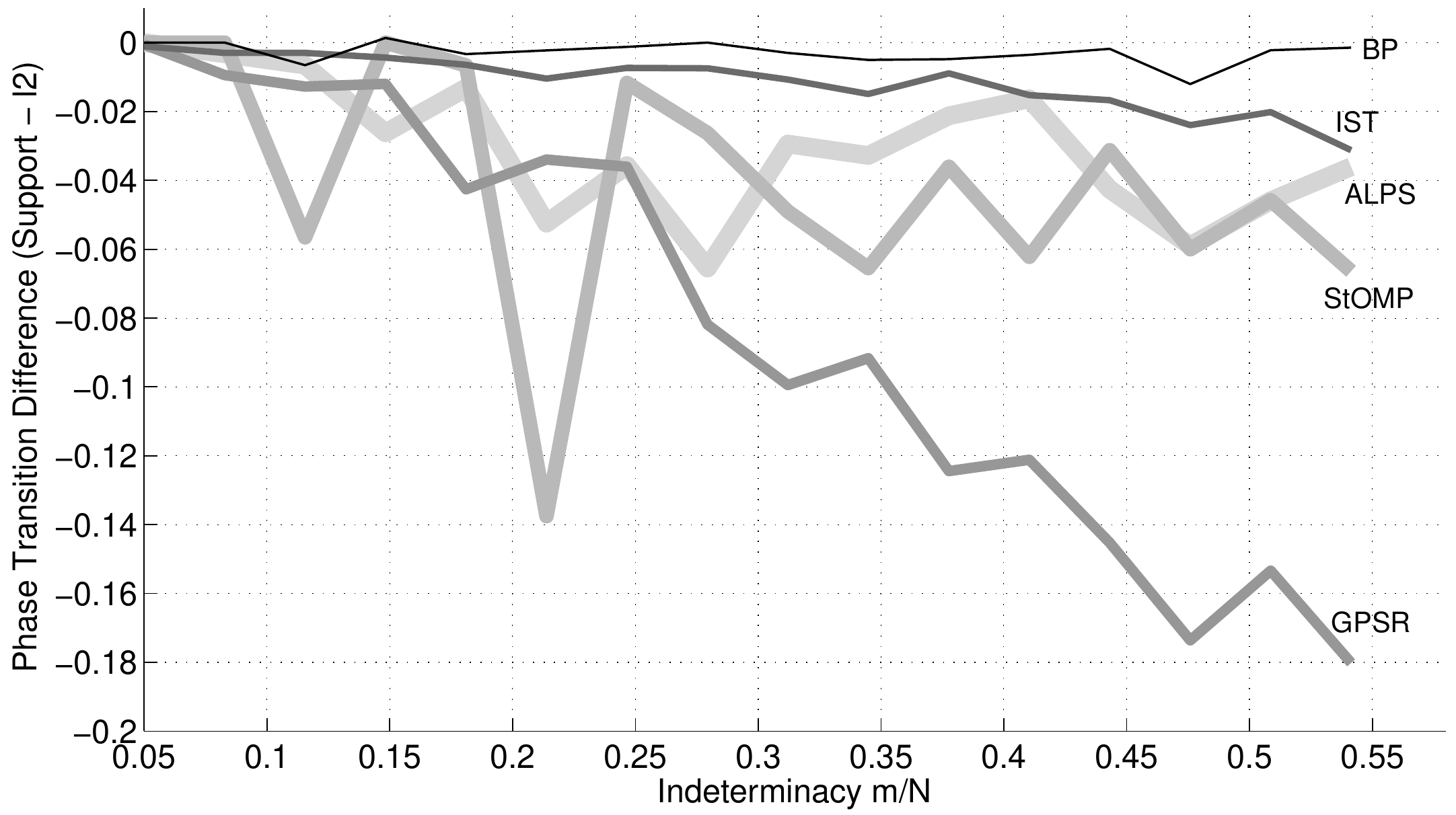}}\\
\subfigure[Uniform]{
\includegraphics[width=0.49\textwidth]{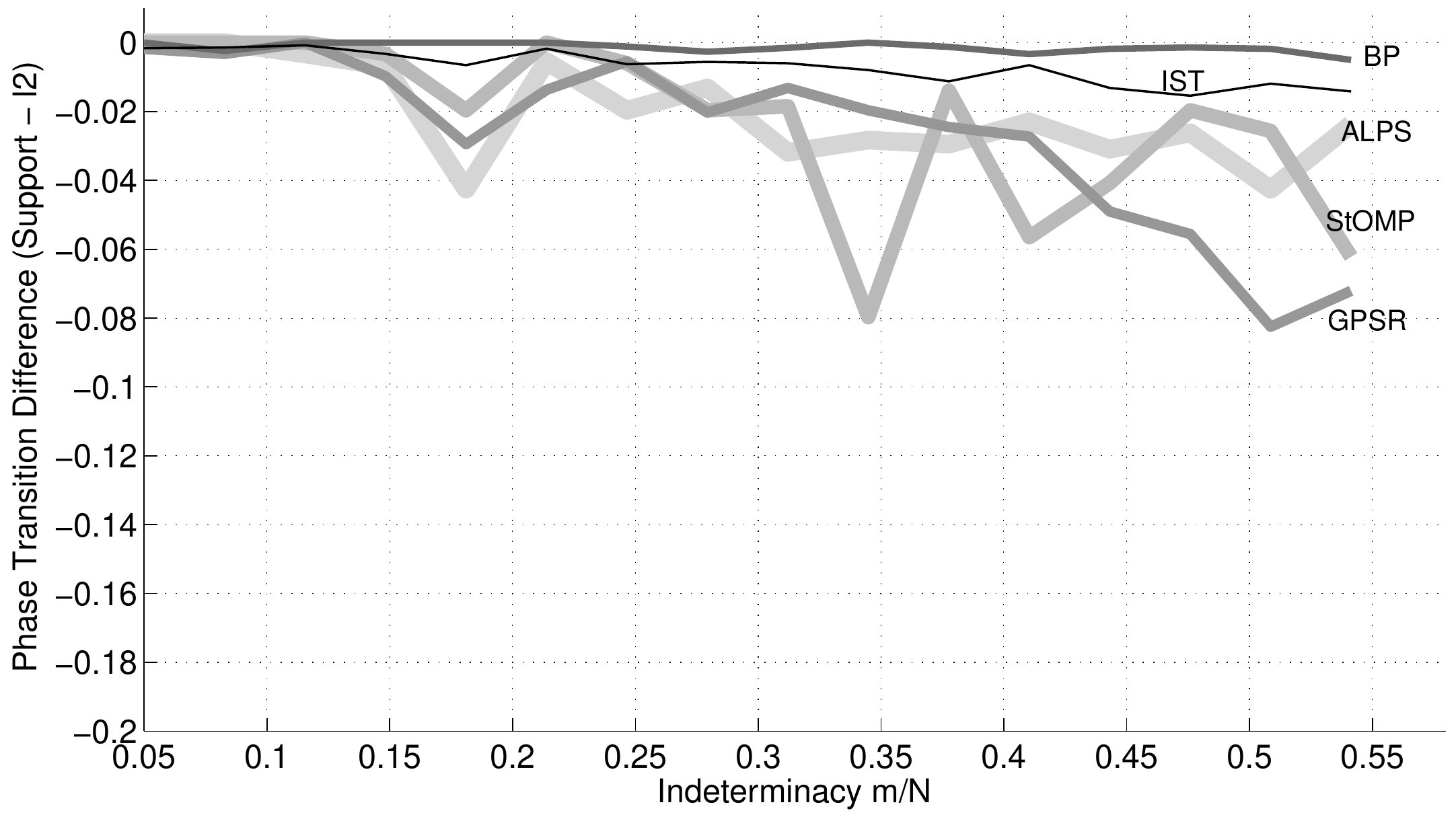}}\hspace{-0.1in}
\subfigure[Bimodal Rayleigh]{
\includegraphics[width=0.49\textwidth]{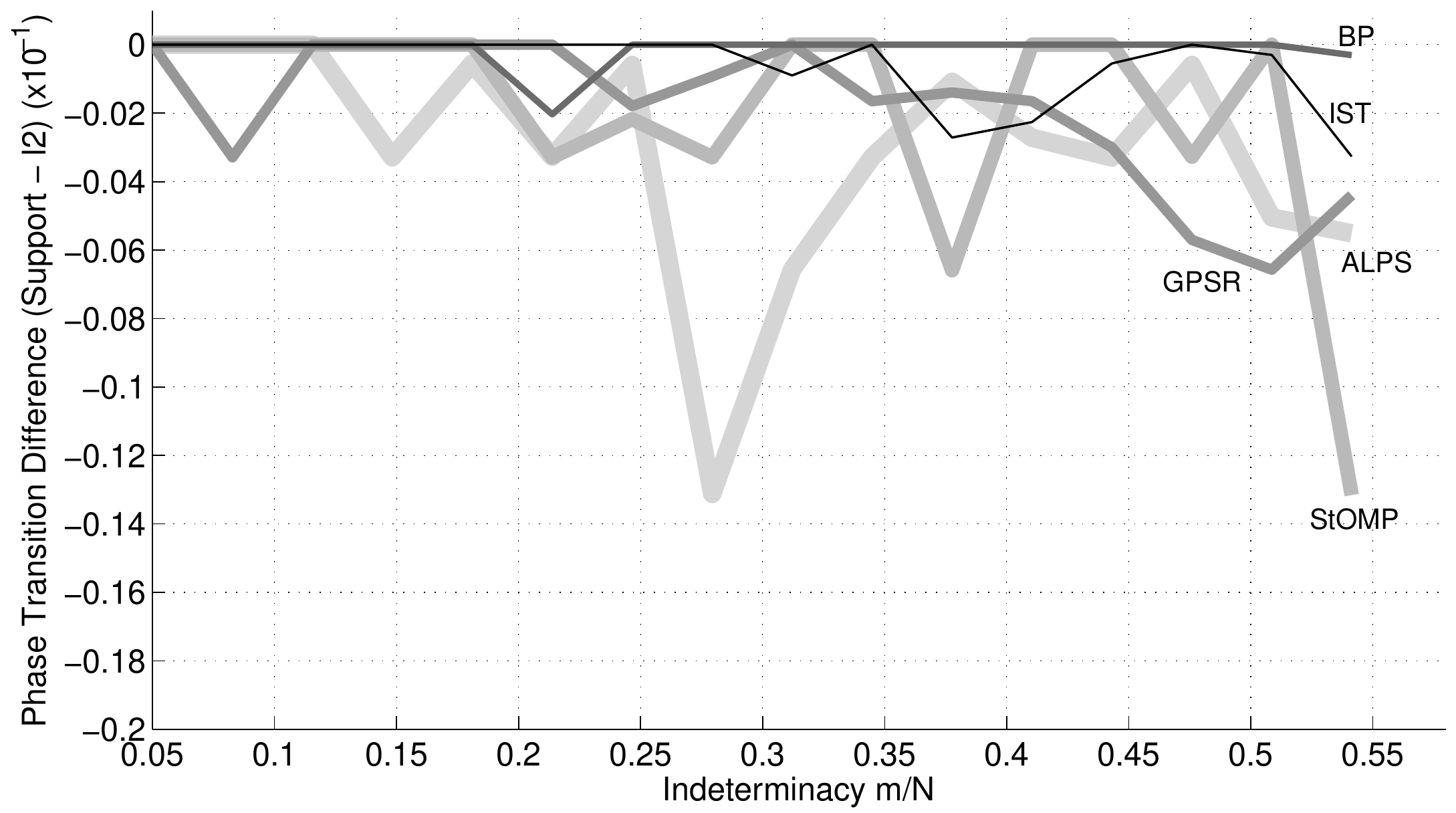}}
\caption{Differences between the empirical phase transitions for 
five recovery algorithms using criterion ($R_{\mathcal{S}}$)
and criterion ($R_{\ell_2}$).
Note different axis scaling in (d).}
\label{fig:Phasedifferencesrecoverycriterion}
\end{figure}

The following experiments test the variability of the phase transitions of
several algorithms depending on these success criteria,
and the distribution underlying the sparse signals.
Figure \ref{fig:Phasedifferencesrecoverycriterion} 
shows the differences in empirical phase transitions of five 
recovery algorithms using ($R_{\mathcal{S}}$) or (\ref{eq:successcriterion1}).
Since ($R_{\mathcal{S}}$) implies (\ref{eq:successcriterion1}),
the empirical phase transition of the latter will always be equal to or greater 
than that of the former.
The empirical phase transitions difference is zero across all problem indeterminacies
when there is no difference between these two criteria.
Figure \ref{fig:Phasedifferencesrecoverycriterion} shows 
a significant dependence of the empirical phase transition
and the success criteria for four sparse vector distributions.
For all other algorithms, and the three other distributions
(Bernoulli, bimodal uniform, and bimodal Gaussian),
the differences are nearly always zero.

I do not completely know the reason why only these five algorithms
out of the 15 I test show significant differences in their empirical phase transitions,
or why GPSR and StOMP appear the most volatile of these algorithms.
It could be that they are better than the others at estimating
the large non-zero components at the expense of modeling the small ones.
However, this experiment clearly reveals that for sparse vectors distributed with
probability density concentrated near zero, criterion (\ref{eq:successcriterion1}) does allow
many small components to pass detection without consequence.
The more probability density is distributed around zero,
the more criterion (\ref{eq:successcriterion1}) is likely to be satisfied
while criterion (\ref{eq:successcriterion2}) is violated.
It is clear from this experiment that we must take caution when
judging the performance of recovery algorithms 
by a success criterion that can be very lax.
In the following experiments,
I use criterion (\ref{eq:successcriterion2}) 
to measure and compare the success of the algorithms
since it also implies (\ref{eq:successcriterion1}).

\subsection{Sparse Vector Distributions and their Effects}
\begin{figure}[tb]
\centering
\subfigure[BP, AMP]{
\includegraphics[width=0.49\textwidth]{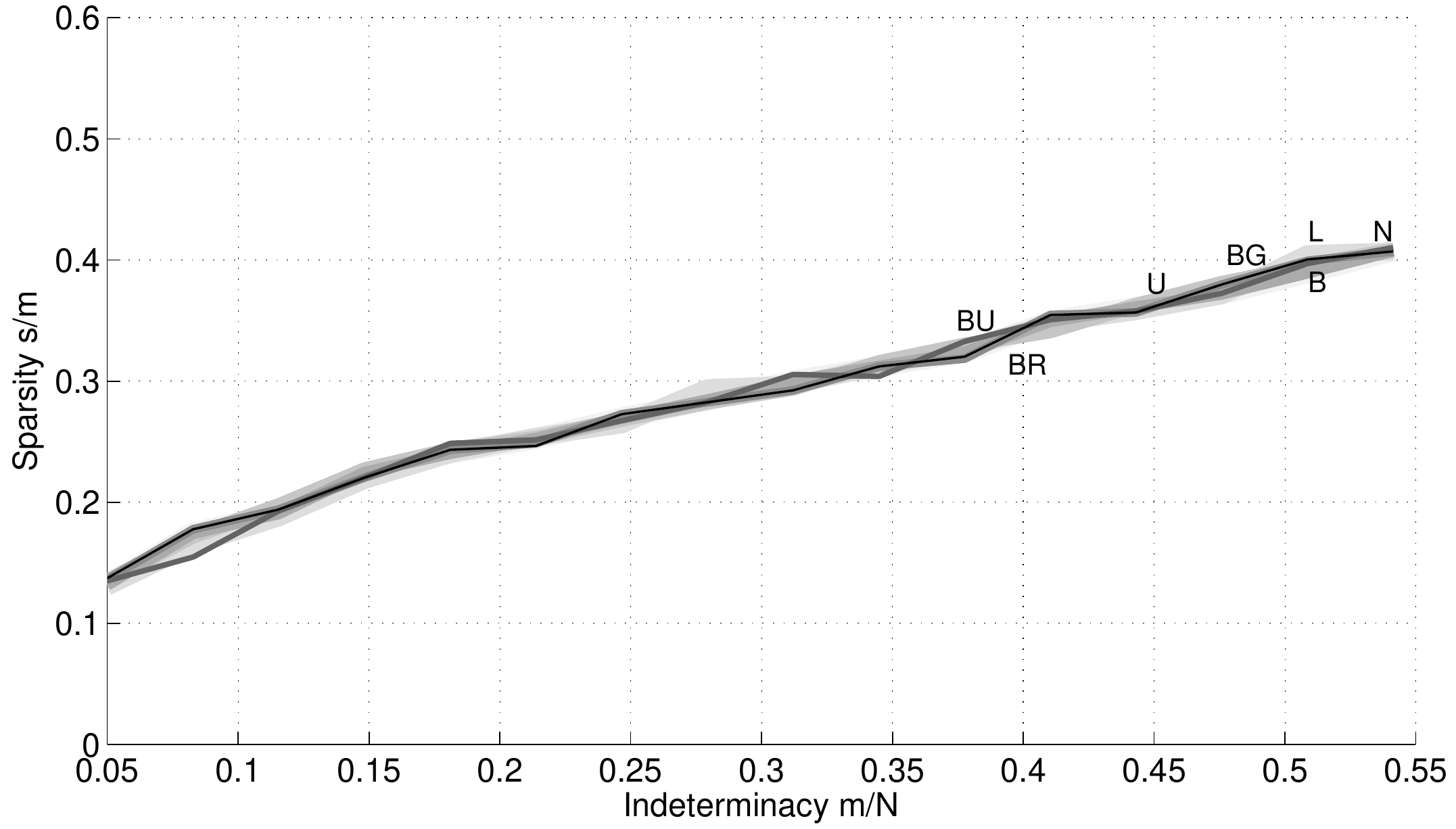}}\hspace{-0.1in}
\subfigure[Recommended TST]{
\includegraphics[width=0.49\textwidth]{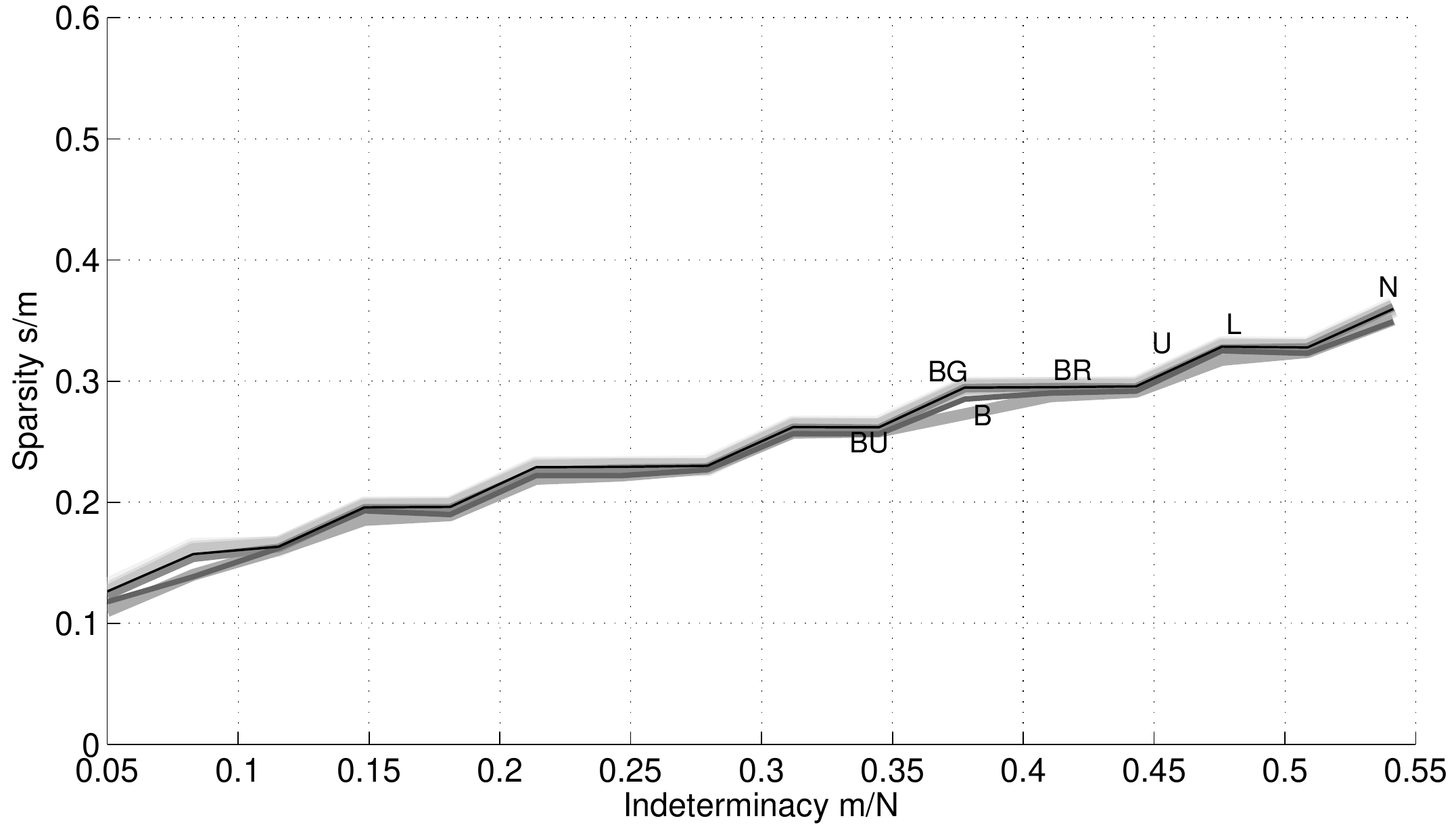}}\\ \vspace{-0.1in}

\subfigure[Recommended IST]{
\includegraphics[width=0.49\textwidth]{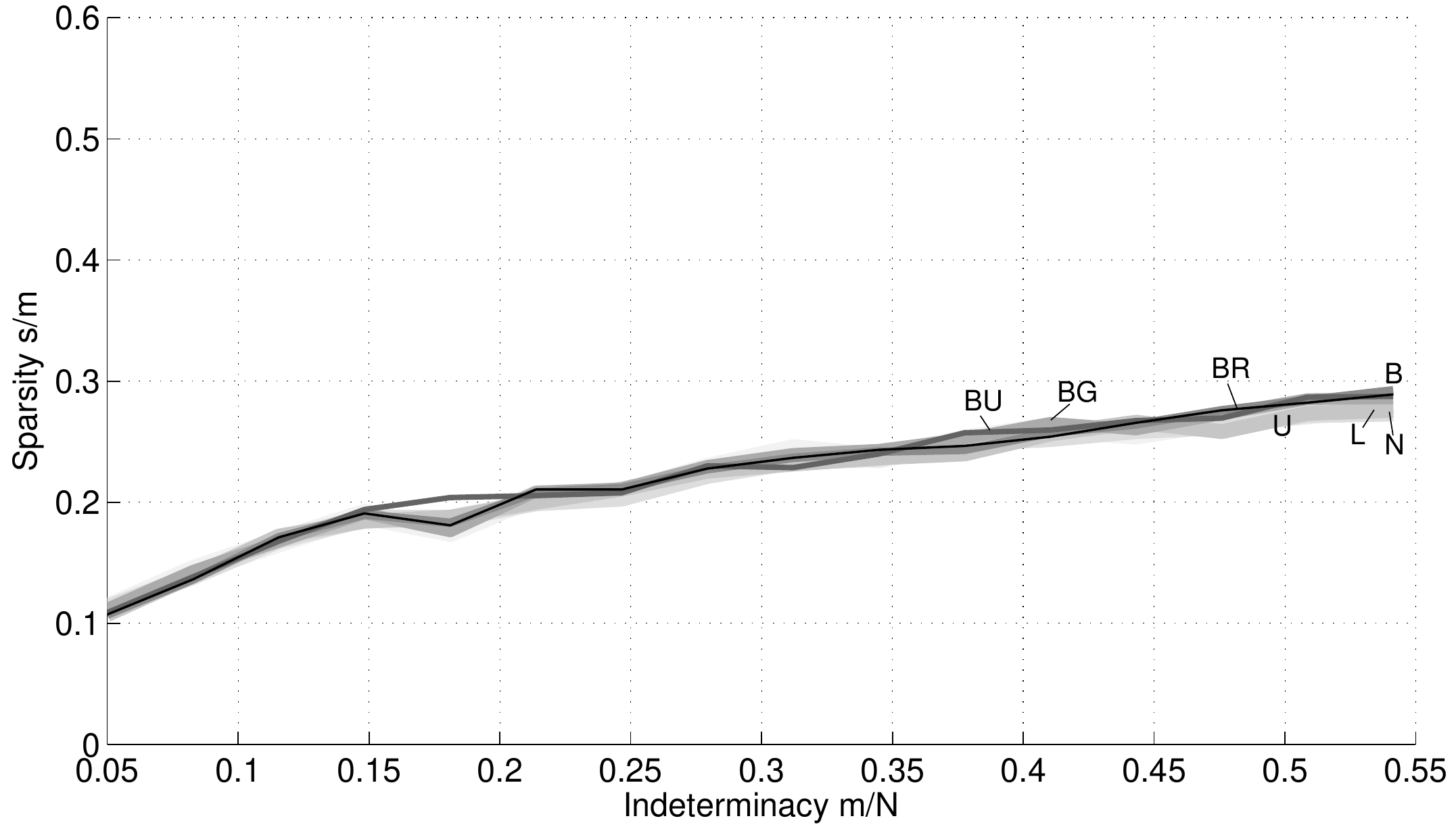}}\hspace{-0.1in}
\subfigure[IRl1]{
\includegraphics[width= 0.49\textwidth]{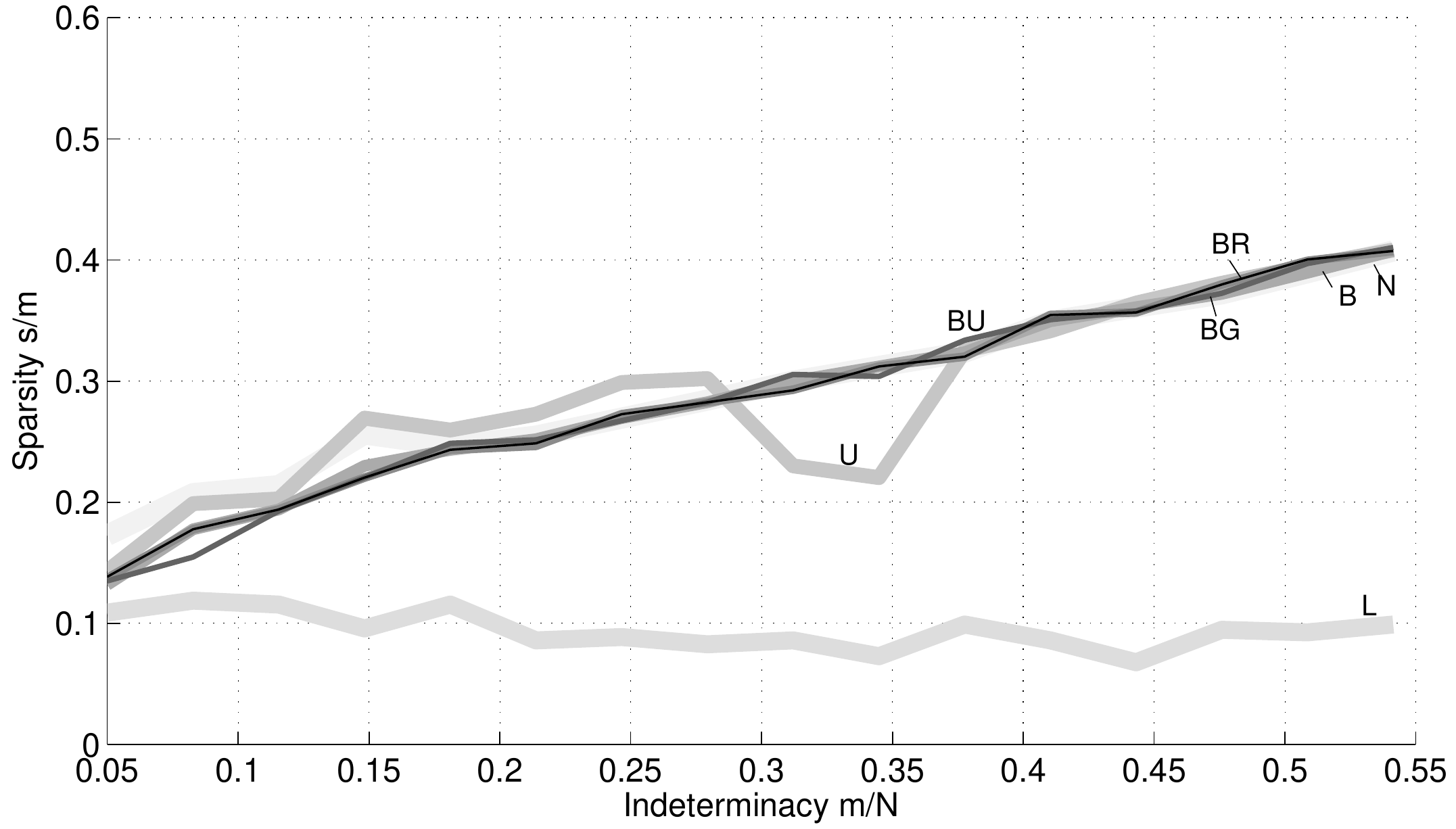}}\\ \vspace{-0.1in}

\subfigure[GPSR]{
\includegraphics[width=0.49\textwidth]{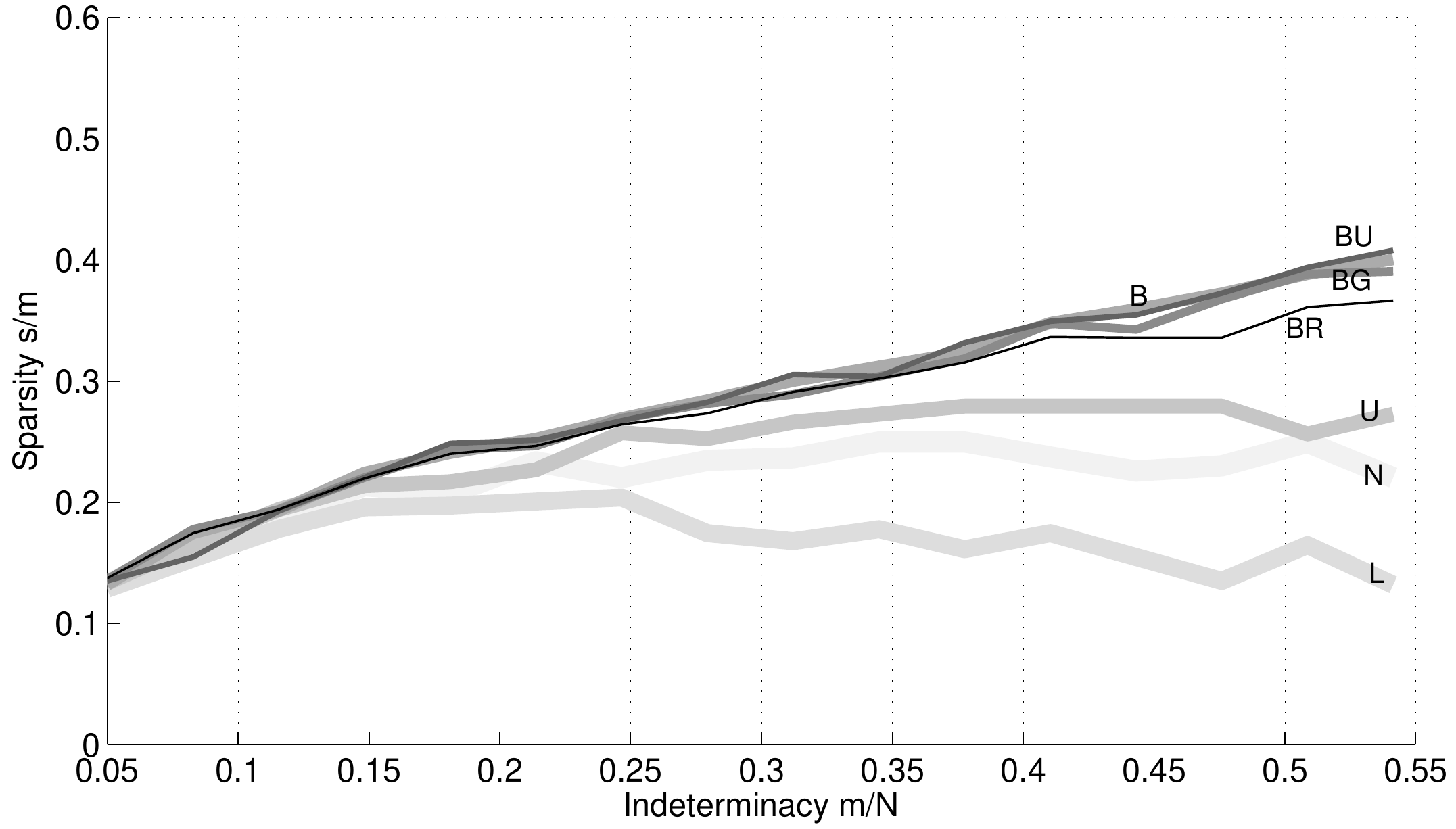}}\hspace{-0.1in}
\subfigure[ROMP]{
\includegraphics[width=0.49\textwidth]{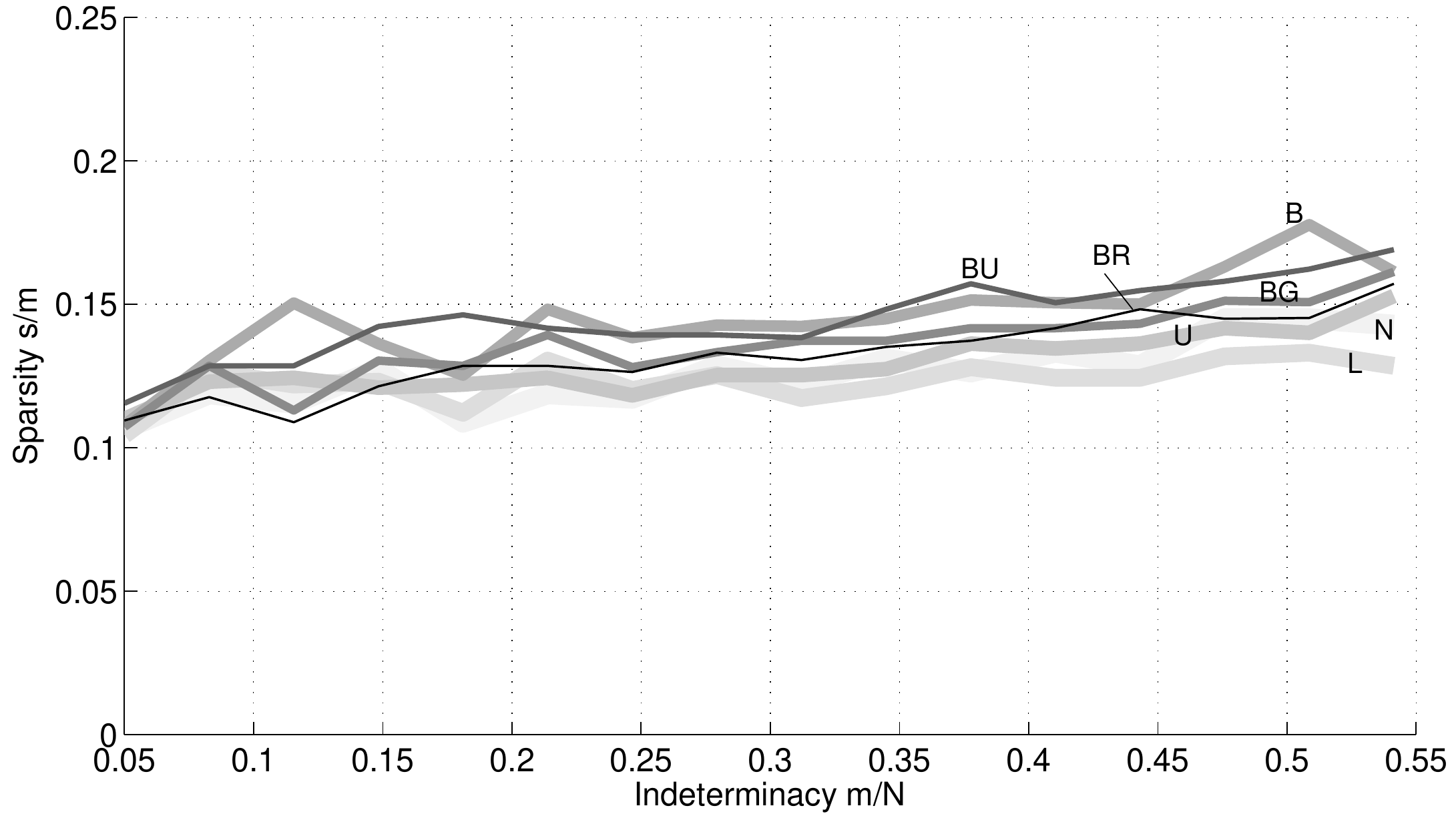}}\\ \vspace{-0.1in}

\subfigure[Recommended IHT]{
\includegraphics[width= 0.49\textwidth]{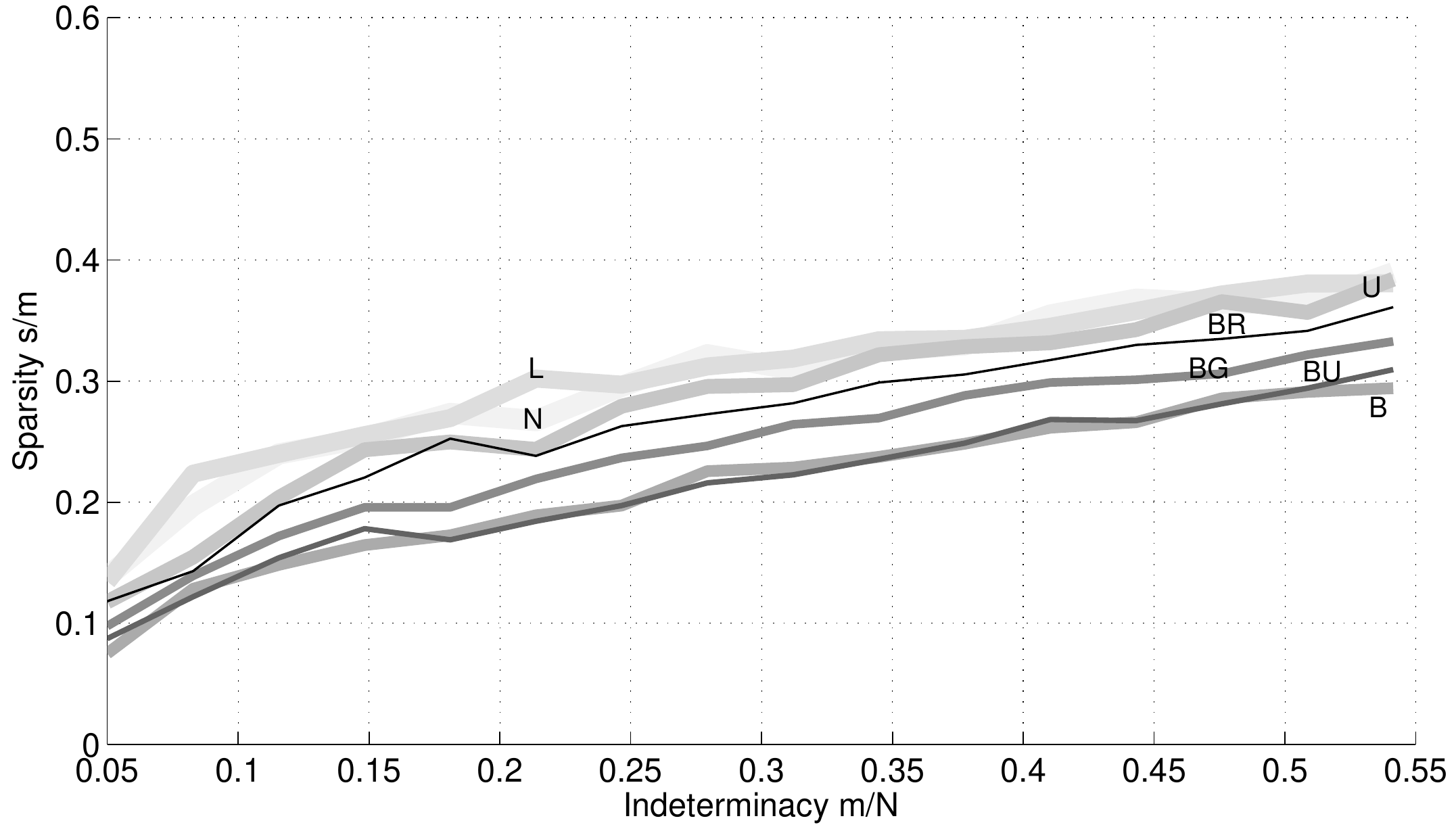}}\hspace{-0.1in}
\subfigure[ALPS]{
\includegraphics[width= 0.49\textwidth]{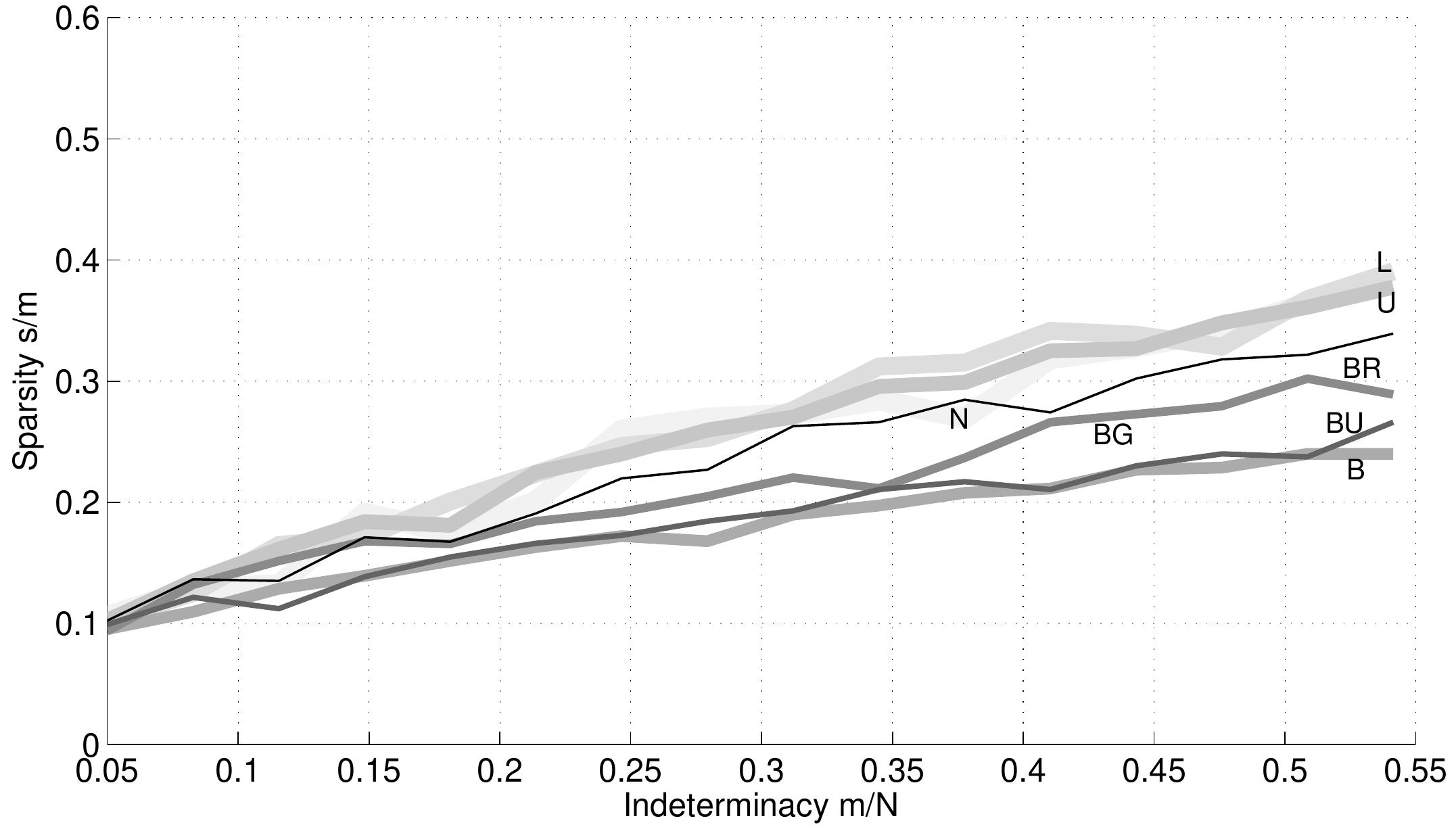}}

\caption{Empirical phase transitions using criterion (\ref{eq:successcriterion2}) 
of nine recovery algorithms for a variety of sparse vector distributions:
Normal (N), Laplacian (L), Uniform (U), Bernoulli (B),
Bimodal Gaussian (BG), Bimodal Uniform (BU), Bimodal Rayleigh (BR).
Note different y-scaling for ROMP in (f).}
\label{fig:phasevsdistributions1}
\end{figure}

\begin{figure}[tb]
\centering
\subfigure[StOMP]{
\includegraphics[width=0.49\textwidth]{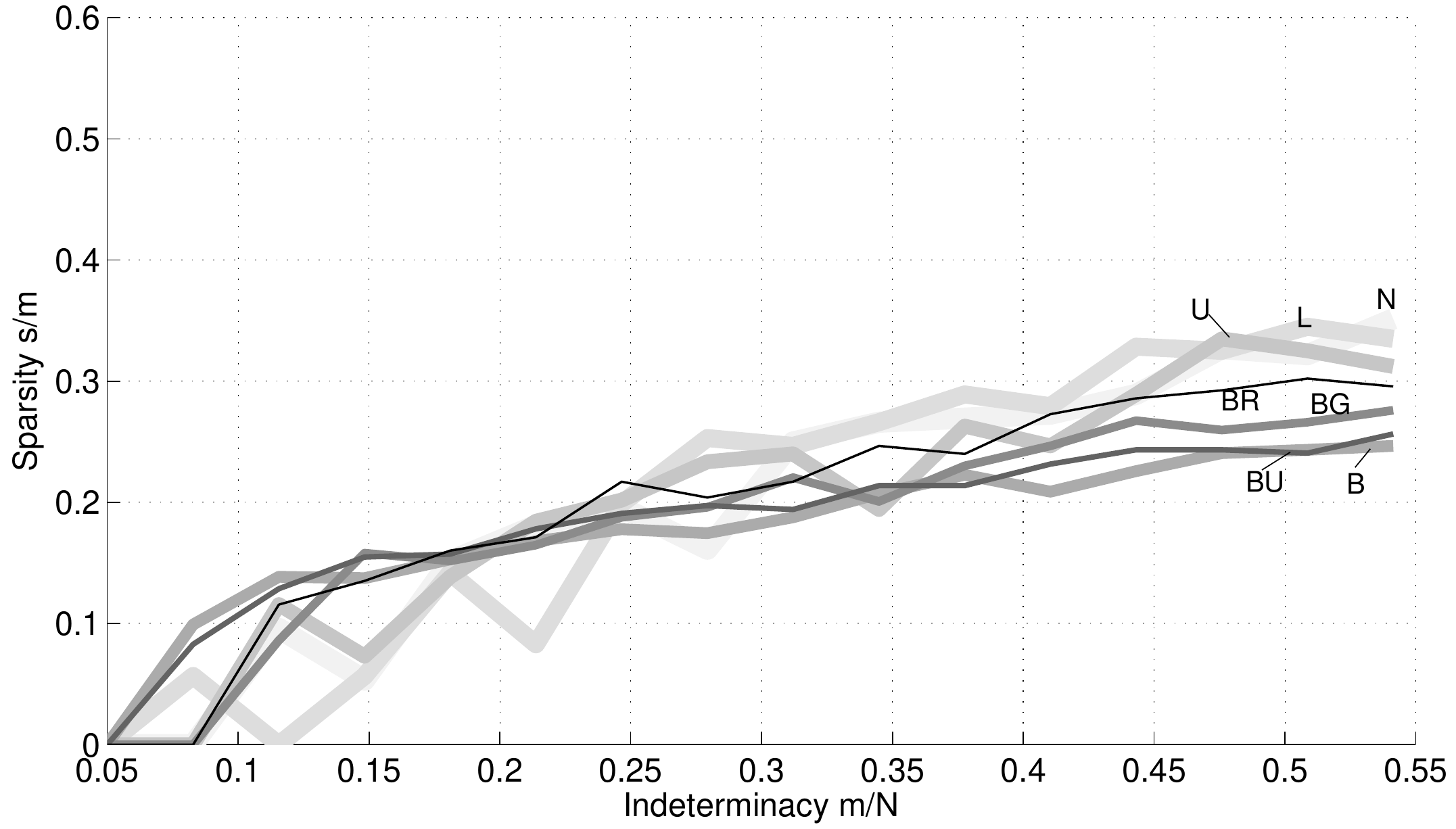}}\hspace{-0.1in}
\subfigure[CoSaMP]{
\includegraphics[width= 0.49\textwidth]{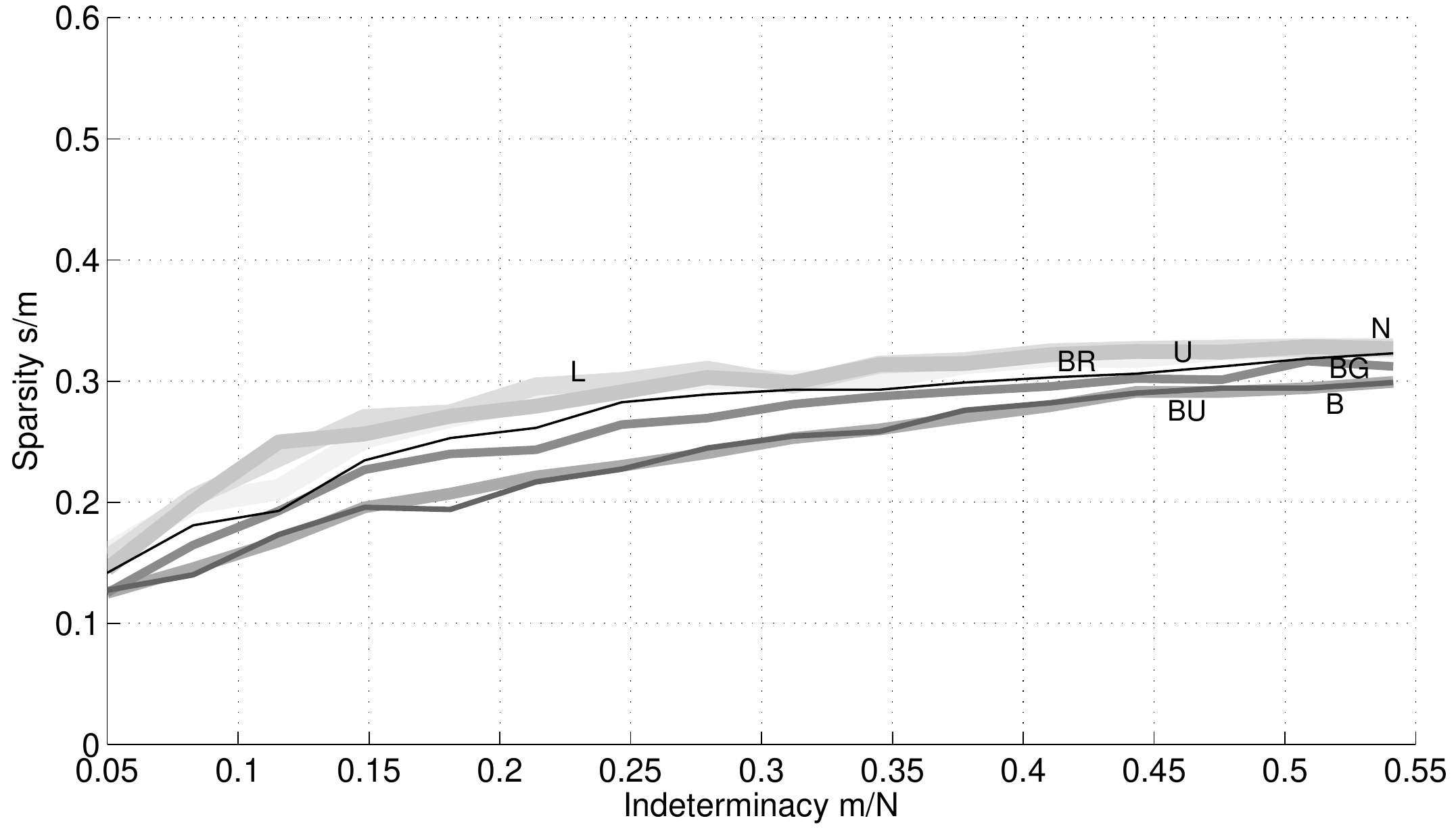}}\\ \vspace{-0.1in}

\subfigure[SP]{
\includegraphics[width=0.49\textwidth]{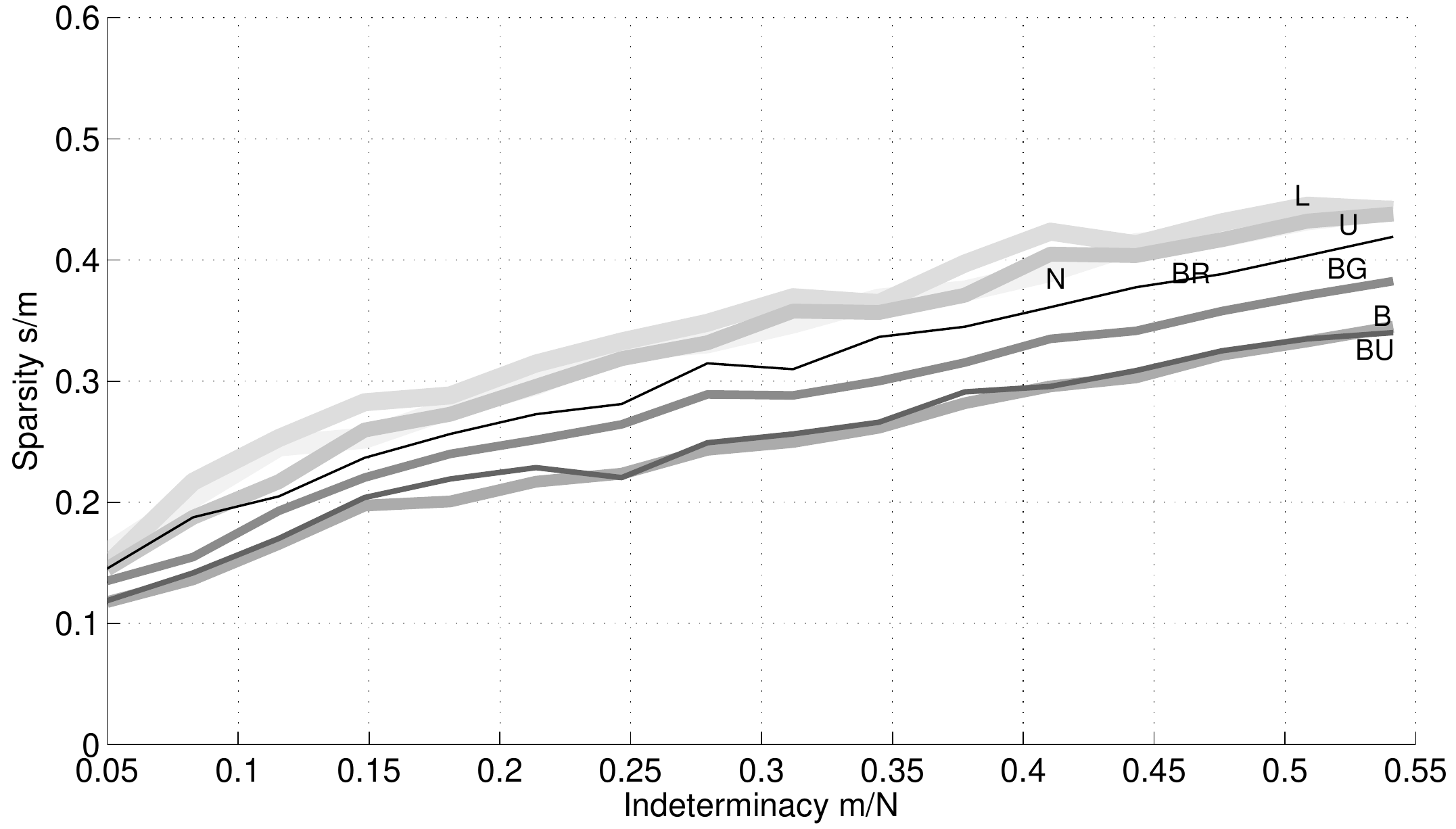}} \hspace{-0.1in}
\subfigure[OMP]{
\includegraphics[width= 0.49\textwidth]{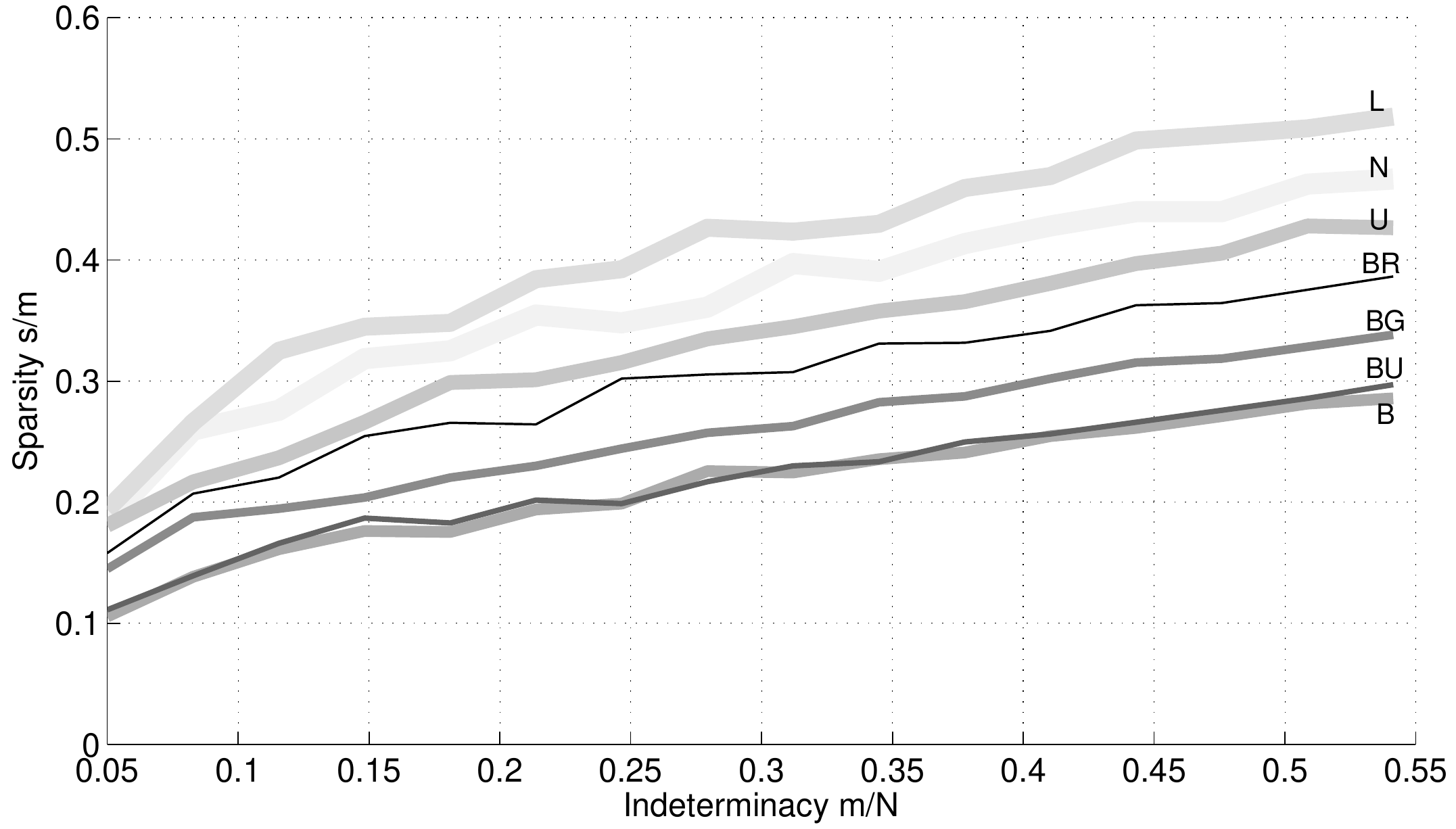}} \\ \vspace{-0.1in} 

\subfigure[PrOMP]{
\includegraphics[width=0.49\textwidth]{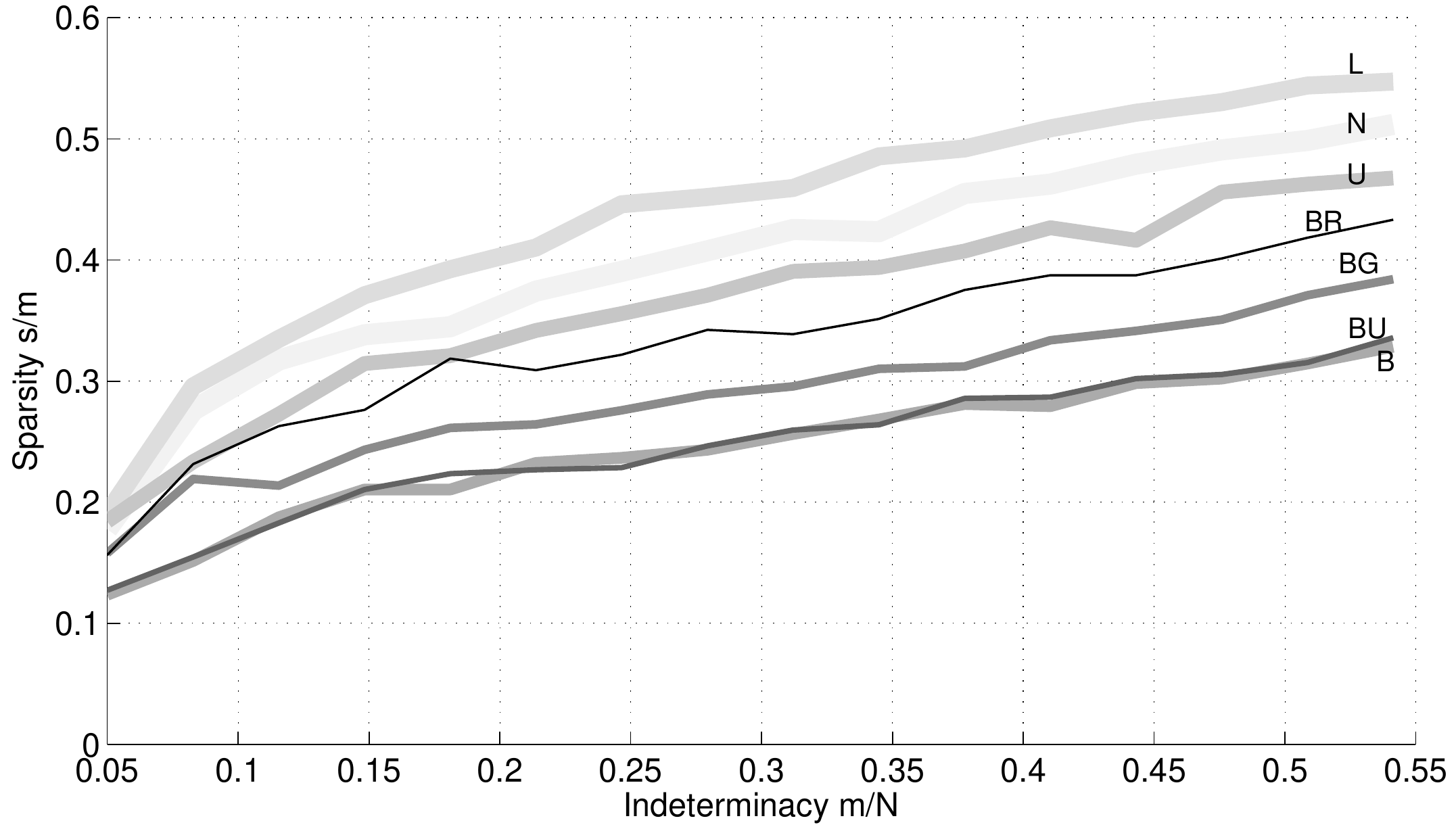}}\hspace{-0.1in}
\subfigure[SL0]{
\includegraphics[width= 0.49\textwidth]{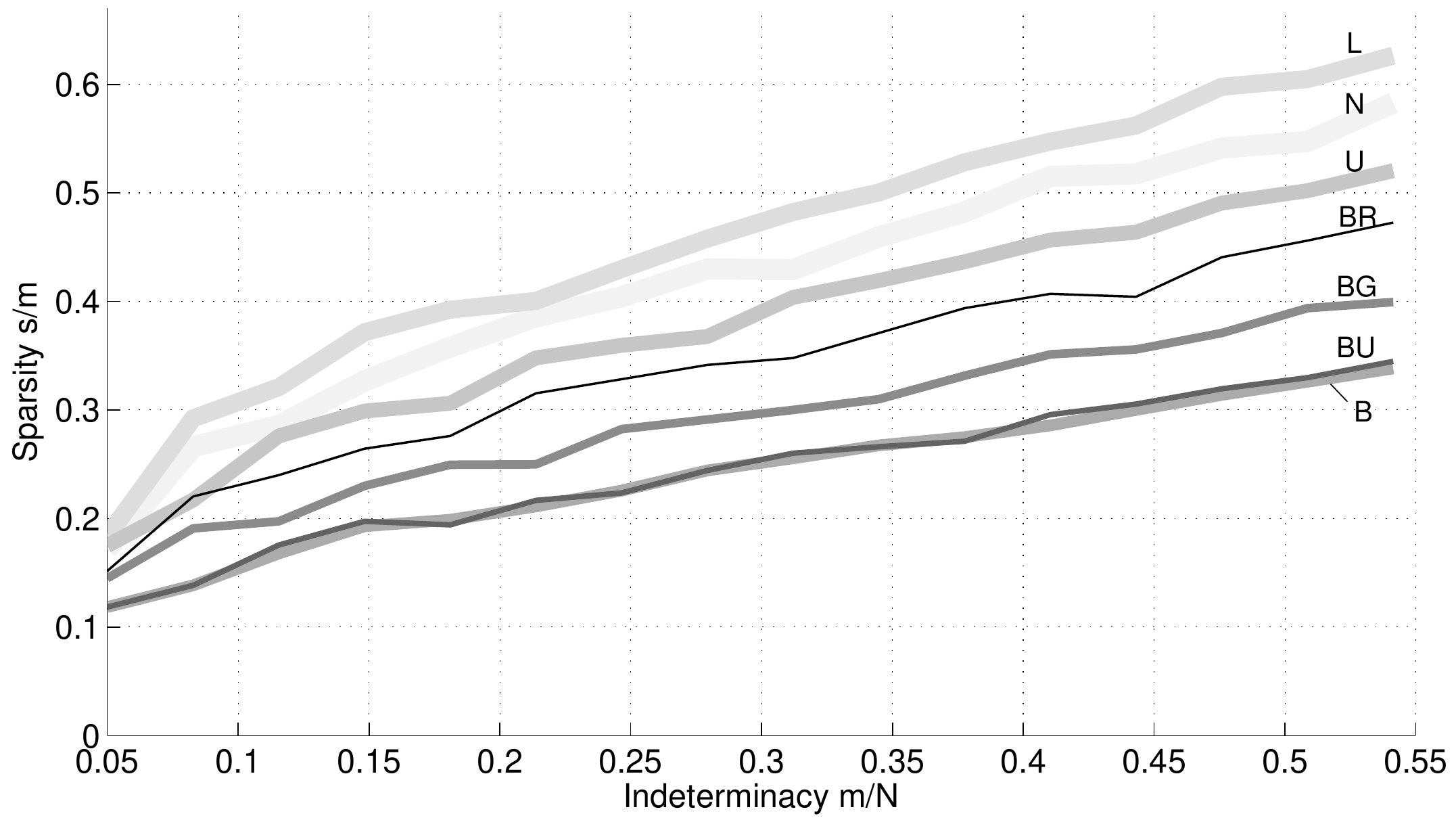}}

\caption{Empirical phase transitions using criterion (\ref{eq:successcriterion2}) 
of six recovery algorithms for a variety of sparse vector distributions:
Normal (N), Laplacian (L), Uniform (U), Bernoulli (B),
Bimodal Gaussian (BG), Bimodal Uniform (BU), Bimodal Rayleigh (BR).
Note different y-scale in (f).}
\label{fig:phasevsdistributions2}
\end{figure}

Figures \ref{fig:phasevsdistributions1} and \ref{fig:phasevsdistributions2}
show the variability of the empirical phase transitions for each algorithm
depending on the sparse vector distributions.
The empirical phase transitions of BP and AMP are exactly the same,
and so I only show one in Figure \ref{fig:phasevsdistributions1}(a).
Clearly, the performance of BP, AMP, and recommended TST and IST, 
appear extremely robust  to the distribution underlying sparse vectors.
In \cite{Donoho2009}, Donoho et al., prove AMP to have this robustness.
ROMP also shows a robustness,
but across all indeterminacies its performance
is relatively poor (note the difference in y-axis scaling).
IRl1 and GPSR shows the same robustness to 
Bernoulli and the bimodal distributions;
but they both show a surprisingly poor performance 
for Laplacian distributed vectors, even as the number of measurements increase.
GPSR appears to have poorer empirical phase transitions
as the probability density around zero increases.
I do not yet know why IRl1 fails so poorly for Laplacian vectors.
The specific results for recommended IST, IHT, and TST however, appear very different
to those reported in \cite{Maleki2010},
though the main findings of their work comport with what I show here.

The recovery performance of OMP, PrOMP, and SL0 
varies to the largest degree of all algorithms that I test.
It is clear that these algorithms perform in proportion to
the probability density around zero.
In fact, for eight of the algorithms I test (IHT, ALPS, StOMP, CoSaMP,
SP, OMP, PrOMP, and SL0) we can predict the order of performance
for each distribution by the amount of probability density 
concentrated near zero.
From Fig. \ref{fig:empiricalPDFs} we can see that
these are, in order from most to least concentrated:
Laplacian, Normal, Uniform, 
Bimodal Rayleigh, Bimodal Gaussian, 
Bimodal uniform, and Bernoulli.
This behavior is reversed for only recommended IST, GPSR, and ROMP,
where their performance increases 
the less concentrated probability density is around zero.

\clearpage

\subsection{Comparison of Recovery Algorithms for Each Distribution}
\begin{figure}[htb]
\centering
\subfigure[Bernoulli]{
\includegraphics[width=0.49\textwidth]{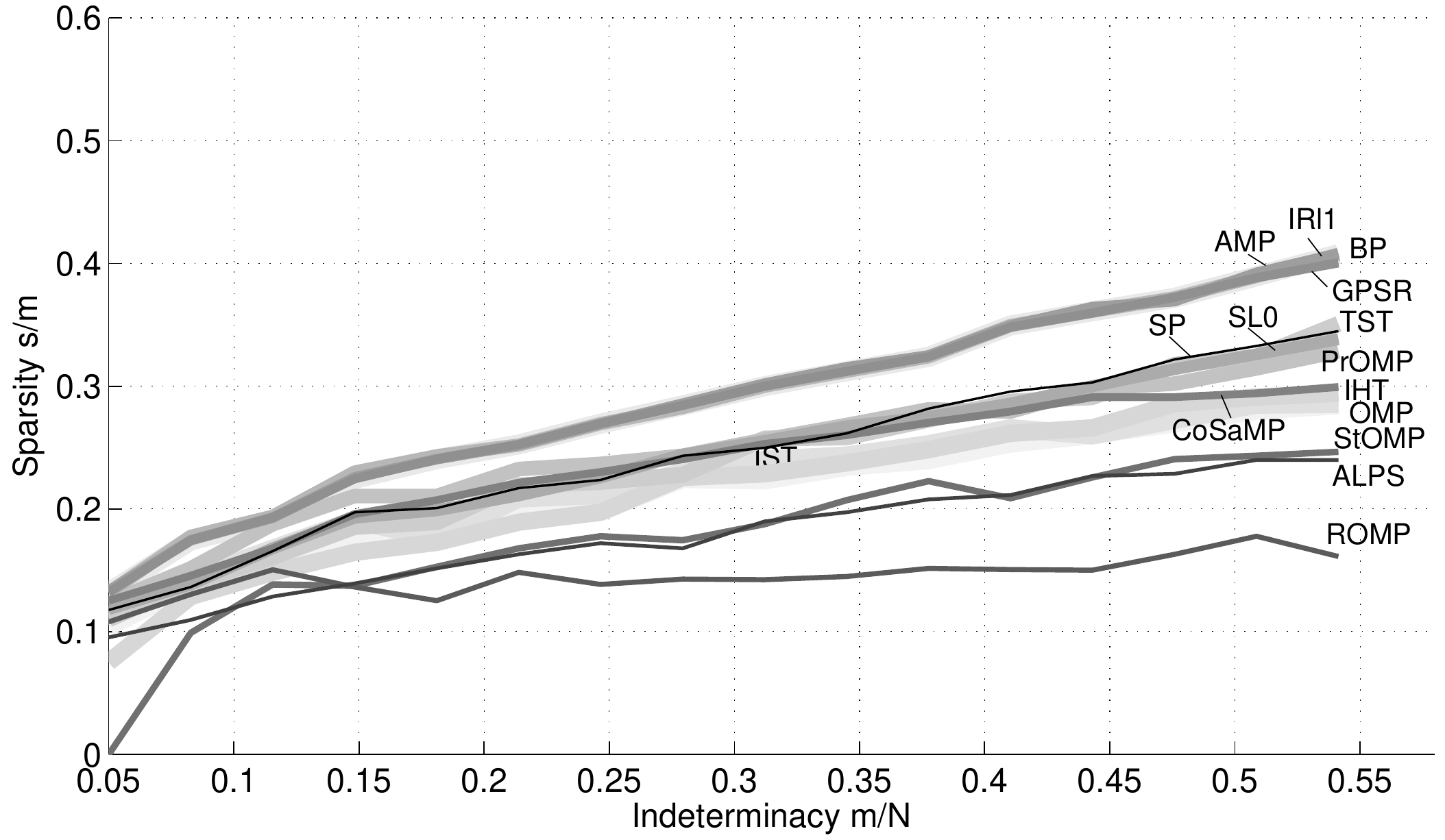}}\hspace{-0.1in}
\subfigure[Bimodal Uniform]{
\includegraphics[width=0.49\textwidth]{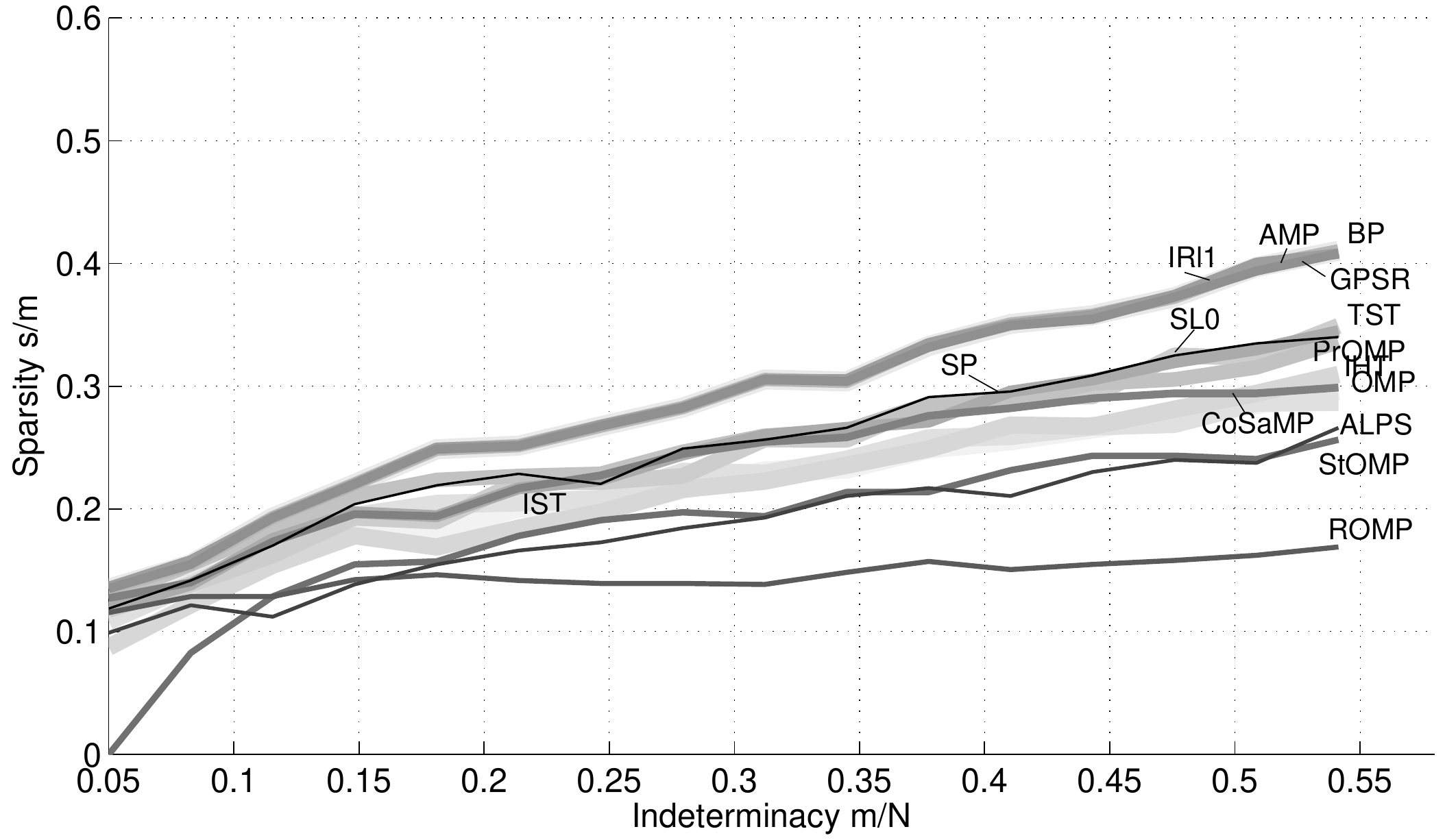}}\\ \vspace{-0.1in}

\subfigure[Bimodal Gaussian]{
\includegraphics[width=0.505\textwidth]{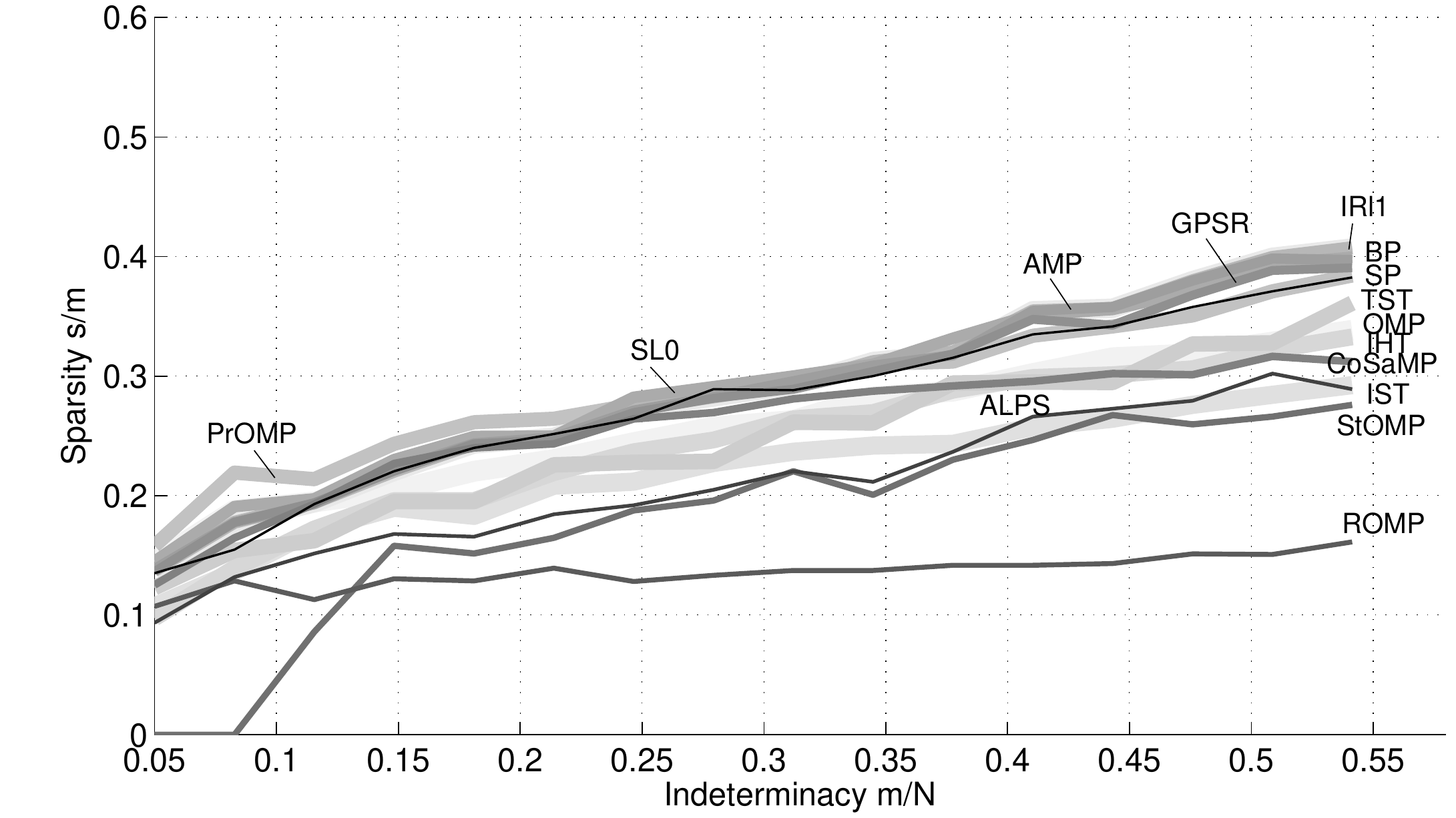}}\hspace{-0.1in} 
\subfigure[Bimodal Rayleigh]{
\includegraphics[width=0.485\textwidth]{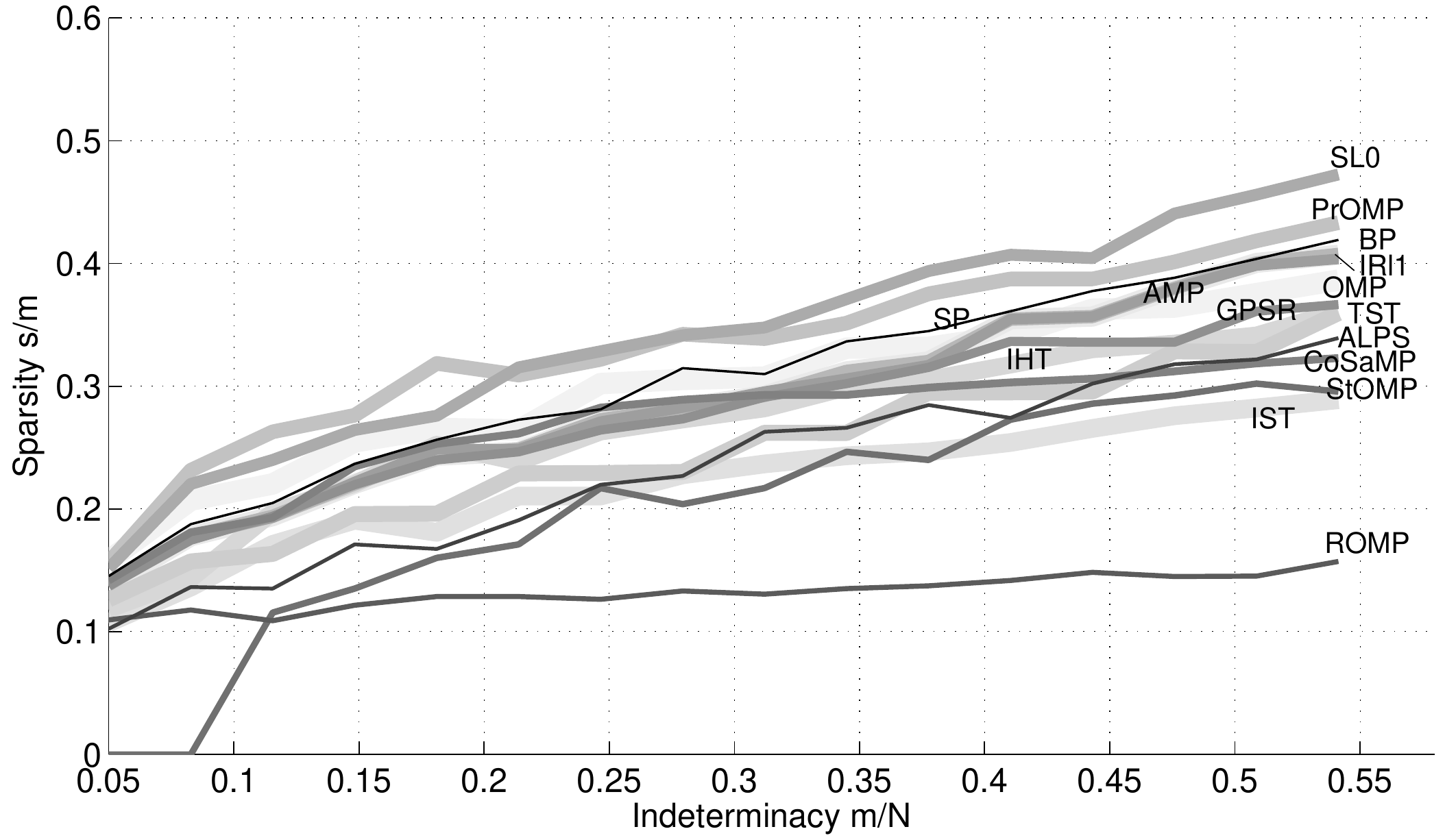}}\\ \vspace{-0.1in}

\subfigure[Uniform]{
\includegraphics[width=0.485\textwidth]{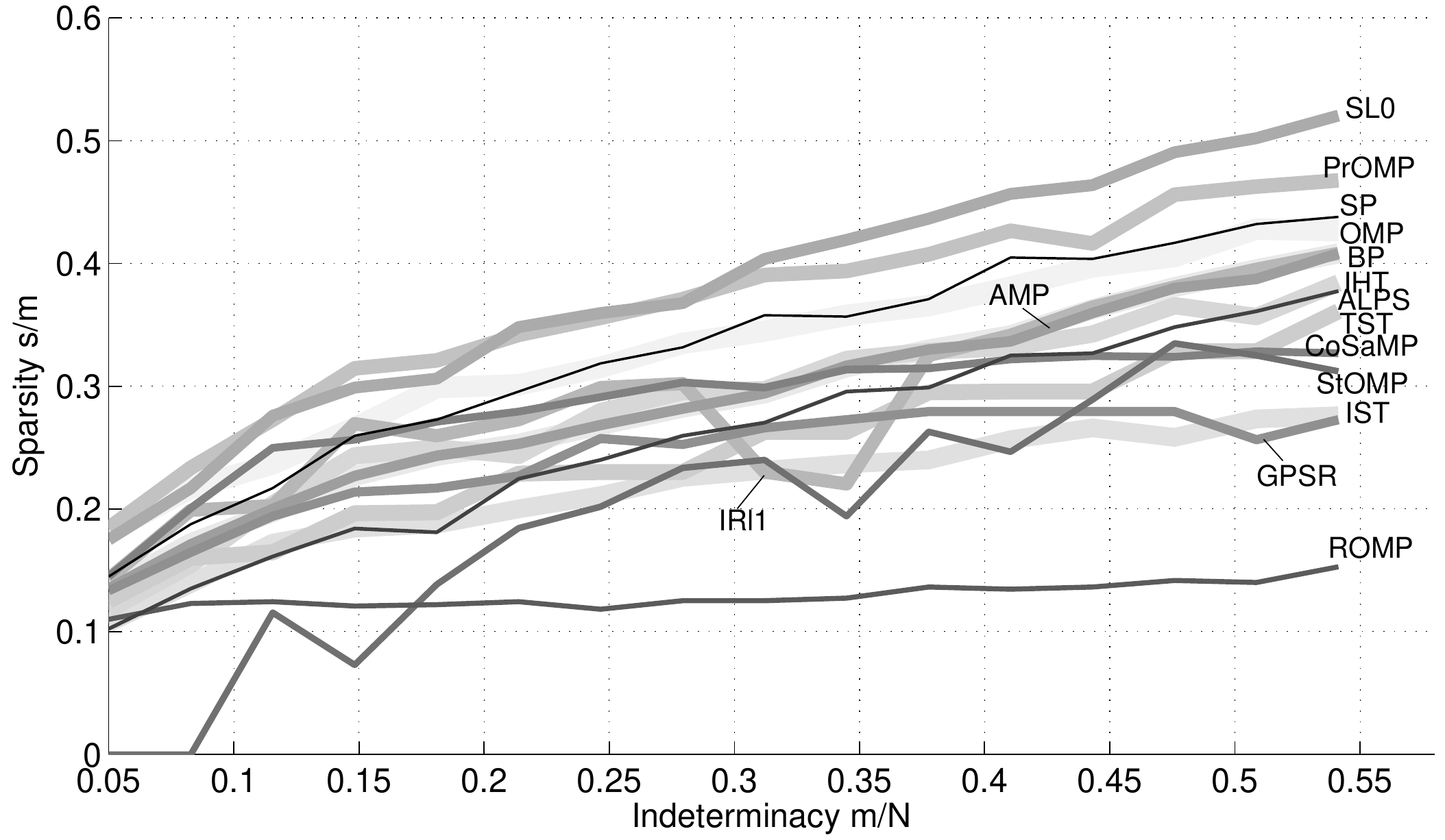}}\hspace{-0.1in}
\subfigure[Normal]{
\includegraphics[width=0.50\textwidth]{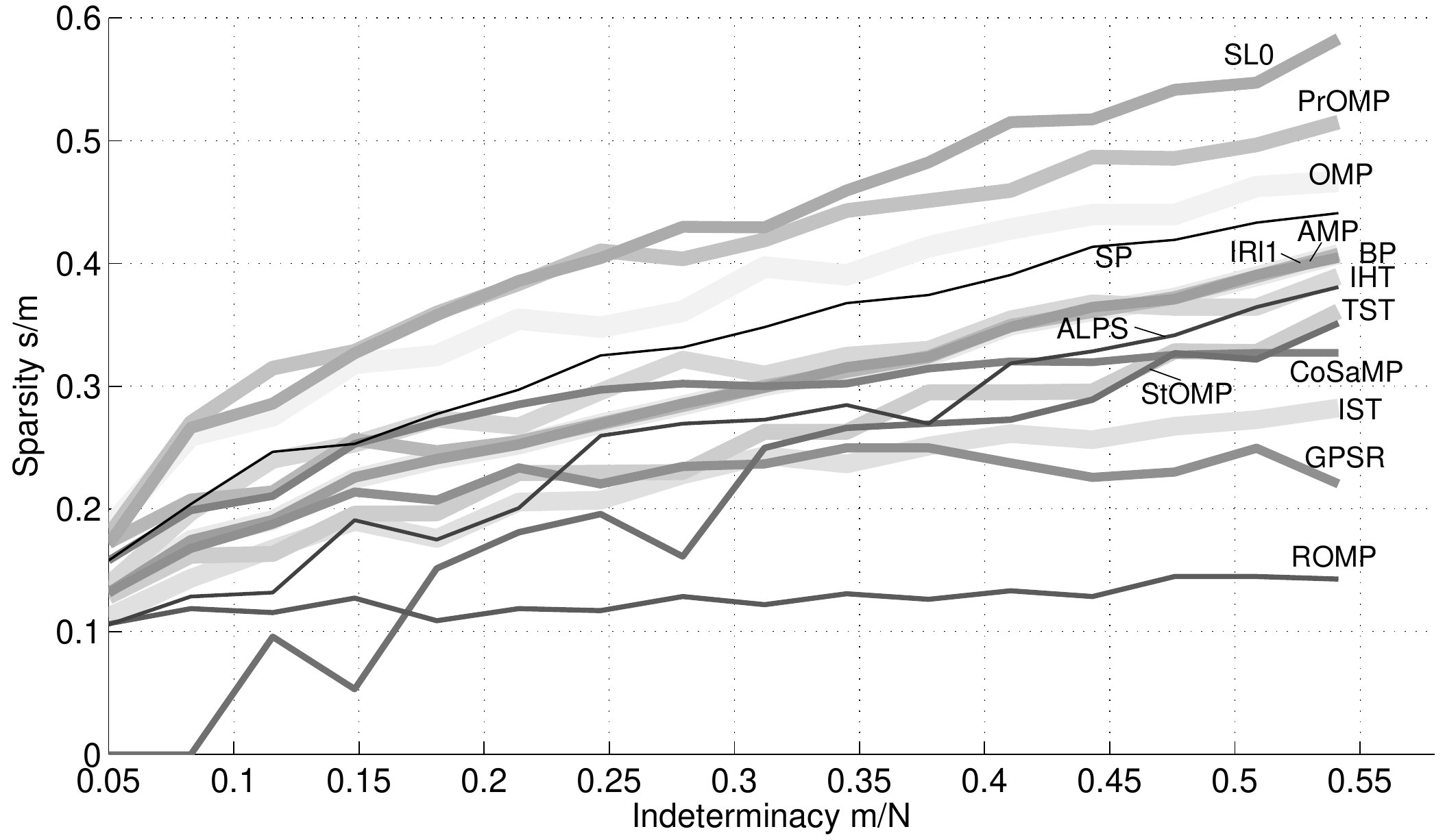}}\\ \vspace{-0.1in}

\subfigure[Laplacian]{
\includegraphics[width=0.49\textwidth]{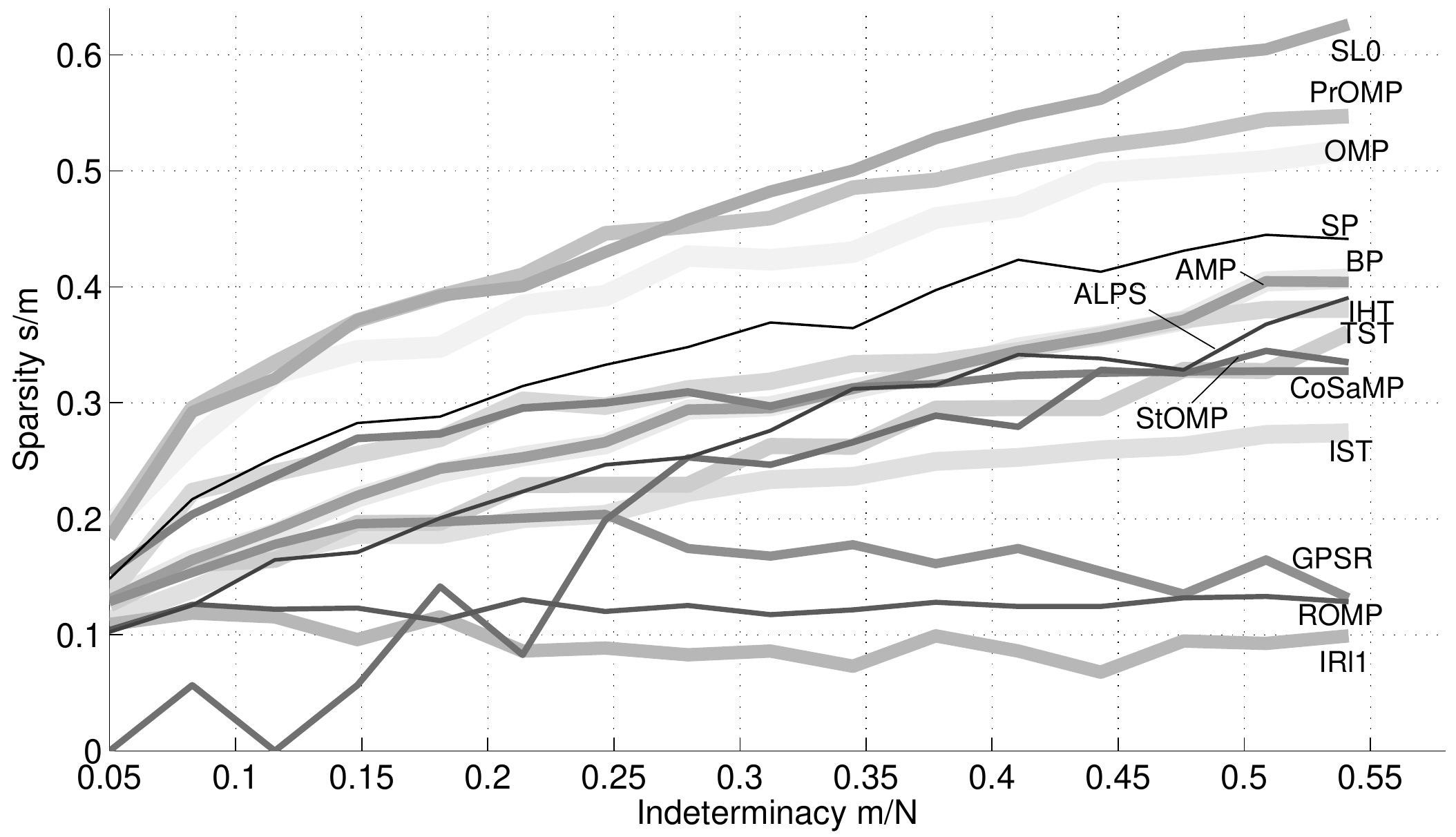}}

\caption{Comparison of phase transitions of
all algorithms I test for each distribution.
Note different y-scale in (g).}
\label{fig:phasevsalgorithms}
\end{figure}

Figure \ref{fig:phasevsalgorithms} shows the same information as
Figs. \ref{fig:phasevsdistributions1} and \ref{fig:phasevsdistributions2},
but compares all fifteen algorithms together for single distributions.
Here we can see that for sparse vectors distributed Bernoulli,
bimodal uniform, and bimodal Gaussian, $\ell_1$-minimization methods (BP, IRl1, GPSR)
and the thresholding approach AMP (which uses soft thresholding
to approximate $\ell_1$ minimization), perform better than all the
greedy methods and the other thresholding methods, and the majorization SL0.
For the other four distributions,
SL0, OMP and/or PrOMP outperform all the other algorithms I test,
with a significantly higher phase transition
in the case of vectors distributed Laplacian.
In every case, AMP and BP perform the same.
We can also see in all cases the phase transition for SP 
is higher than that for recommended TST, and sometimes much higher, 
even though for Maleki and Donoho's recommended algorithm
they find that the $(\alpha, \beta)$ pair that works best is that
that makes it closest in appearance to SP.
As I discuss about recommended TST above though, 
it is only similar to SP, 
and is not guaranteed to behave the same.
Furthermore, and most importantly,
in my simulations I provide SP (as well as CoSaMP, ROMP and ALPS)
with the exact sparsity of the sensed signal,
while recommended TST instead estimates it.
Finally, in the case of sparse vectors distributed Laplacian, Normal, and Uniform,
we can see that recommended IHT performs better than recommended TST,
and sometimes even BP (Normal and Laplacian), 
which is different from how Maleki and Donoho orders their performance
based on recovering sparse vectors distributed Bernoulli \cite{Maleki2010}.

\begin{figure}[htb]
\centering
\includegraphics[width=0.7\textwidth]{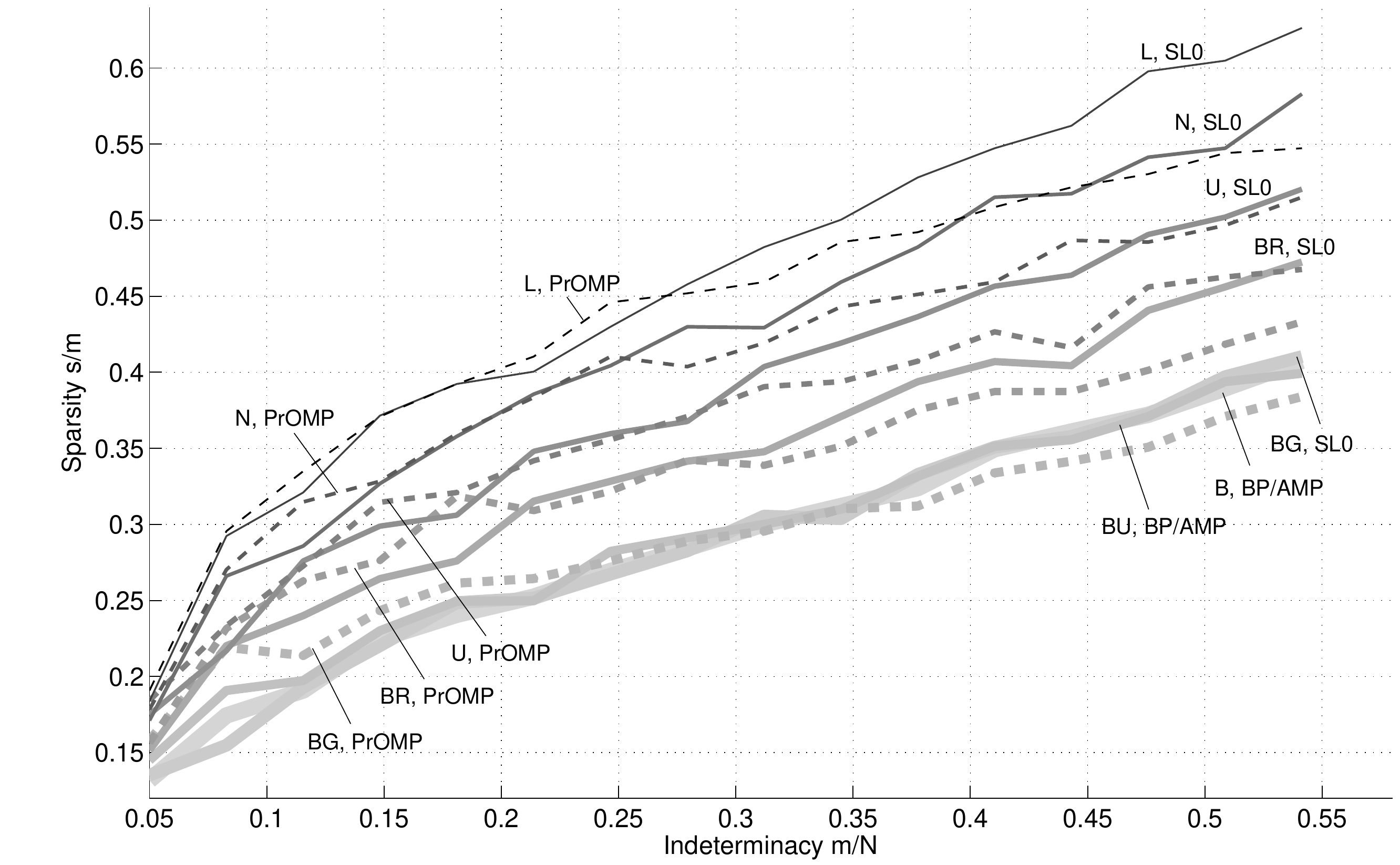}
\caption{Empirical phase transitions of the best performing algorithms
for each distribution.}
\label{fig:bestphase}
\end{figure}

Figure \ref{fig:bestphase} shows the empirical phase transitions
of the best performing algorithms for each distribution.
BP and AMP perform the same for Bernoulli and bimodal uniform distributions.
For the other five distributions,
SL0 performs the best at larger indeterminacies ($\delta > 0.2$),
and PrOMP performs better for these at smaller indeterminacies.
From this graph we also see the large difference
between the recoverability from compressive measurements
of signal distributed Laplacian and Bernoulli.

\clearpage


\subsection{Effects of Distribution on Probability of Exact Recovery}
\begin{figure}[htb]
\centering
\subfigure[Bernoulli]{
\includegraphics[width=0.49\textwidth]{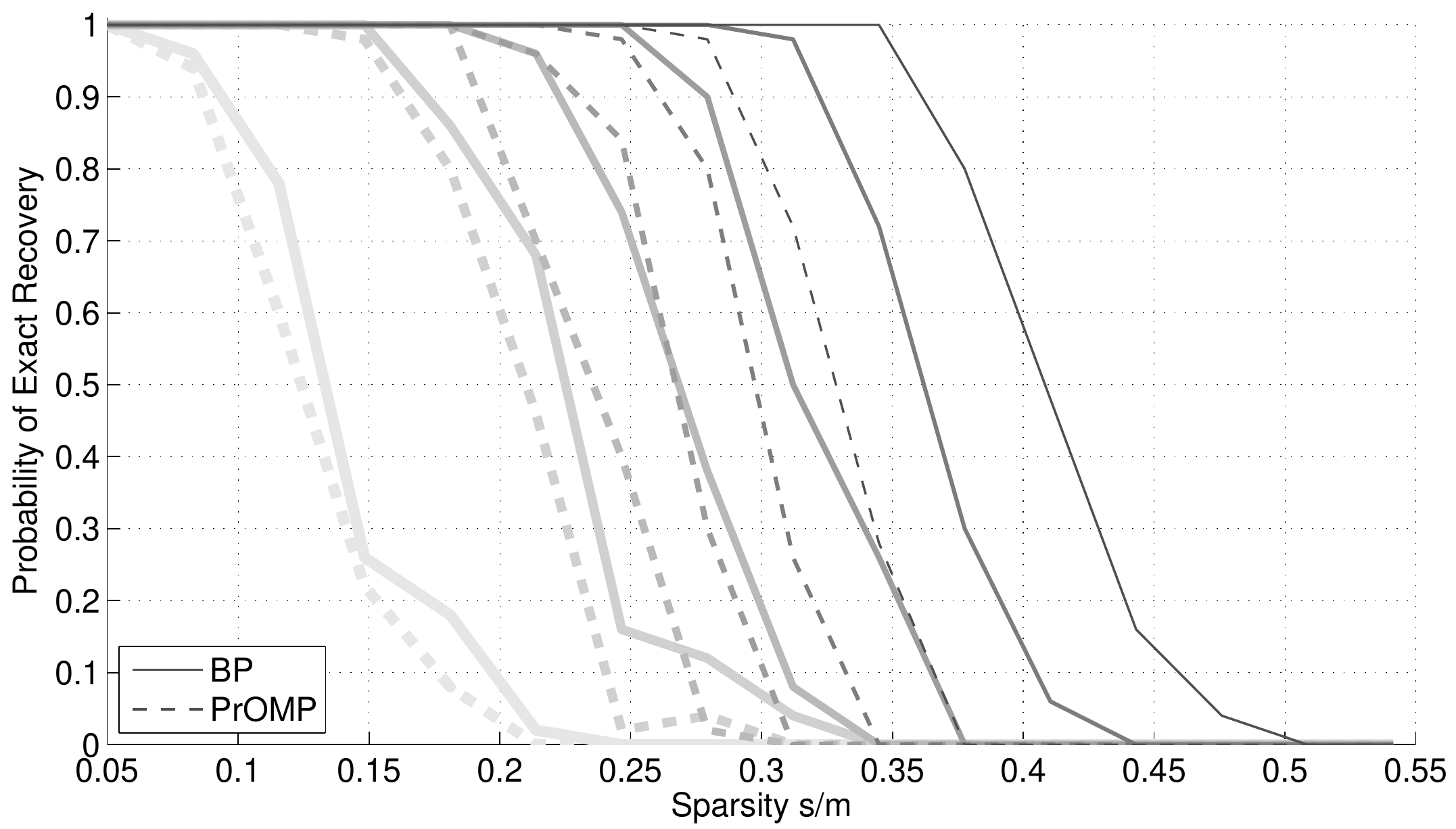}}\hspace{-0.1in}
\subfigure[Bimodal Uniform]{
\includegraphics[width=0.49\textwidth]{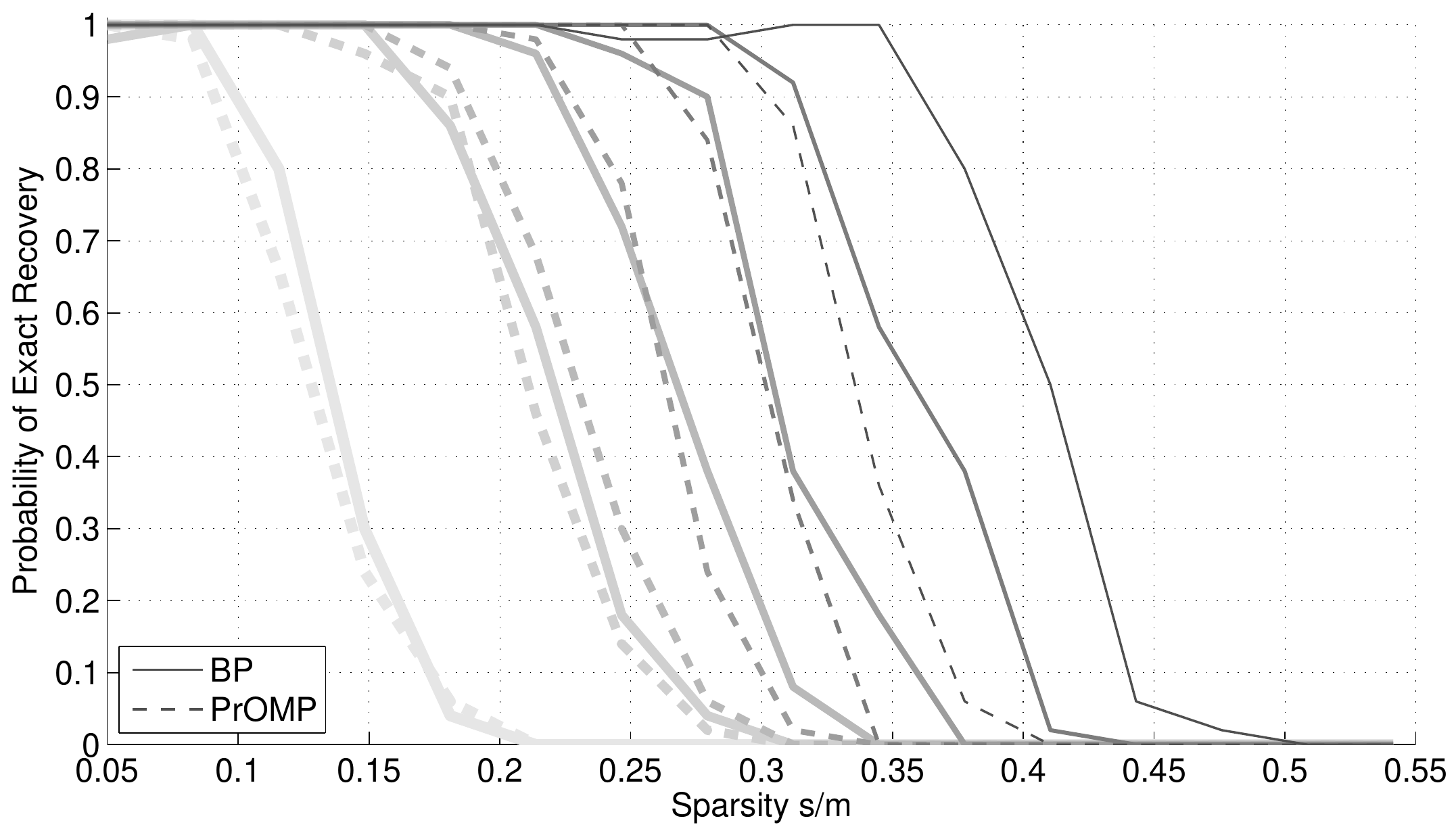}}\\ \vspace{-0.1in}

\subfigure[Bimodal Gaussian]{
\includegraphics[width=0.49\textwidth]{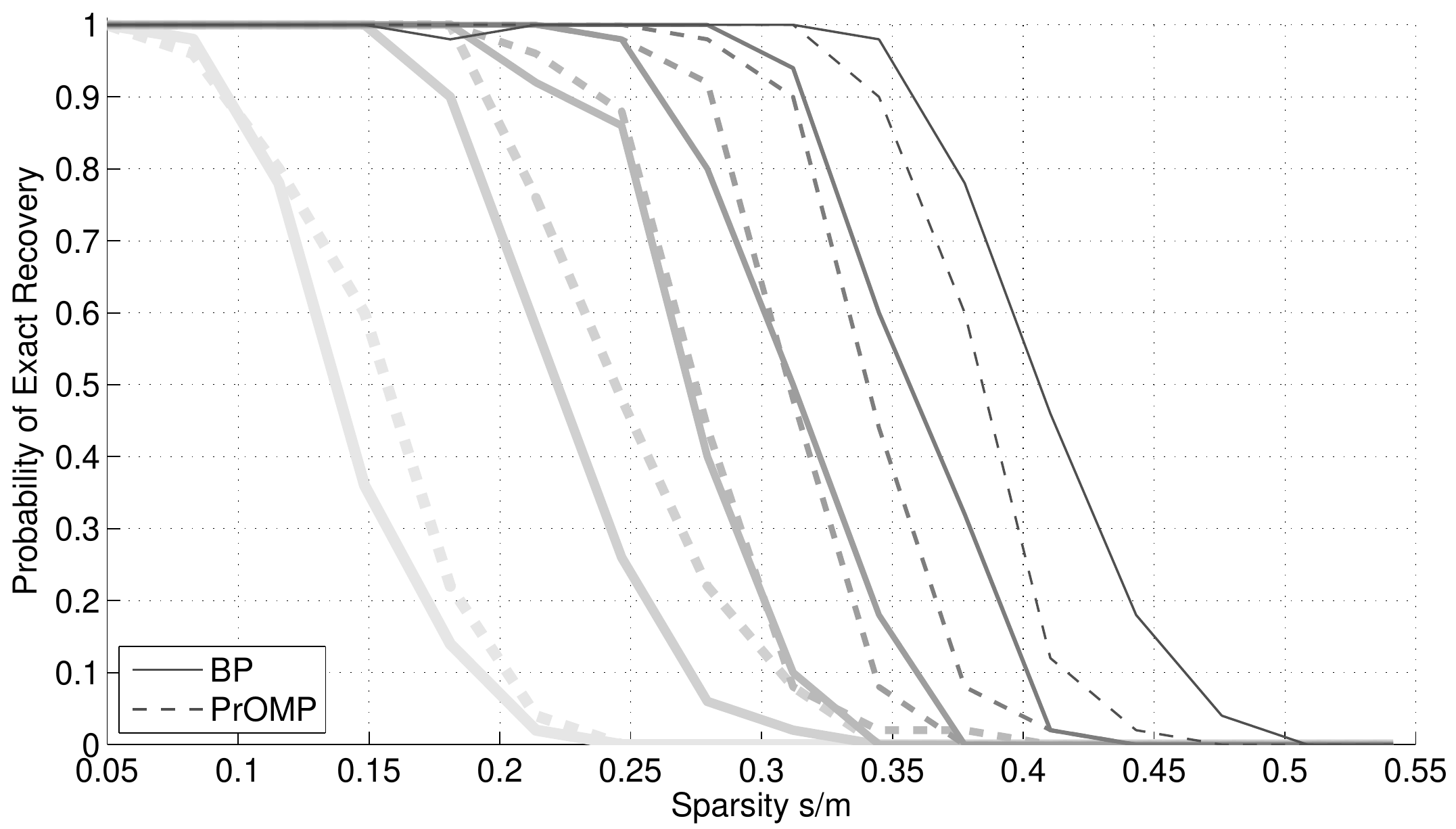}}\hspace{-0.1in}
\subfigure[Uniform]{
\includegraphics[width=0.49\textwidth]{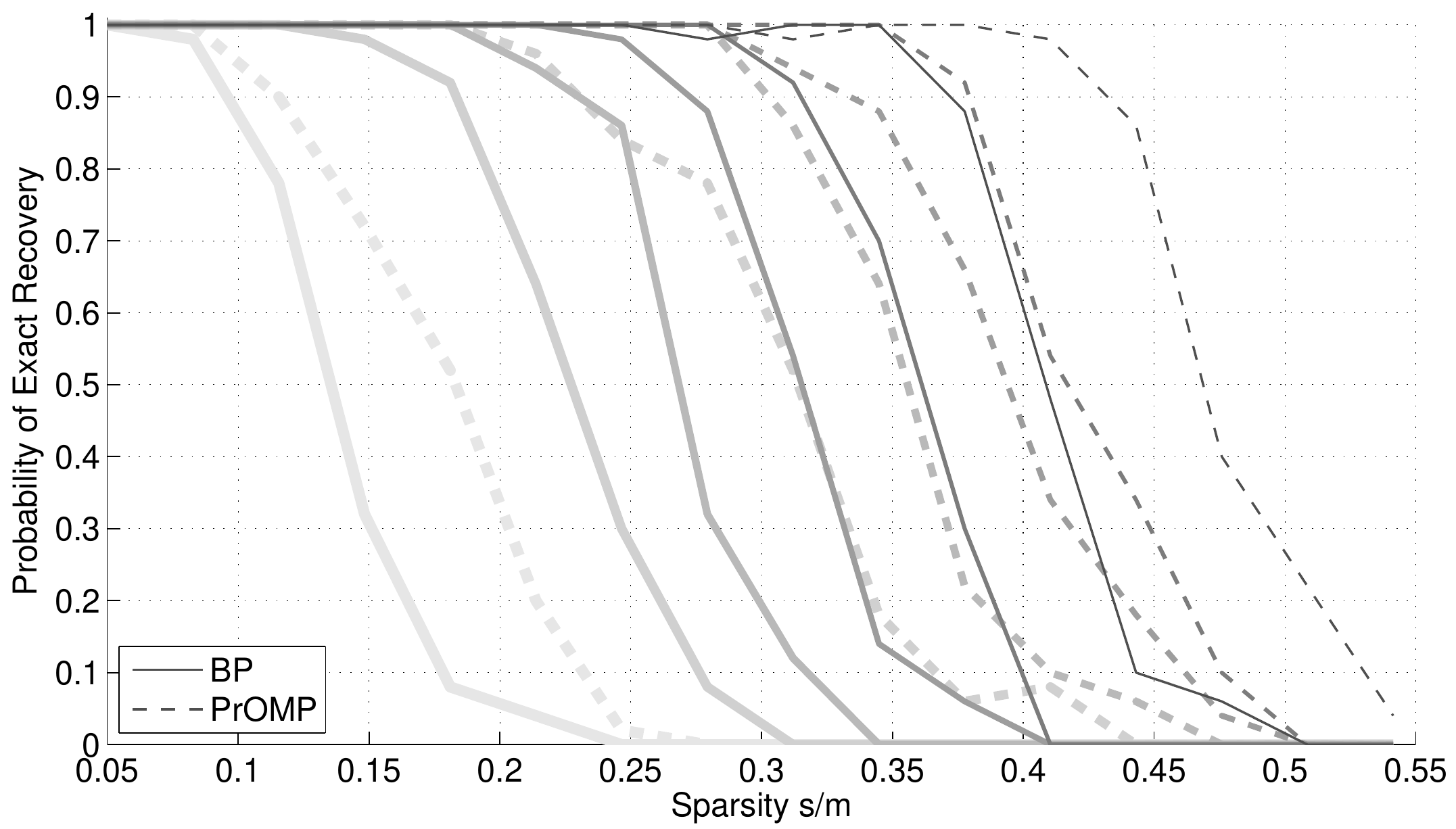}}\\ \vspace{-0.1in}

\subfigure[Bimodal Rayleigh]{
\includegraphics[width=0.49\textwidth]{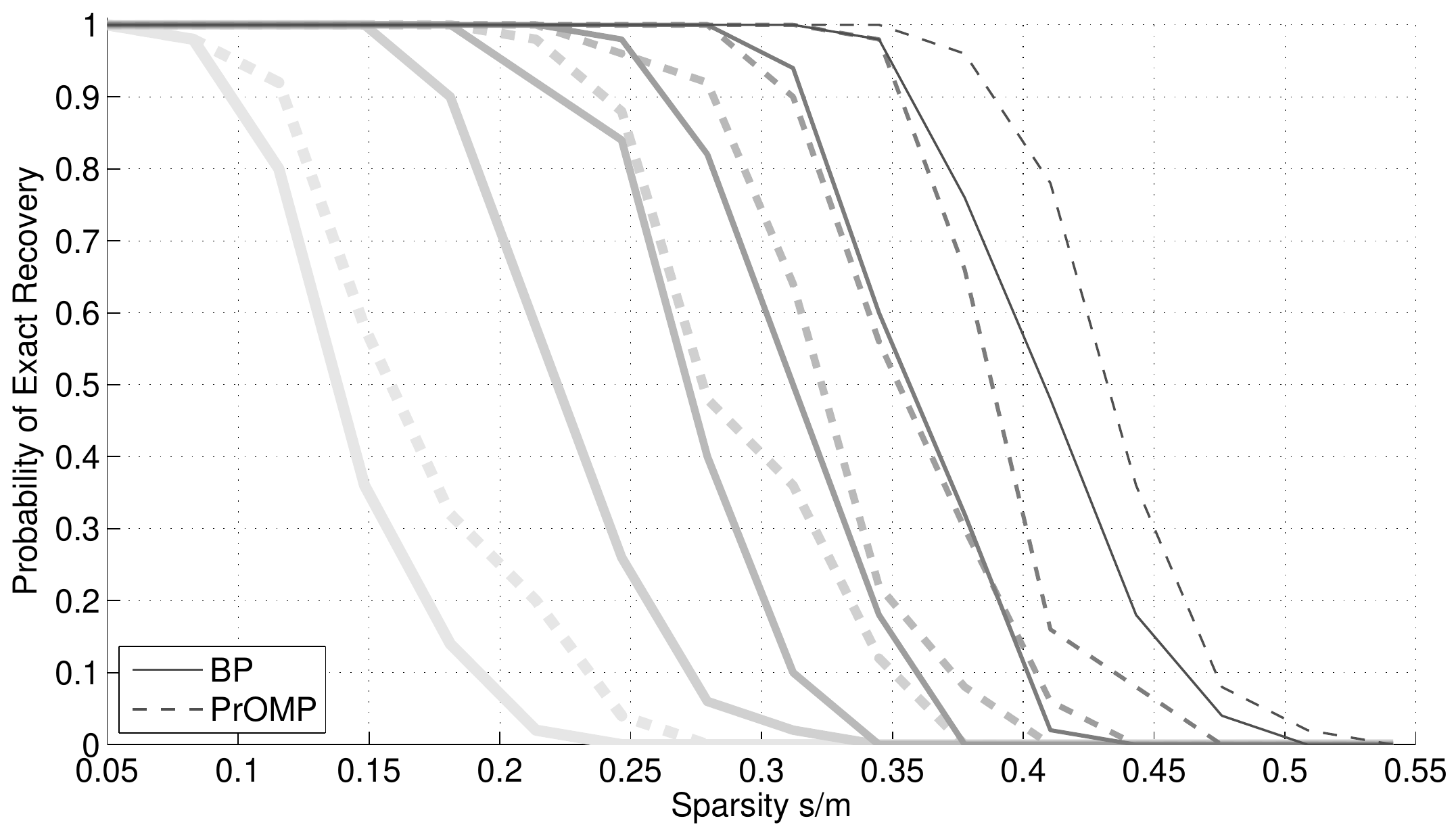}}\hspace{-0.1in}
\subfigure[Normal]{
\includegraphics[width=0.49\textwidth]{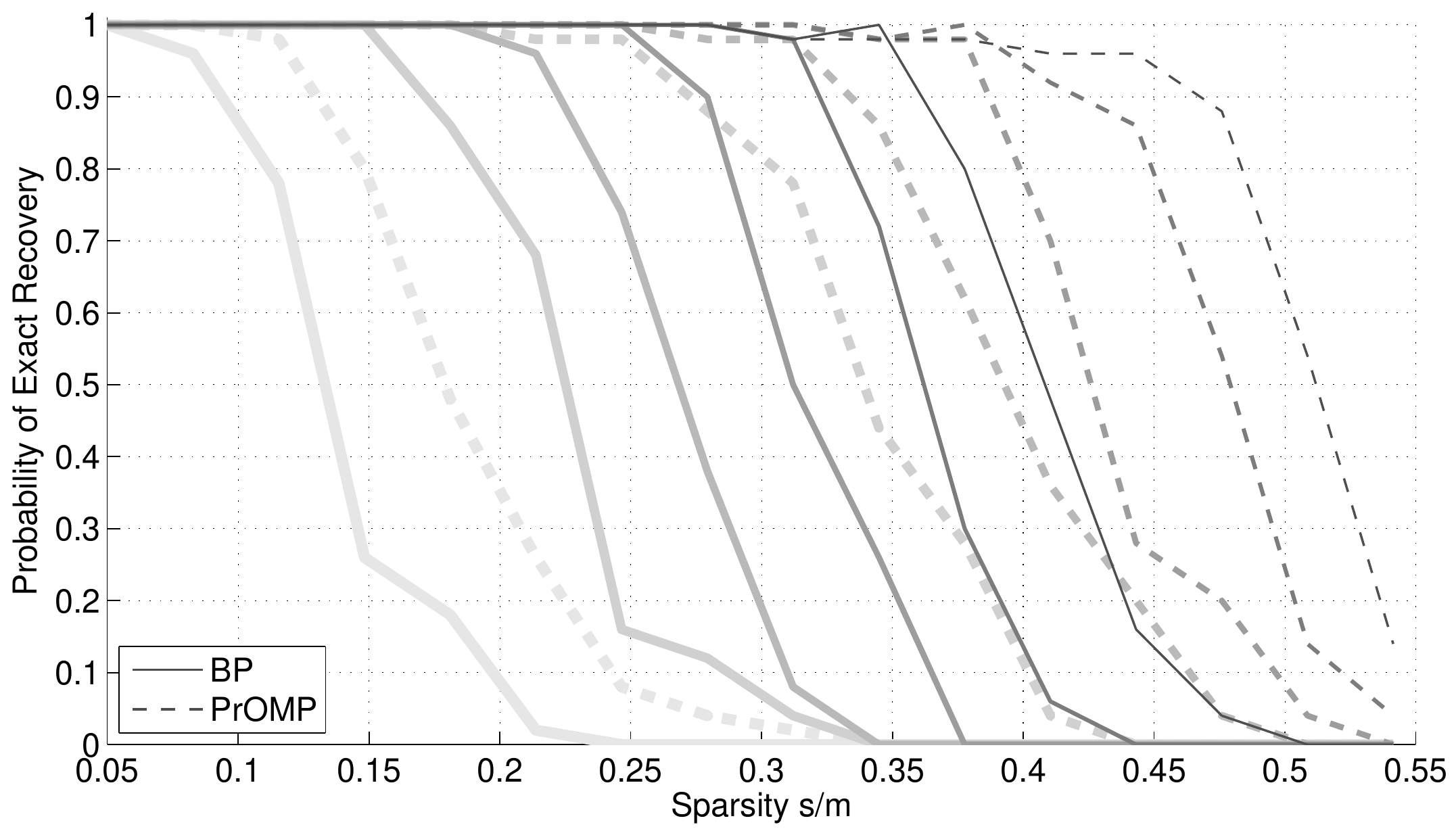}}\\ \vspace{-0.1in}

\subfigure[Laplacian]{
\includegraphics[width=0.49\textwidth]{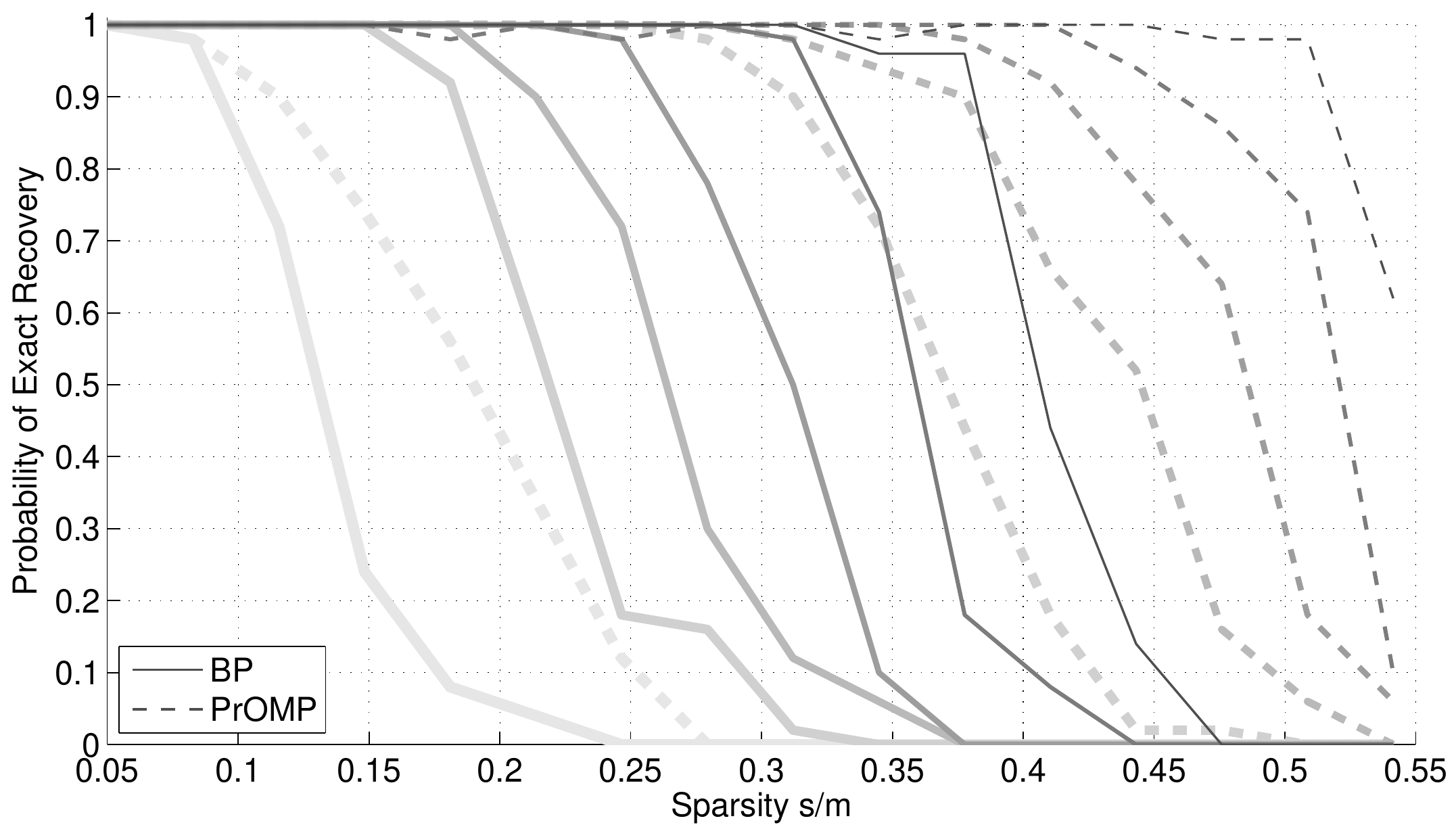}}
\caption{Probability of exact recovery using criterion (\ref{eq:successcriterion2}) 
for BP (solid) and PrOMP (dashed) as a function of problem sparsity
and six problem indeterminacies from thickest to thinest lines: 
\(\delta = m/N = \{0.05, 0.15, 0.25, 0.34, 0.44, 0.54\}\).}
\label{fig:RecoveryProbabilitiesBPPrOMP}
\end{figure}

\begin{figure}[htb]
\centering
\subfigure[Bernoulli]{
\includegraphics[width=0.49\textwidth]{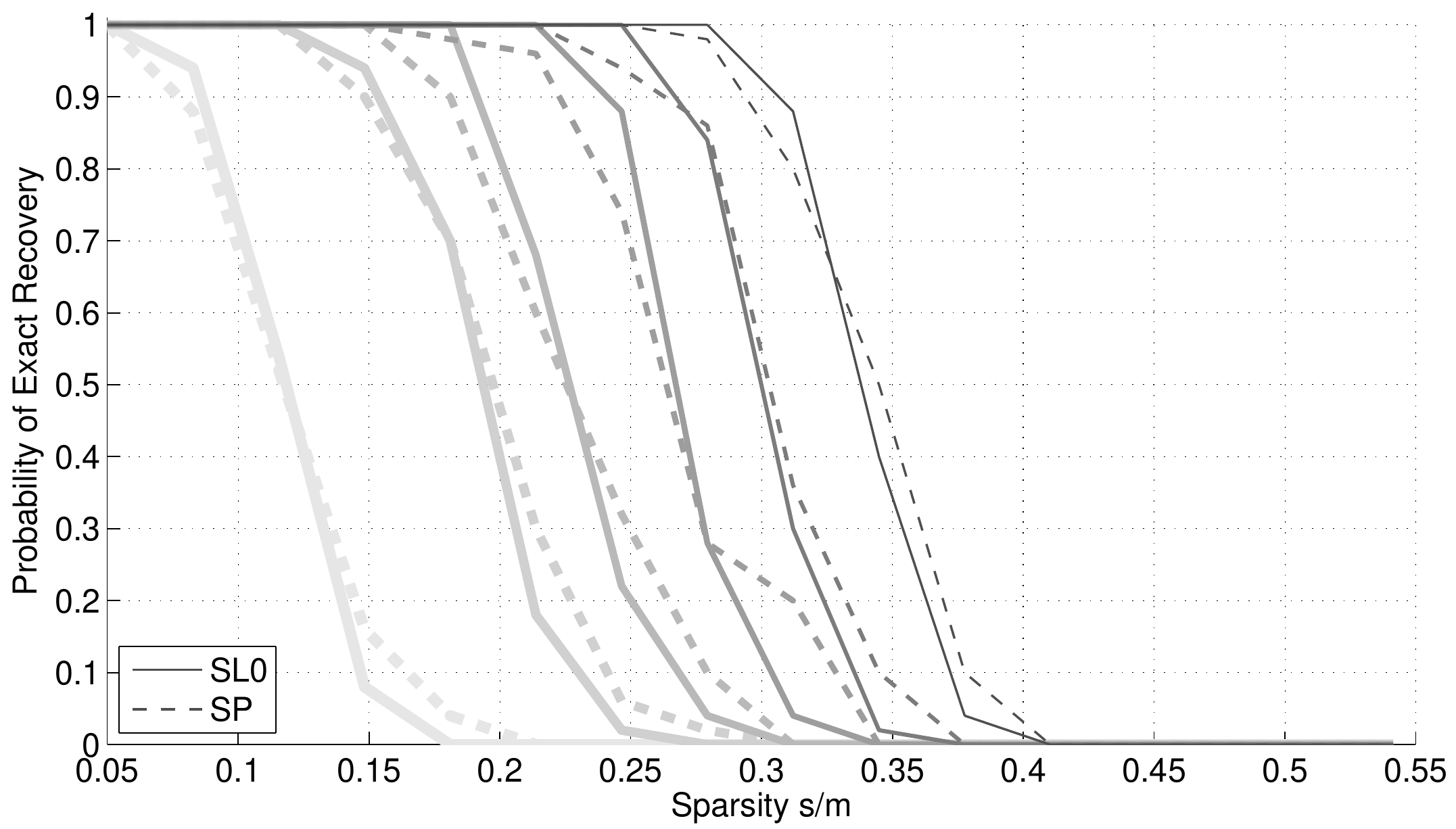}}\hspace{-0.1in}
\subfigure[Bimodal Uniform]{
\includegraphics[width=0.49\textwidth]{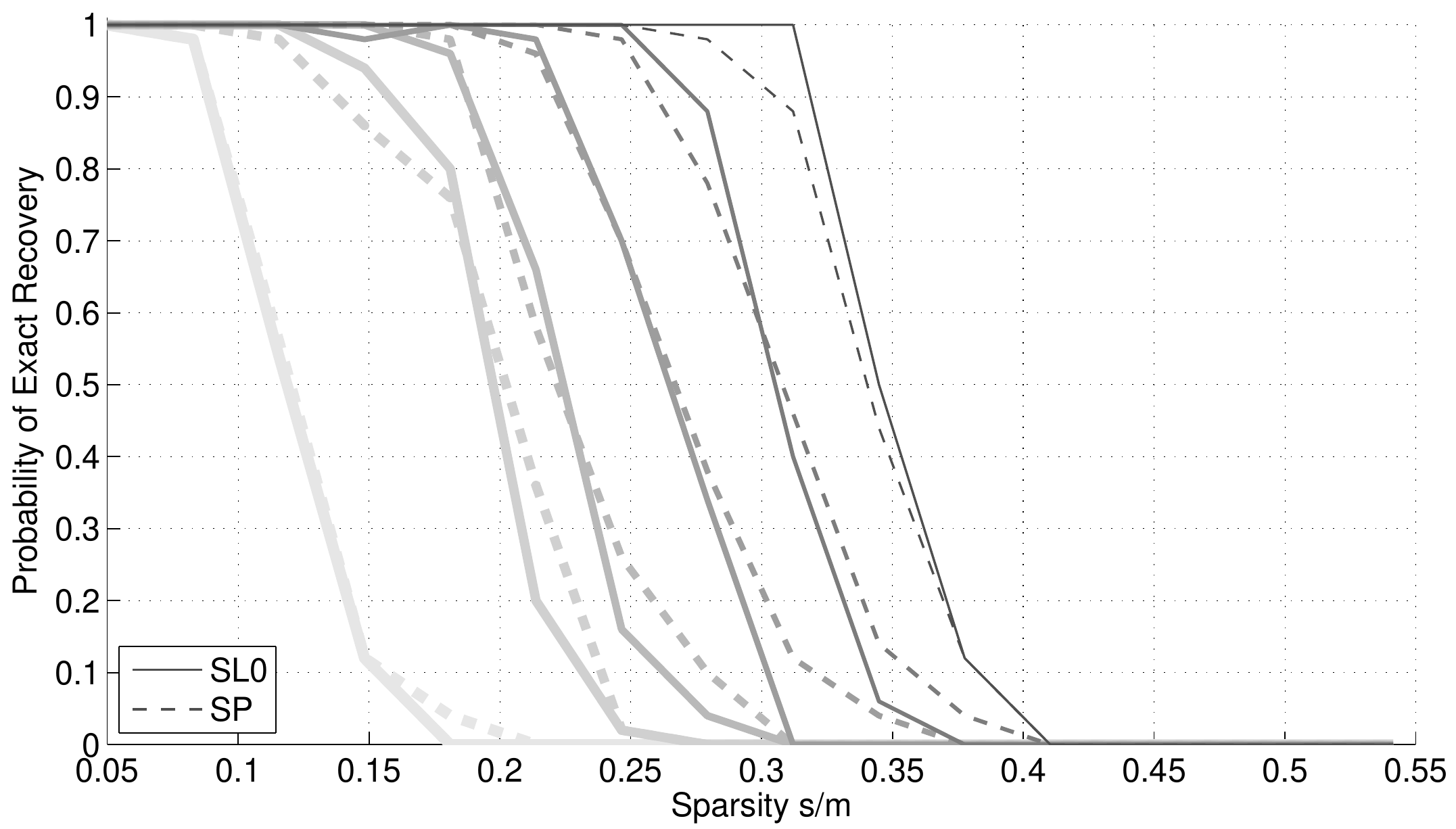}}\\ \vspace{-0.1in}

\subfigure[Bimodal Gaussian]{
\includegraphics[width=0.49\textwidth]{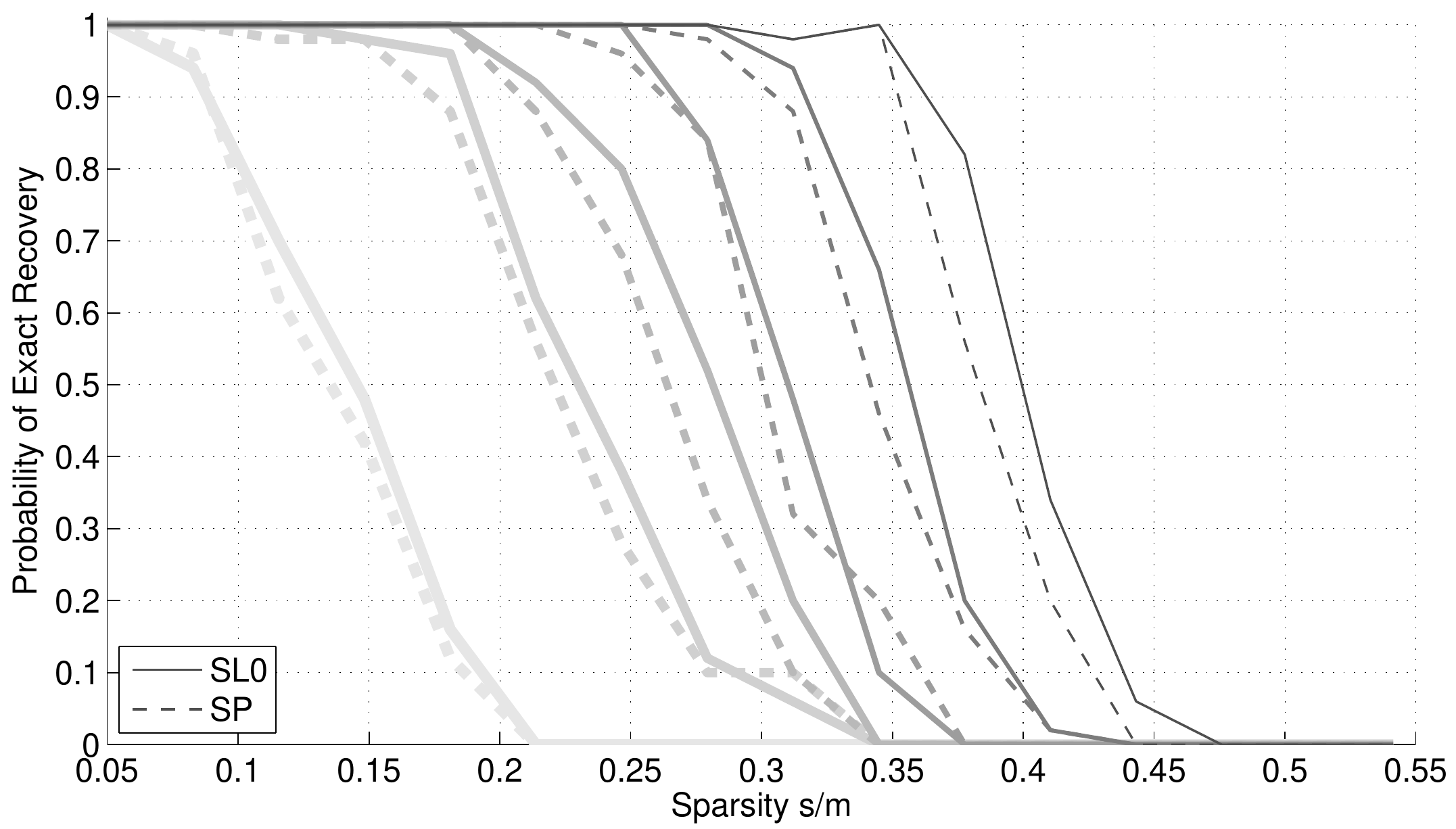}}\hspace{-0.1in}
\subfigure[Uniform]{
\includegraphics[width=0.49\textwidth]{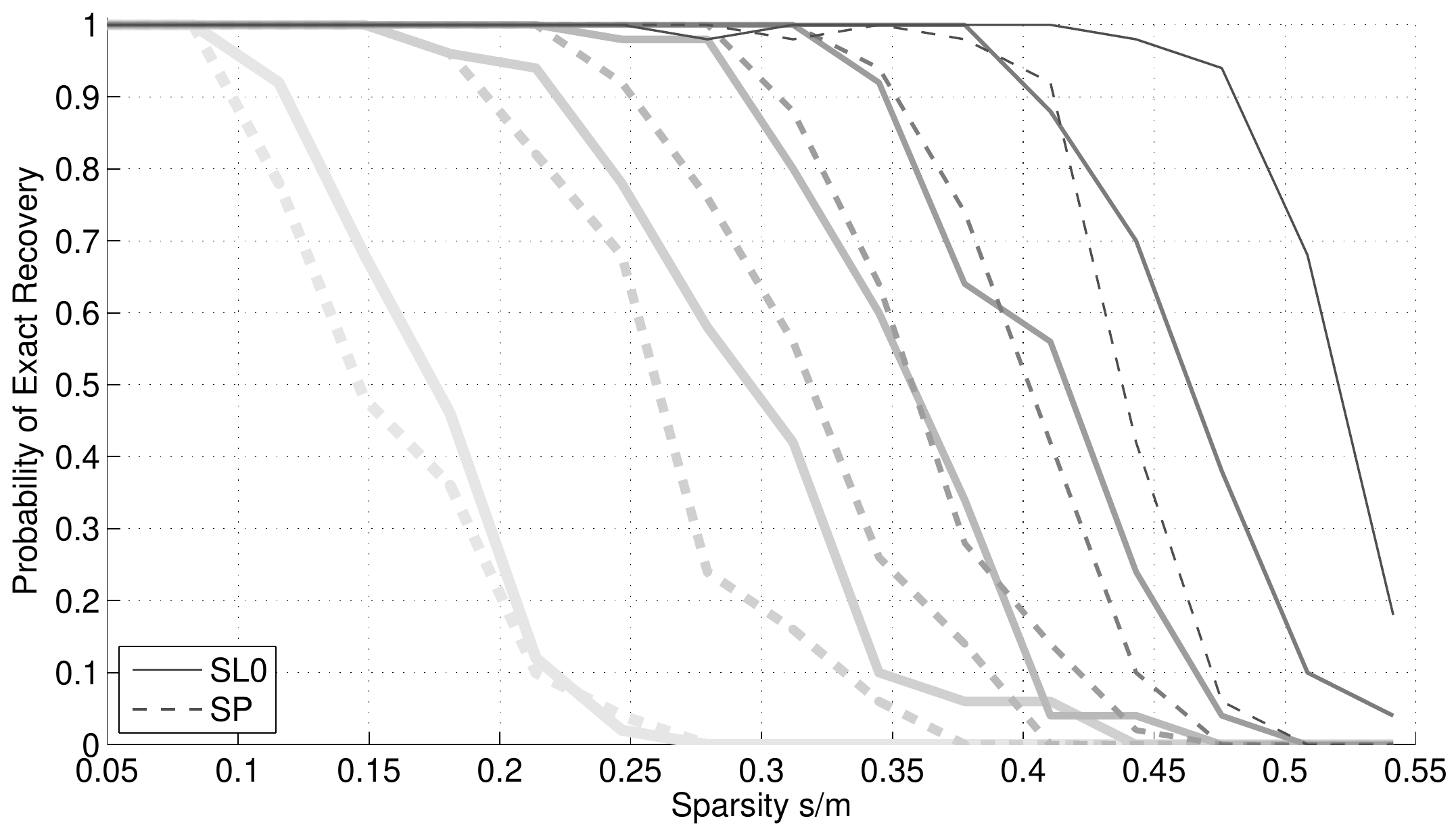}}\\ \vspace{-0.1in}

\subfigure[Bimodal Rayleigh]{
\includegraphics[width=0.49\textwidth]{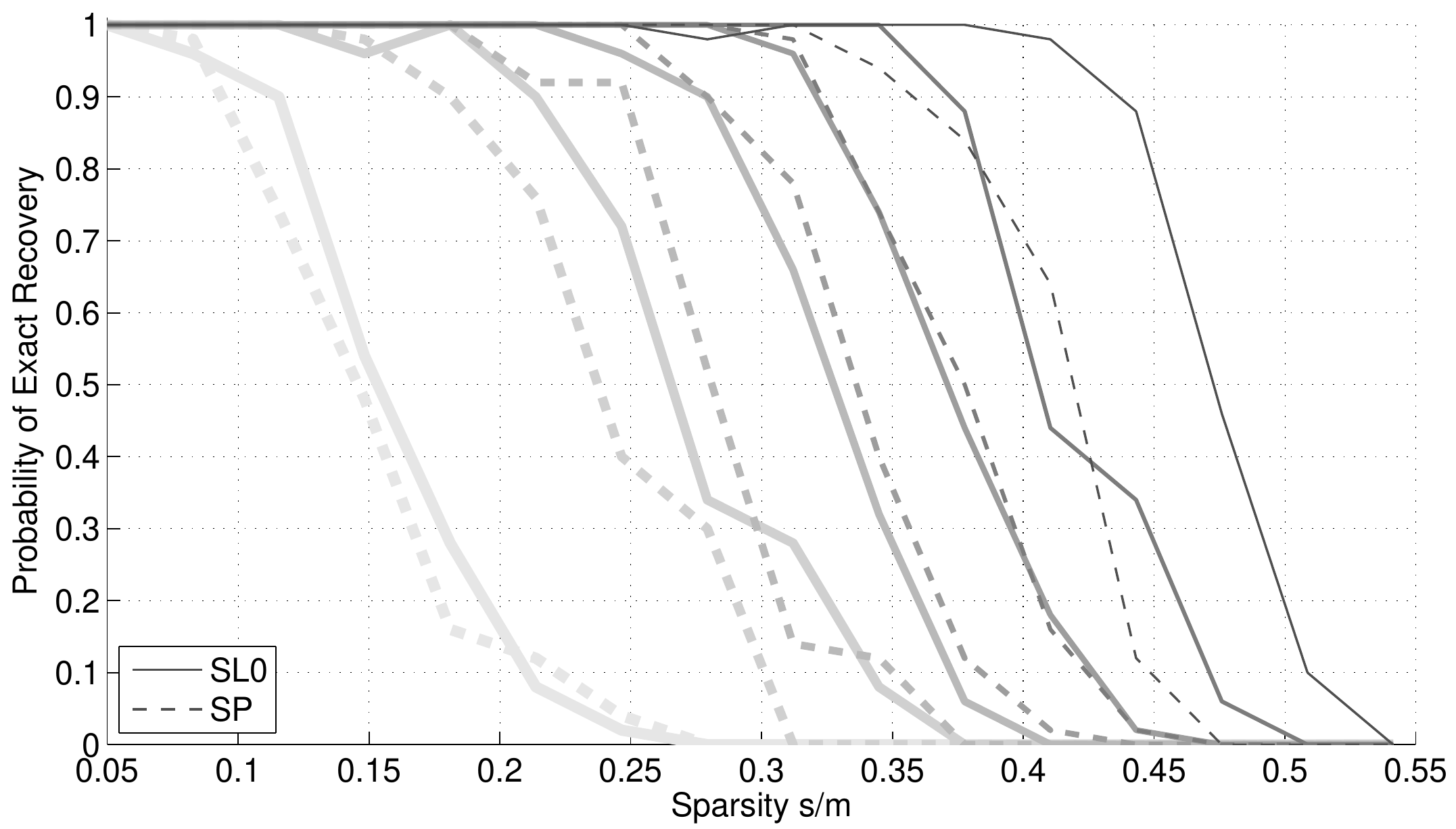}}\hspace{-0.1in}
\subfigure[Normal]{
\includegraphics[width=0.49\textwidth]{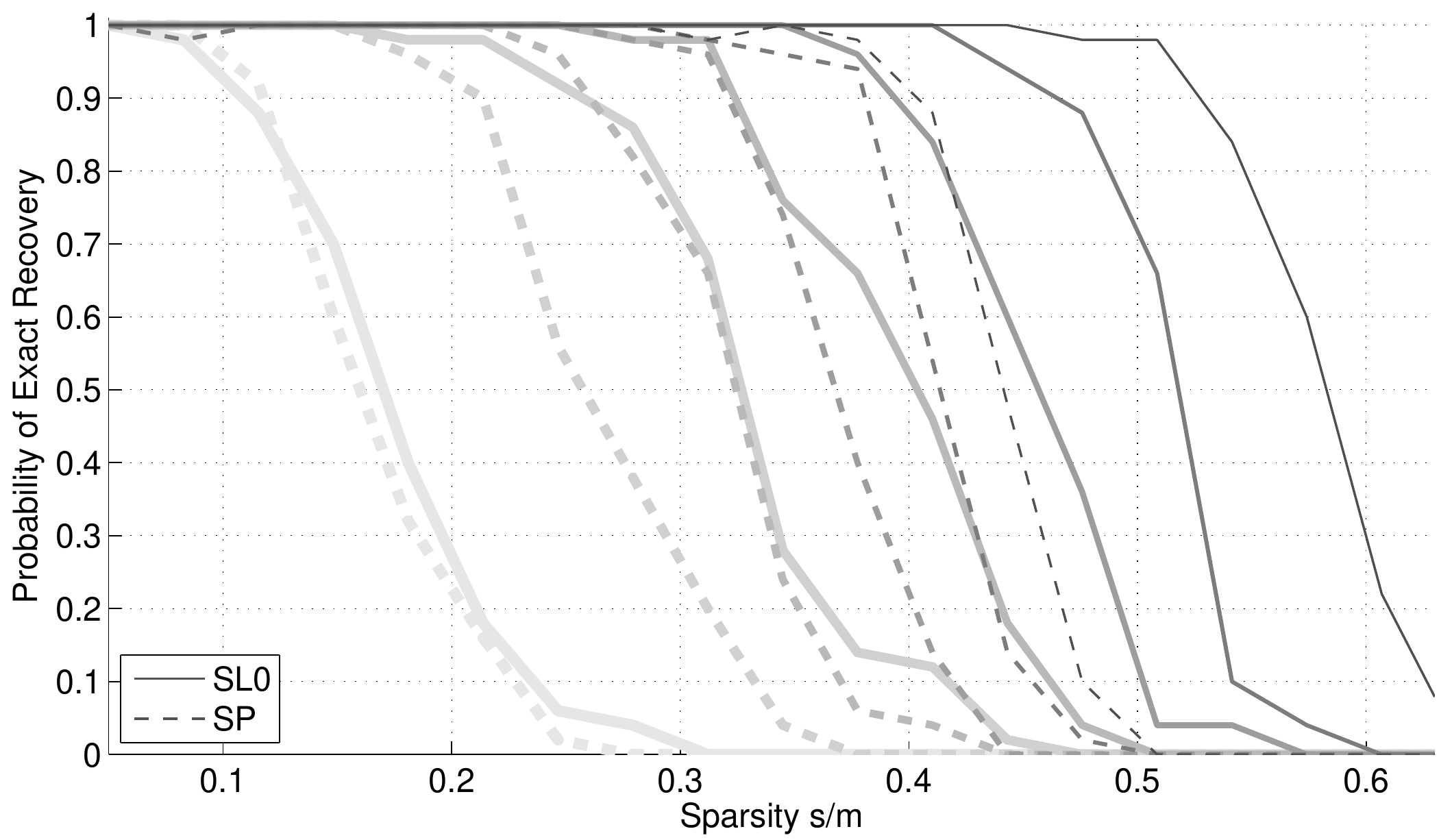}}\\ \vspace{-0.1in}

\subfigure[Laplacian]{
\includegraphics[width=0.49\textwidth]{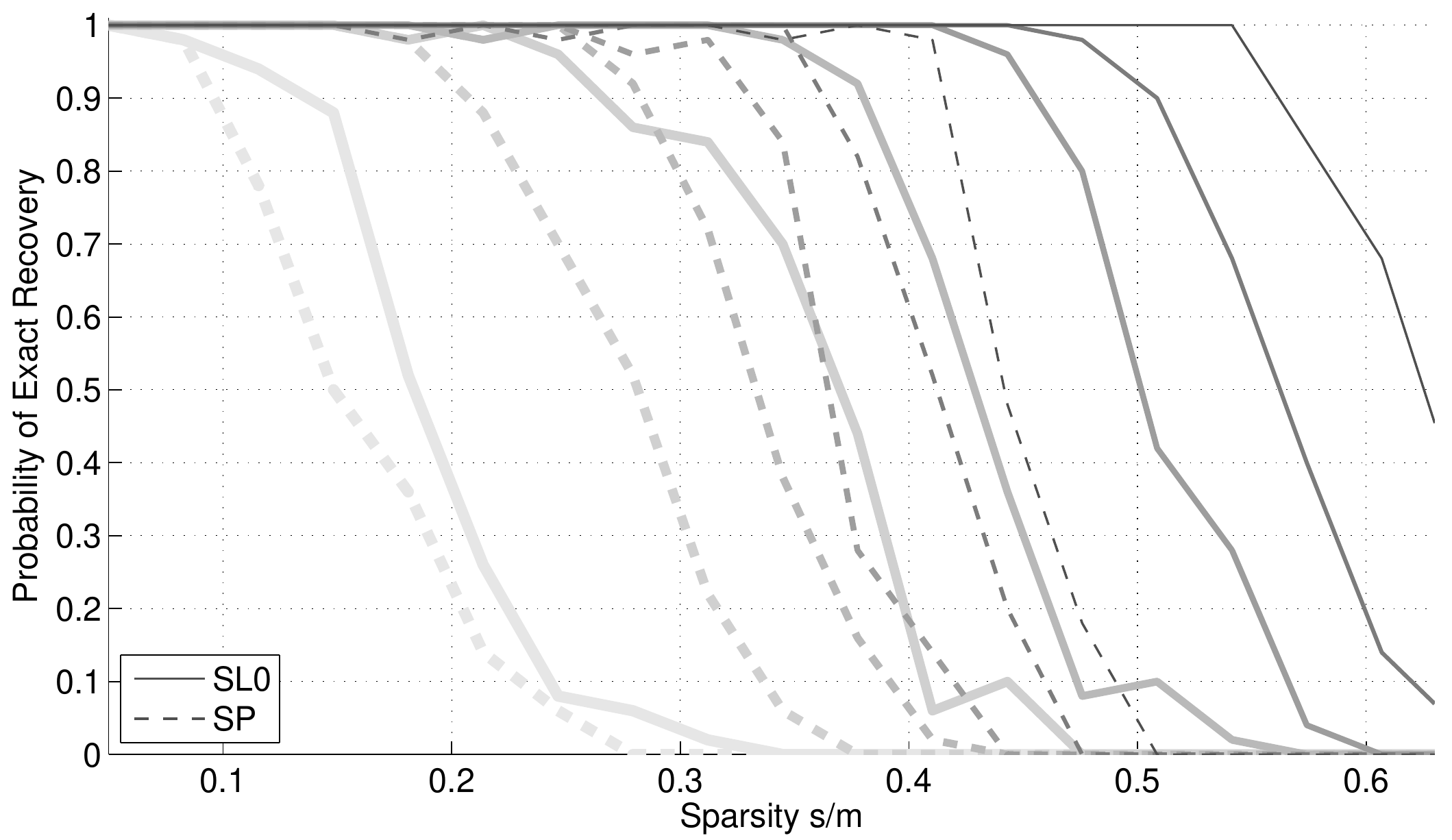}}
\caption{Probability of exact recovery using criterion (\ref{eq:successcriterion2}) 
for SL0 (solid) and SP (dashed) as a function of problem sparsity
and six problem indeterminacies from thickest to thinest lines: 
\(\delta = m/N = \{0.05, 0.15, 0.25, 0.34, 0.44, 0.54\}\).
Note change in y-scale in (f) and (g).}
\label{fig:RecoveryProbabilitiesSL0SP}
\end{figure}

\begin{figure}[htb]
\centering
\subfigure[Bernoulli]{
\includegraphics[width=0.49\textwidth]{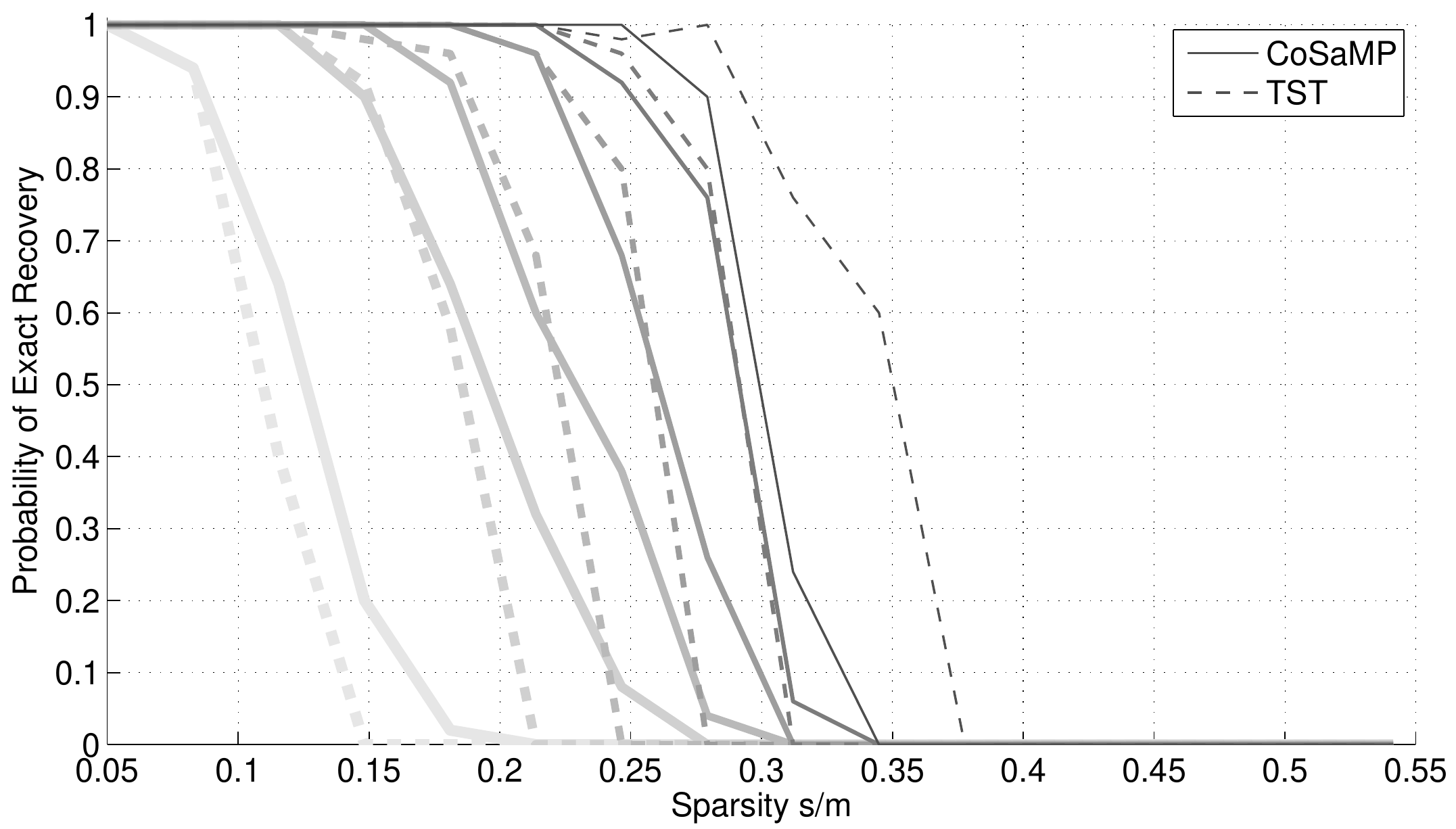}}\hspace{-0.1in}
\subfigure[Bimodal Uniform]{
\includegraphics[width=0.49\textwidth]{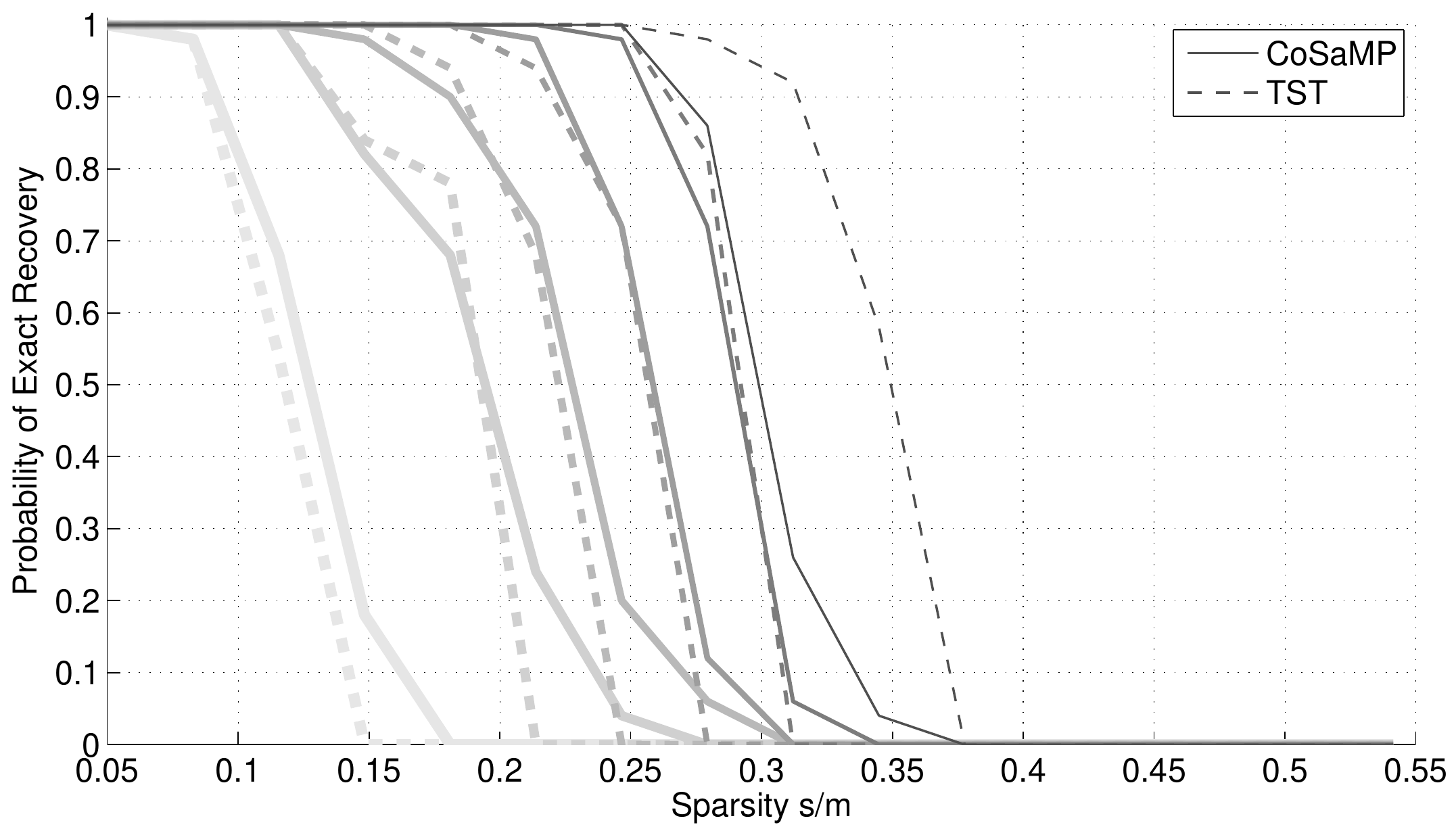}}\\ \vspace{-0.1in}

\subfigure[Bimodal Gaussian]{
\includegraphics[width=0.49\textwidth]{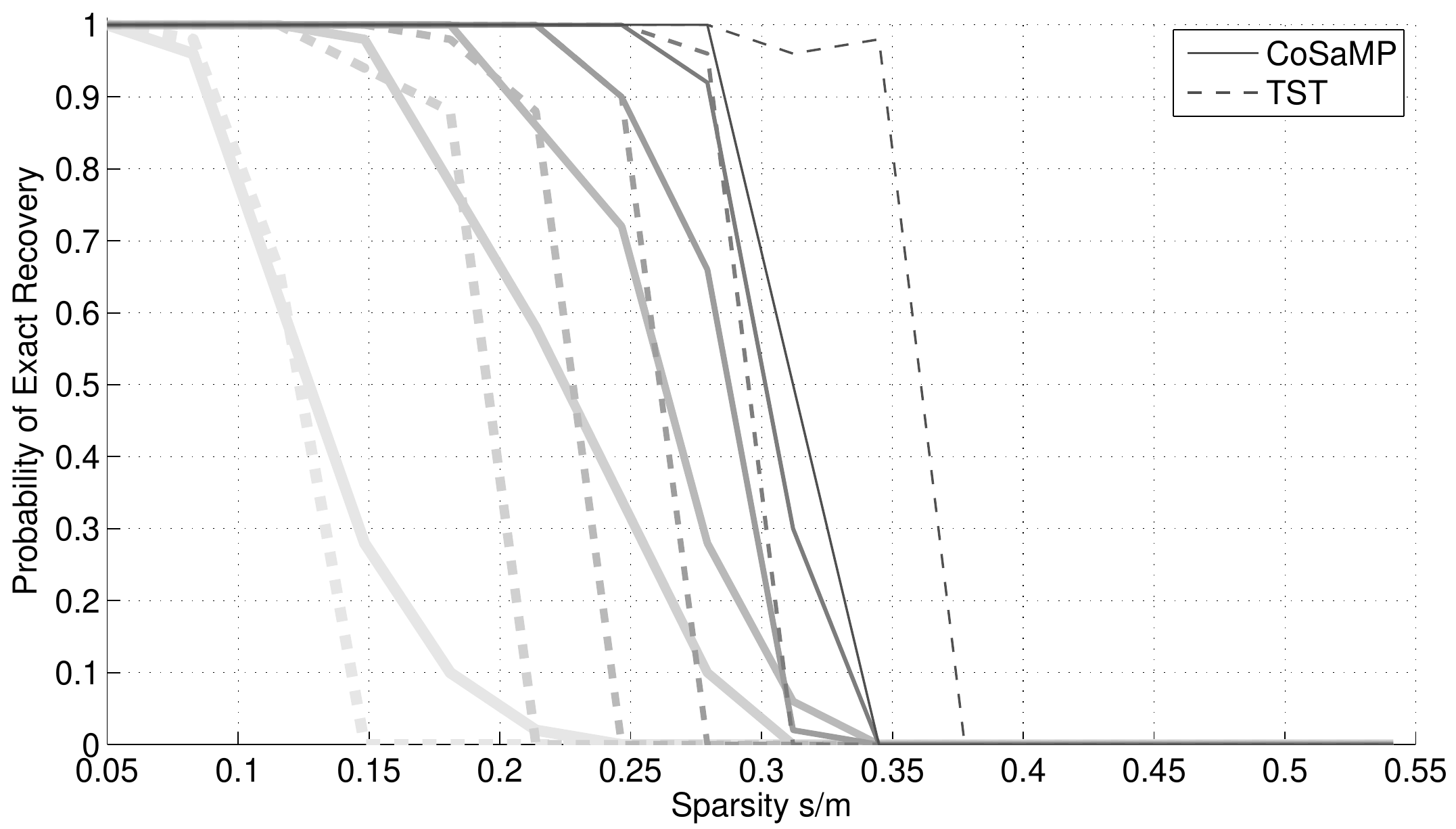}}\hspace{-0.1in}
\subfigure[Uniform]{
\includegraphics[width=0.49\textwidth]{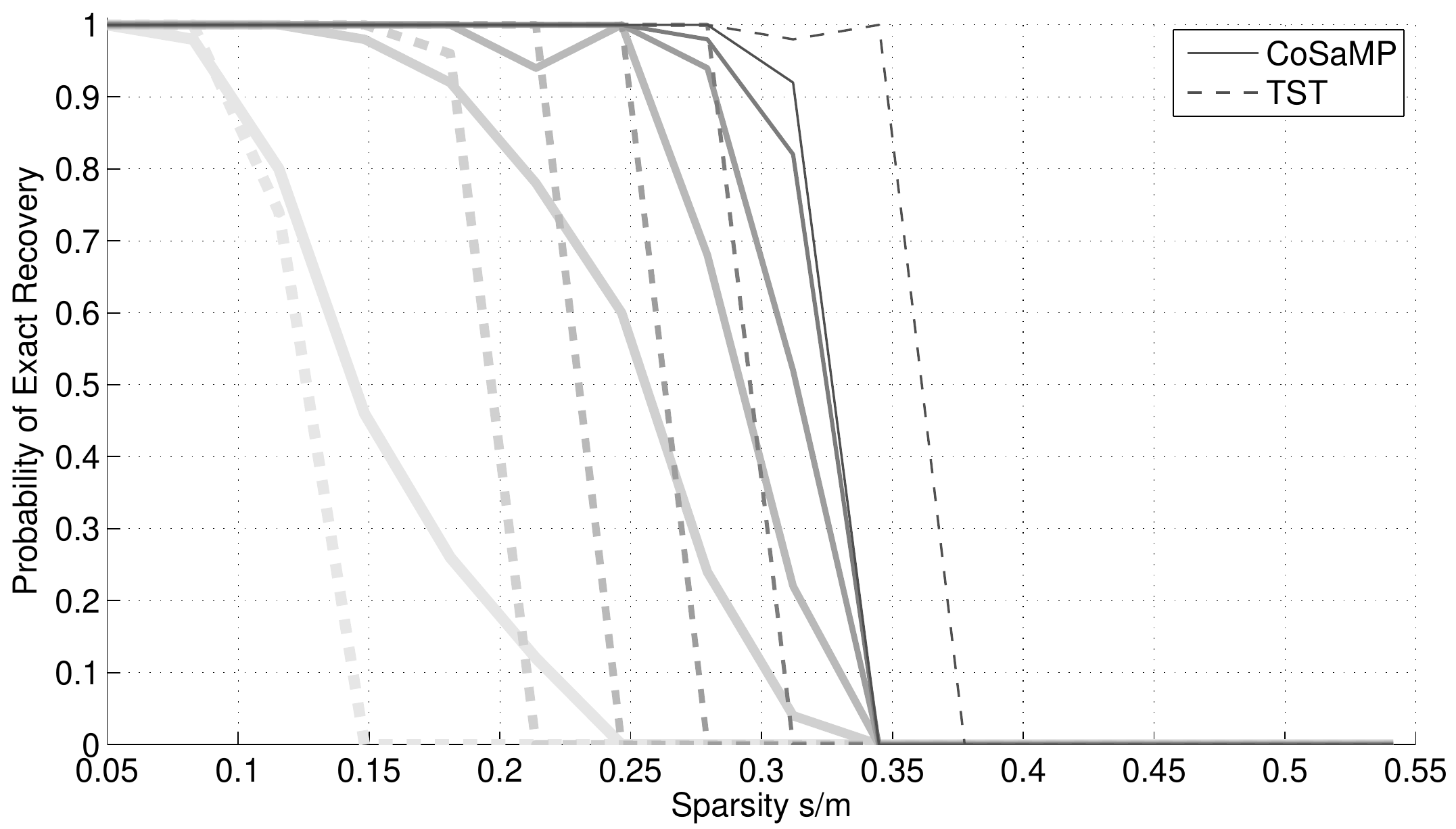}}\\ \vspace{-0.1in}

\subfigure[Bimodal Rayleigh]{
\includegraphics[width=0.49\textwidth]{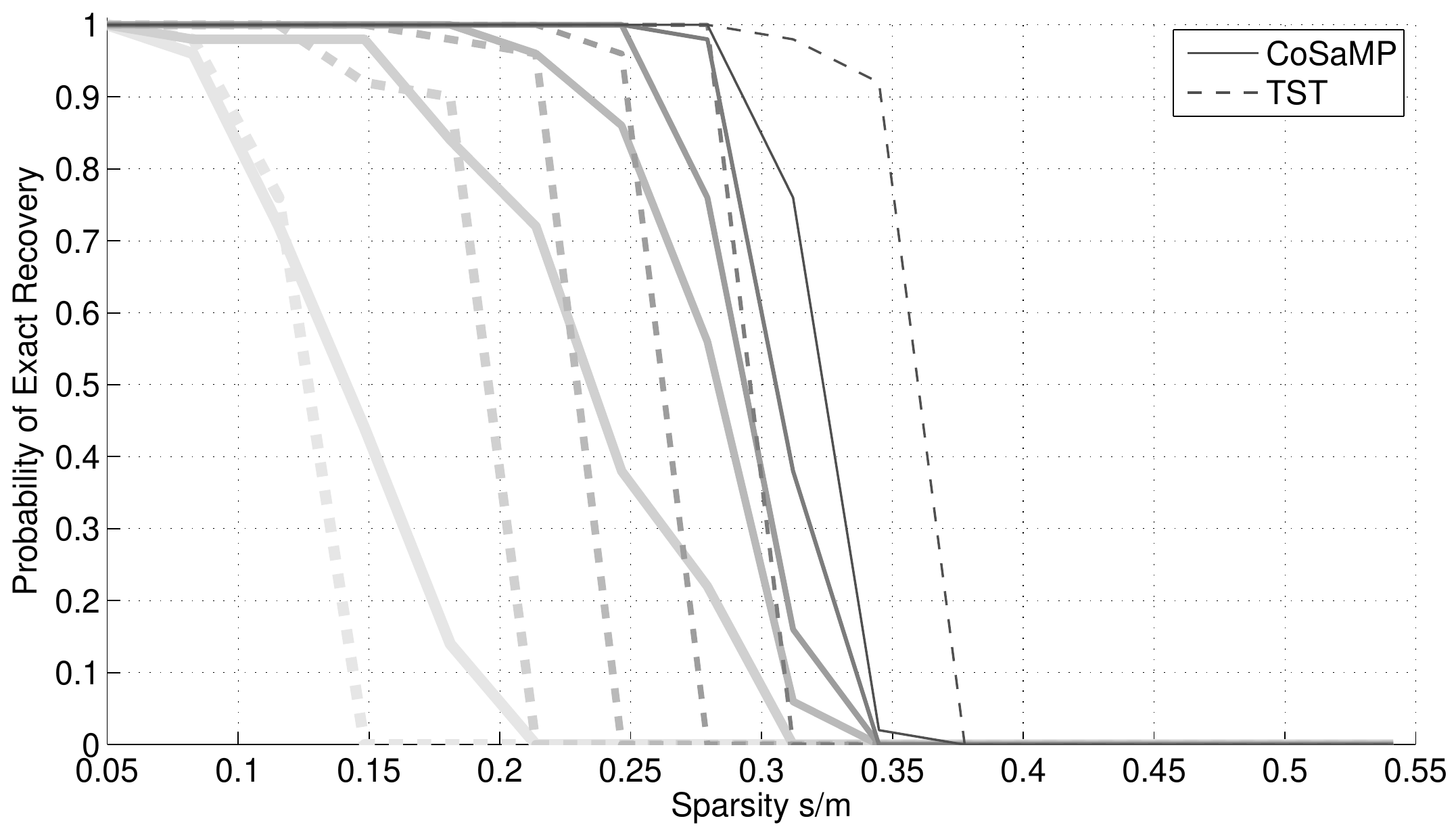}}\hspace{-0.1in}
\subfigure[Normal]{
\includegraphics[width=0.49\textwidth]{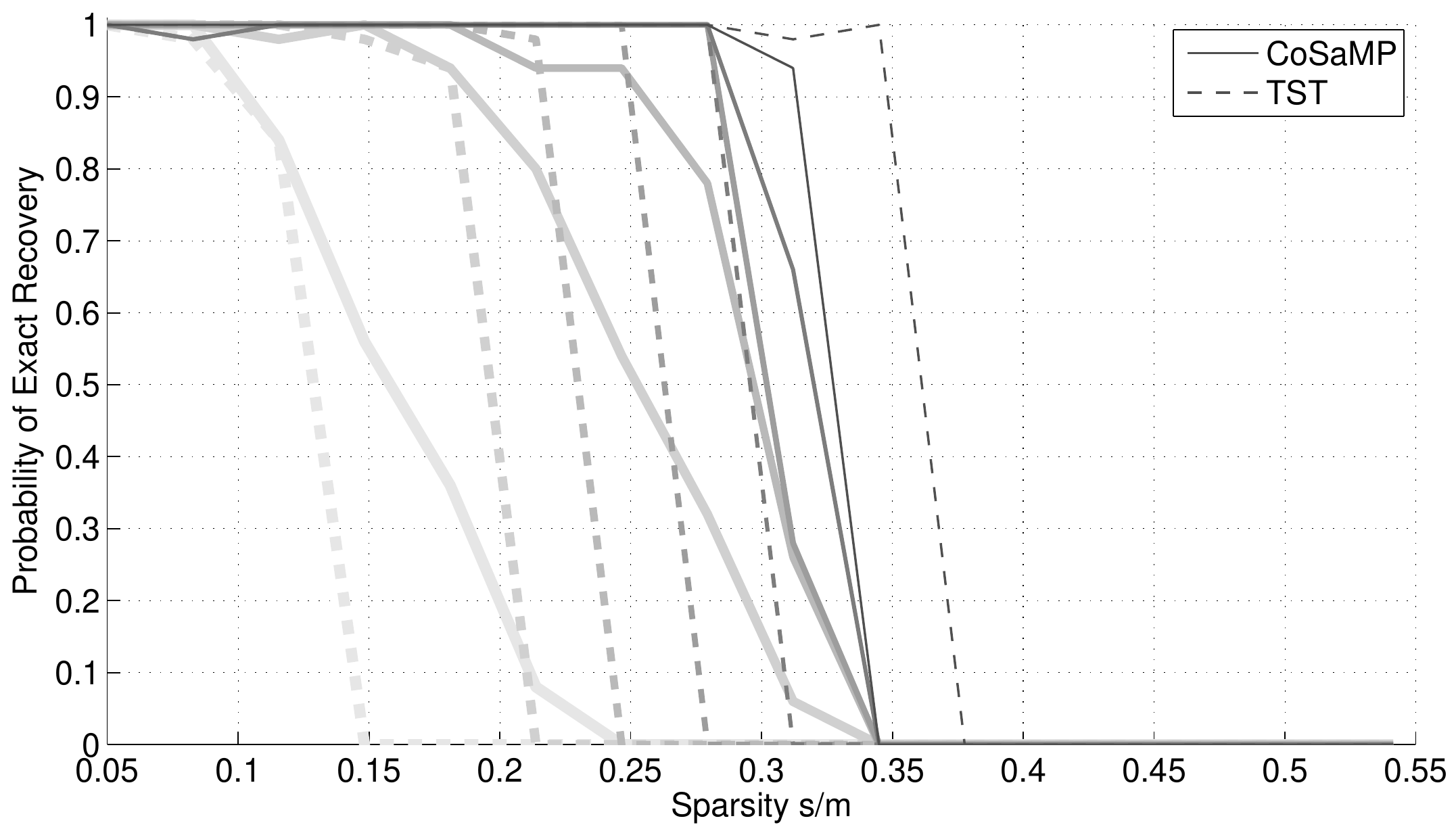}}\\ \vspace{-0.1in}

\subfigure[Laplacian]{
\includegraphics[width=0.49\textwidth]{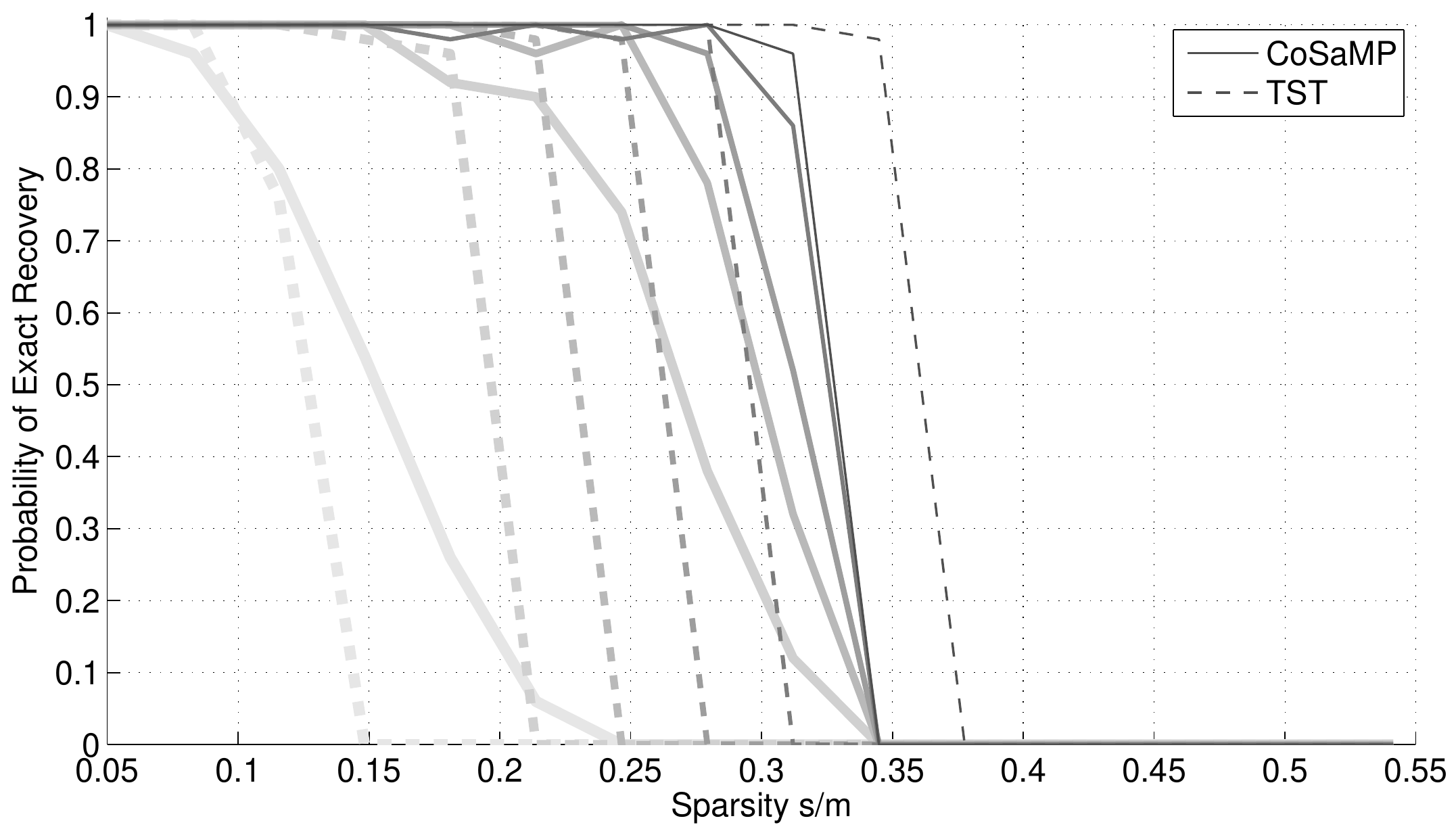}}
\caption{Probability of exact recovery using criterion (\ref{eq:successcriterion2}) 
for CoSaMP (solid) and TST (dashed) as a function of problem sparsity
and six problem indeterminacies from thickest to thinest lines: 
\(\delta = m/N = \{0.05, 0.15, 0.25, 0.34, 0.44, 0.54\}\).}
\label{fig:RecoveryProbabilitiesCoSaMPTST}
\end{figure}

\begin{figure}[htb]
\centering
\subfigure[Bernoulli]{
\includegraphics[width=0.49\textwidth]{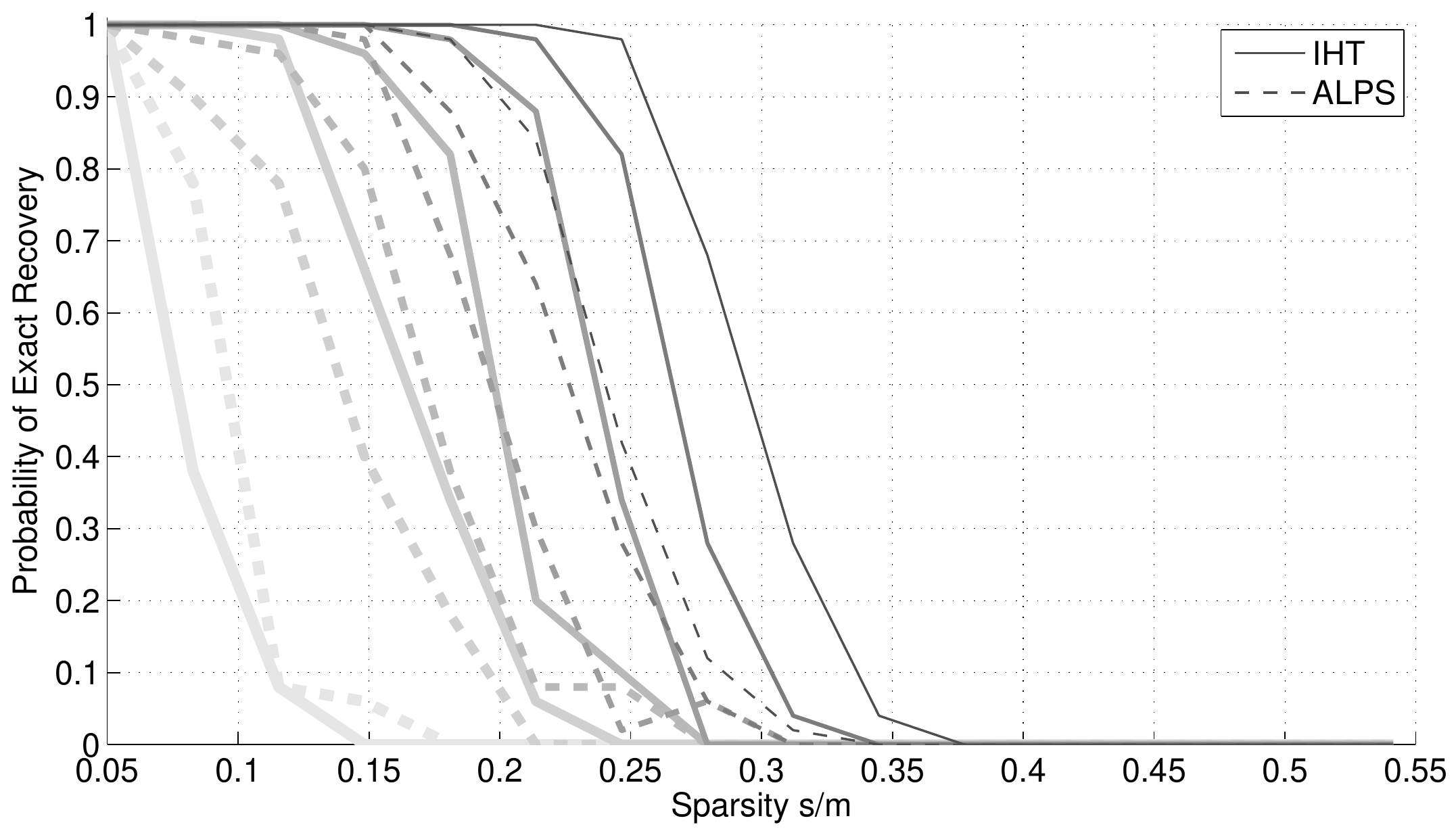}}\hspace{-0.1in}
\subfigure[Bimodal Uniform]{
\includegraphics[width=0.49\textwidth]{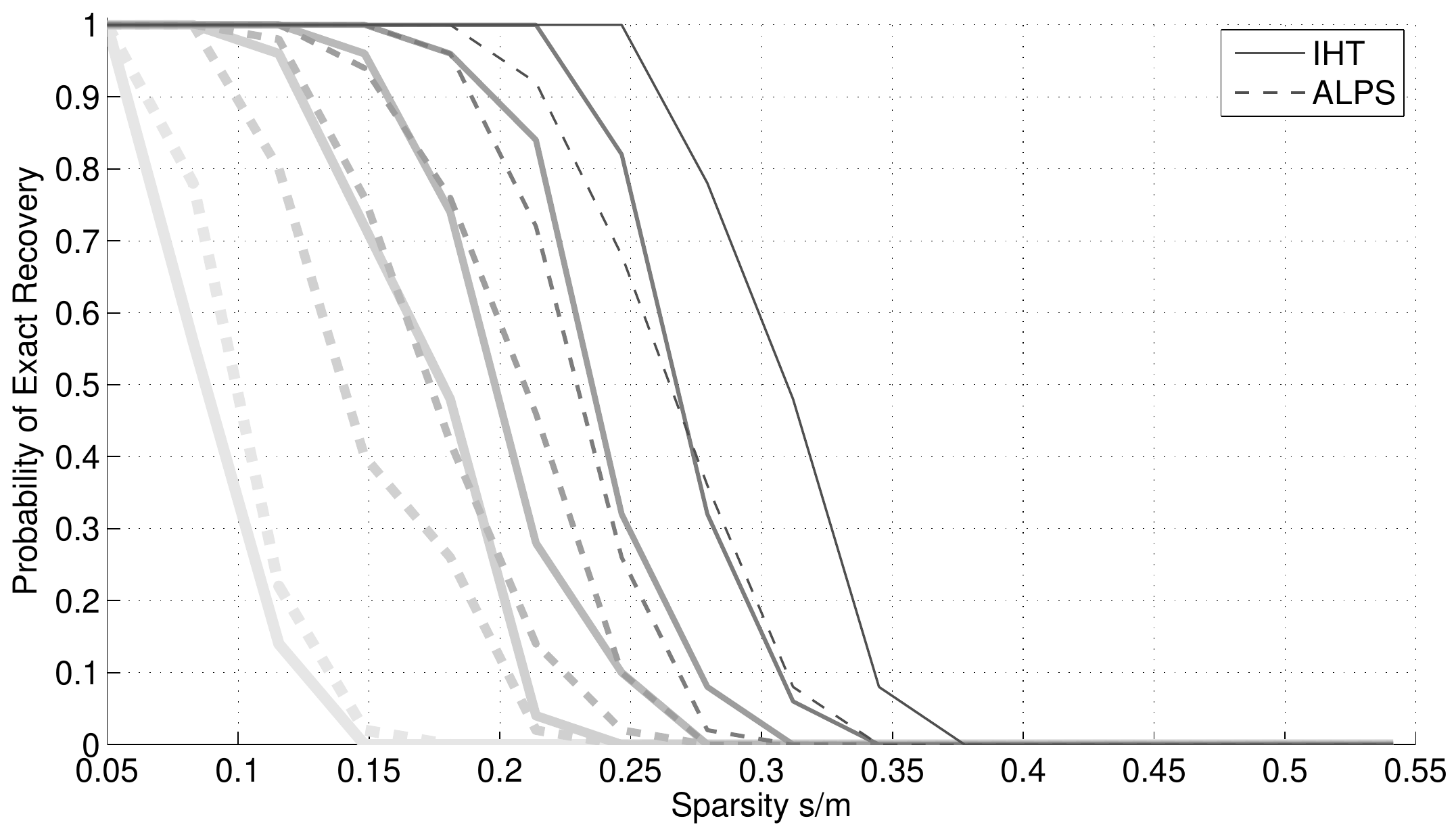}}\\ \vspace{-0.1in}

\subfigure[Bimodal Gaussian]{
\includegraphics[width=0.49\textwidth]{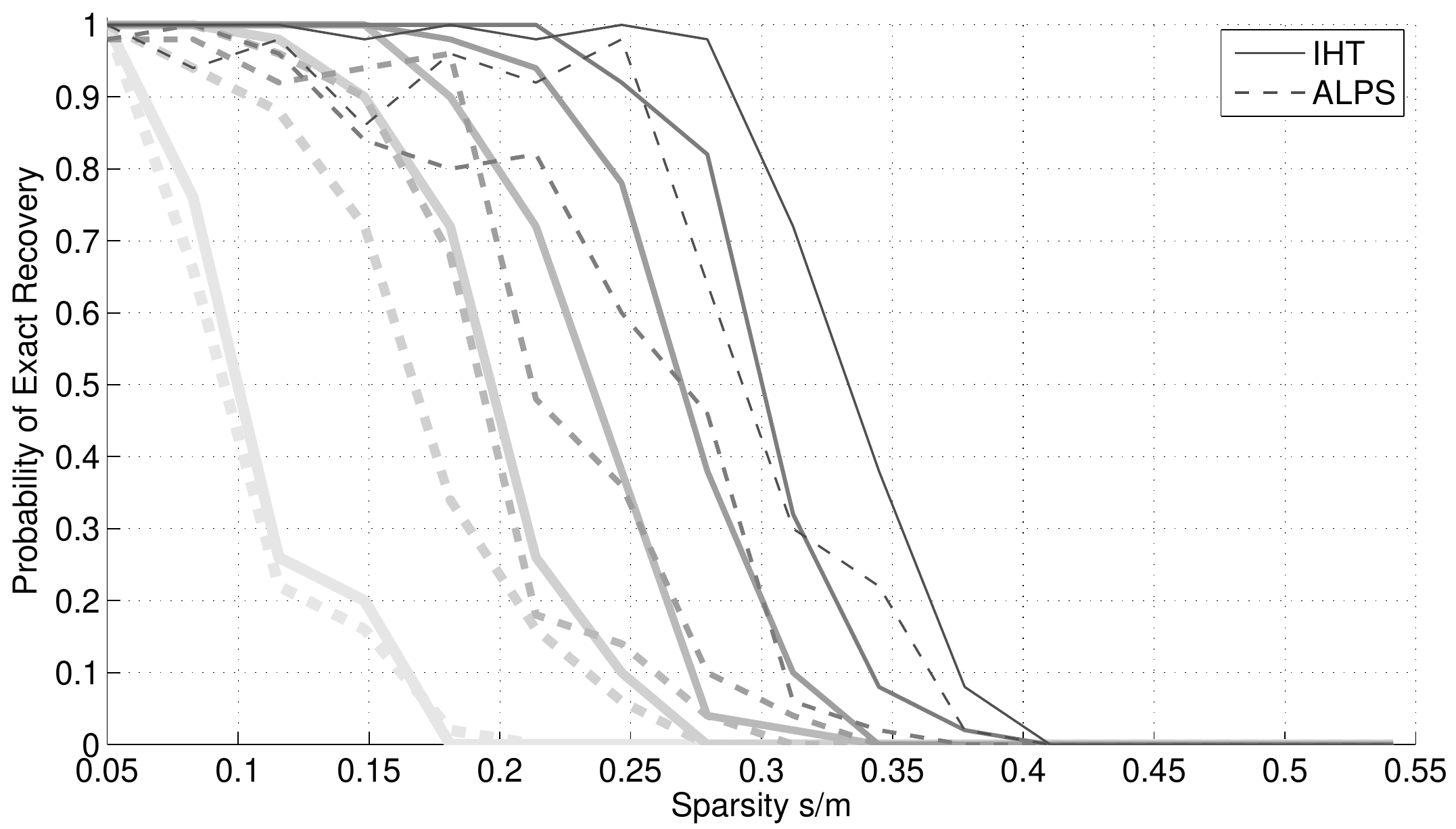}}\hspace{-0.1in}
\subfigure[Uniform]{
\includegraphics[width=0.49\textwidth]{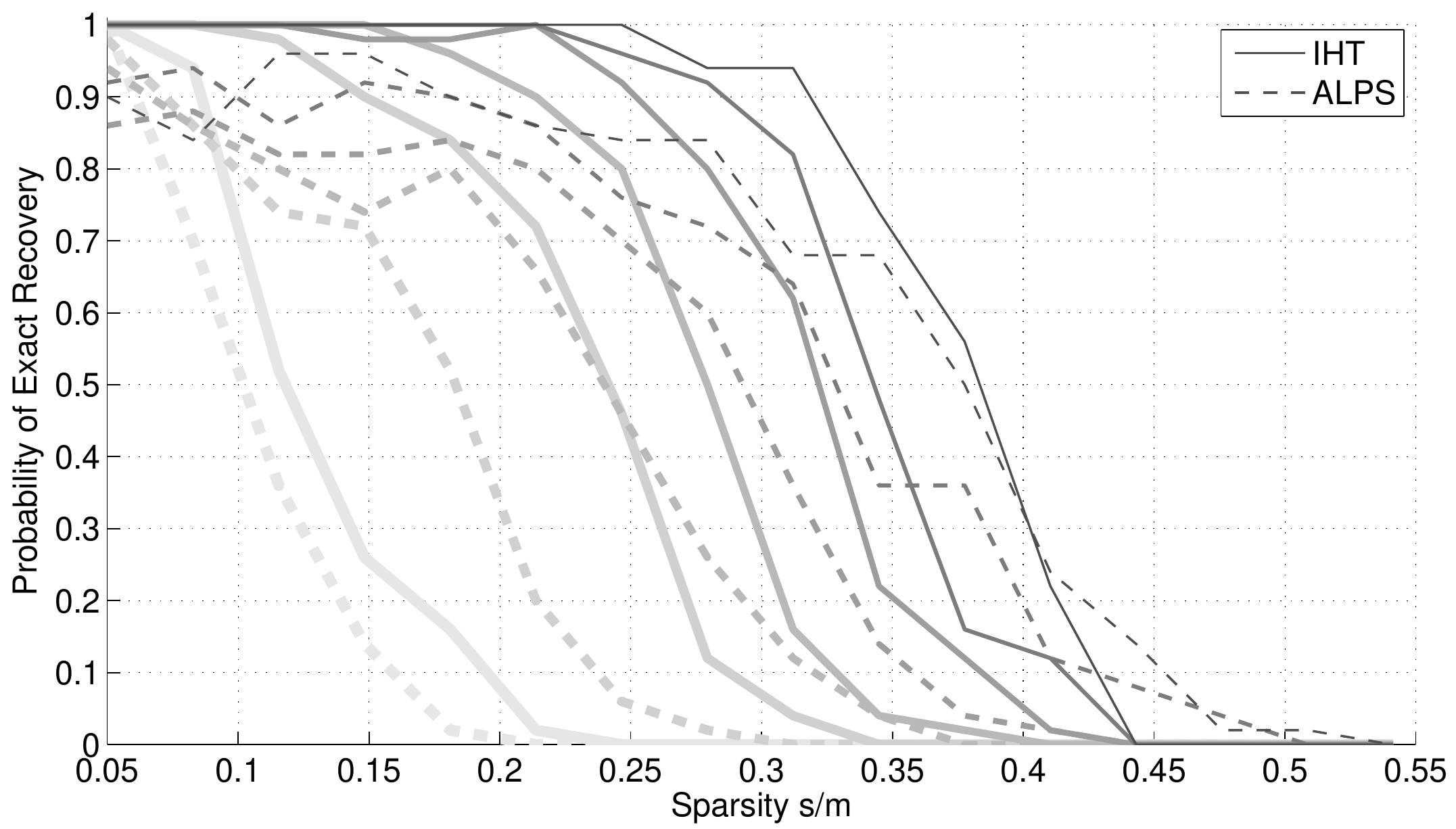}}\\ \vspace{-0.1in}

\subfigure[Bimodal Rayleigh]{
\includegraphics[width=0.49\textwidth]{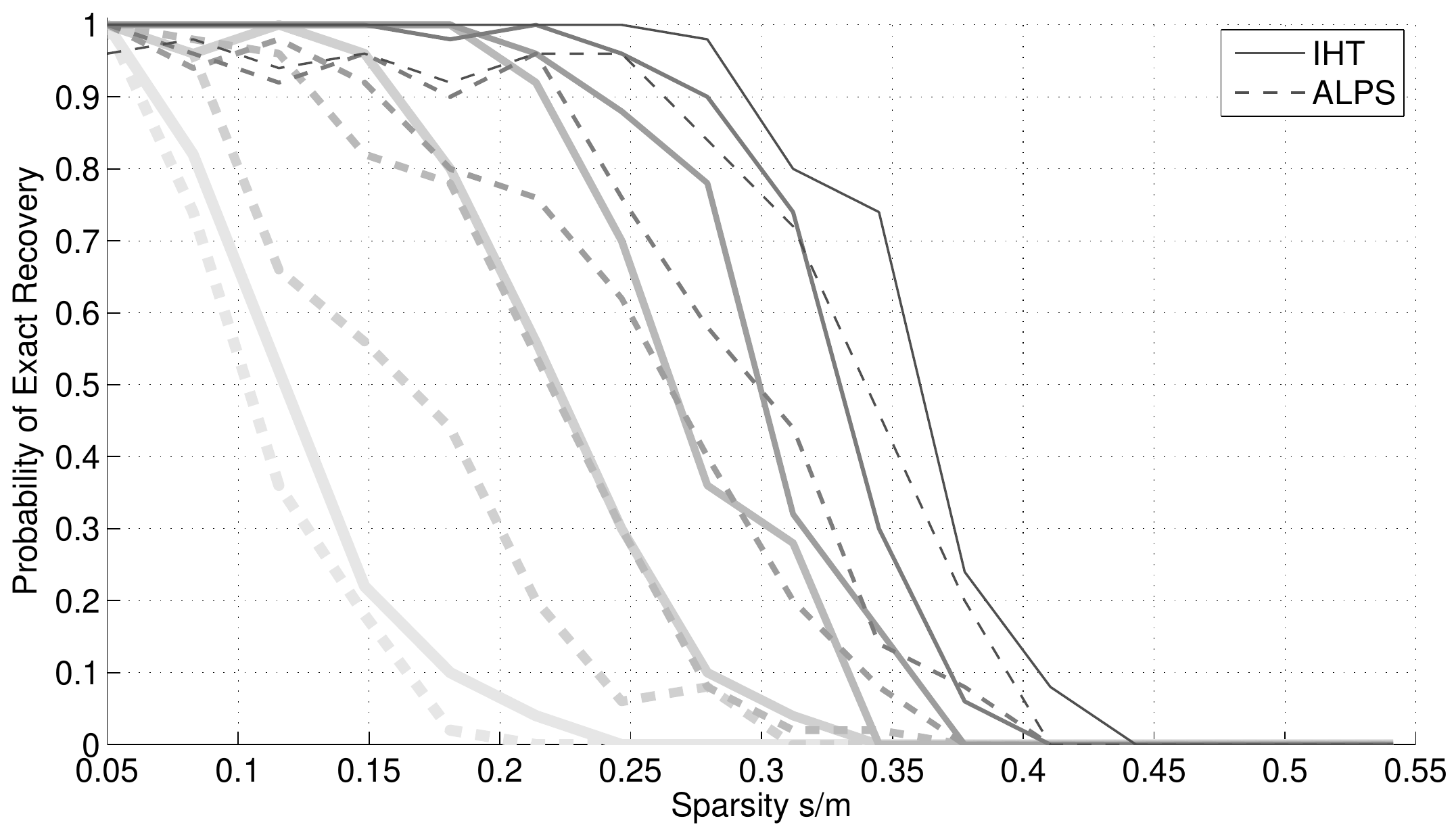}}\hspace{-0.1in}
\subfigure[Normal]{
\includegraphics[width=0.49\textwidth]{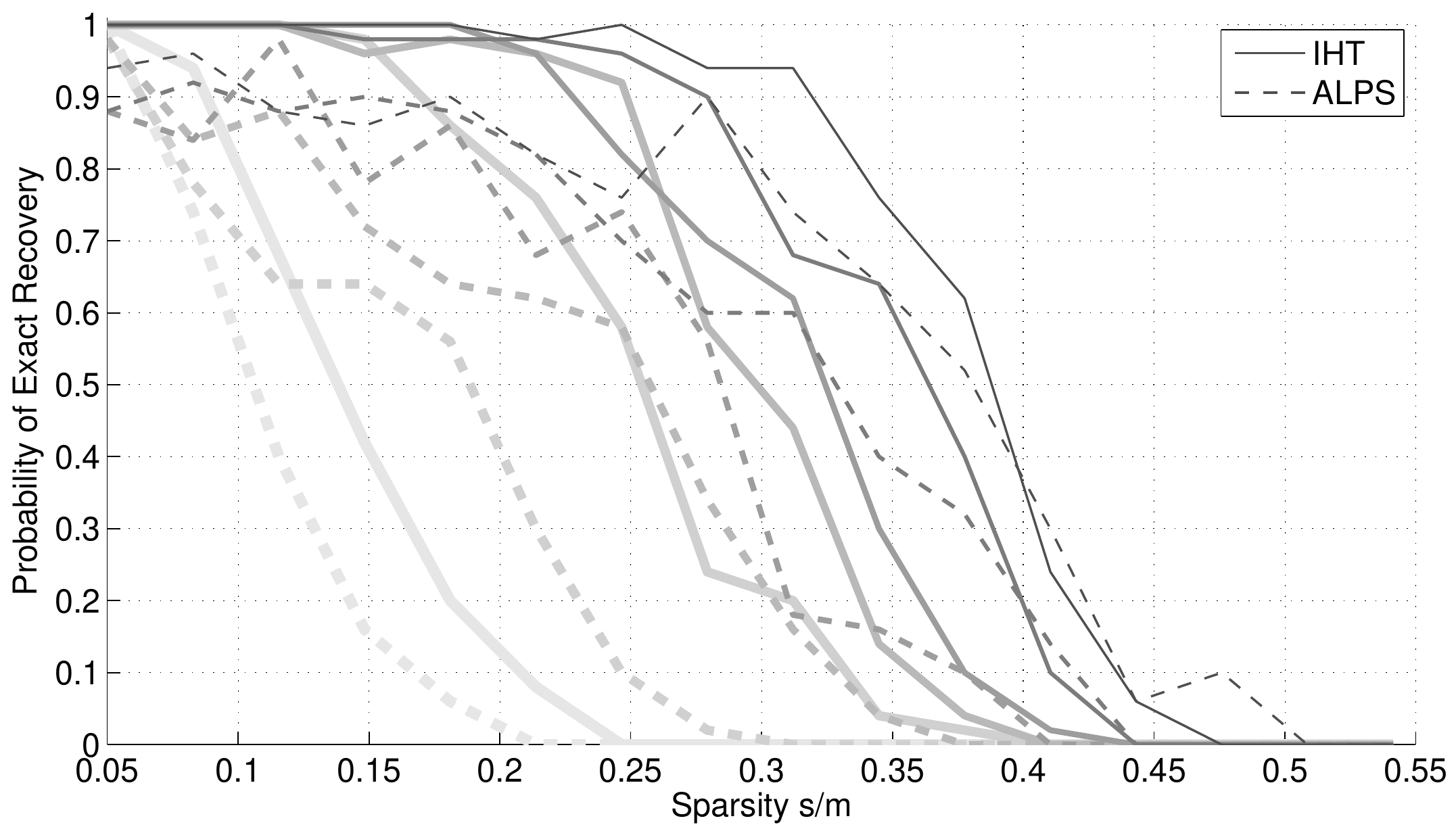}}\\ \vspace{-0.1in}

\subfigure[Laplacian]{
\includegraphics[width=0.49\textwidth]{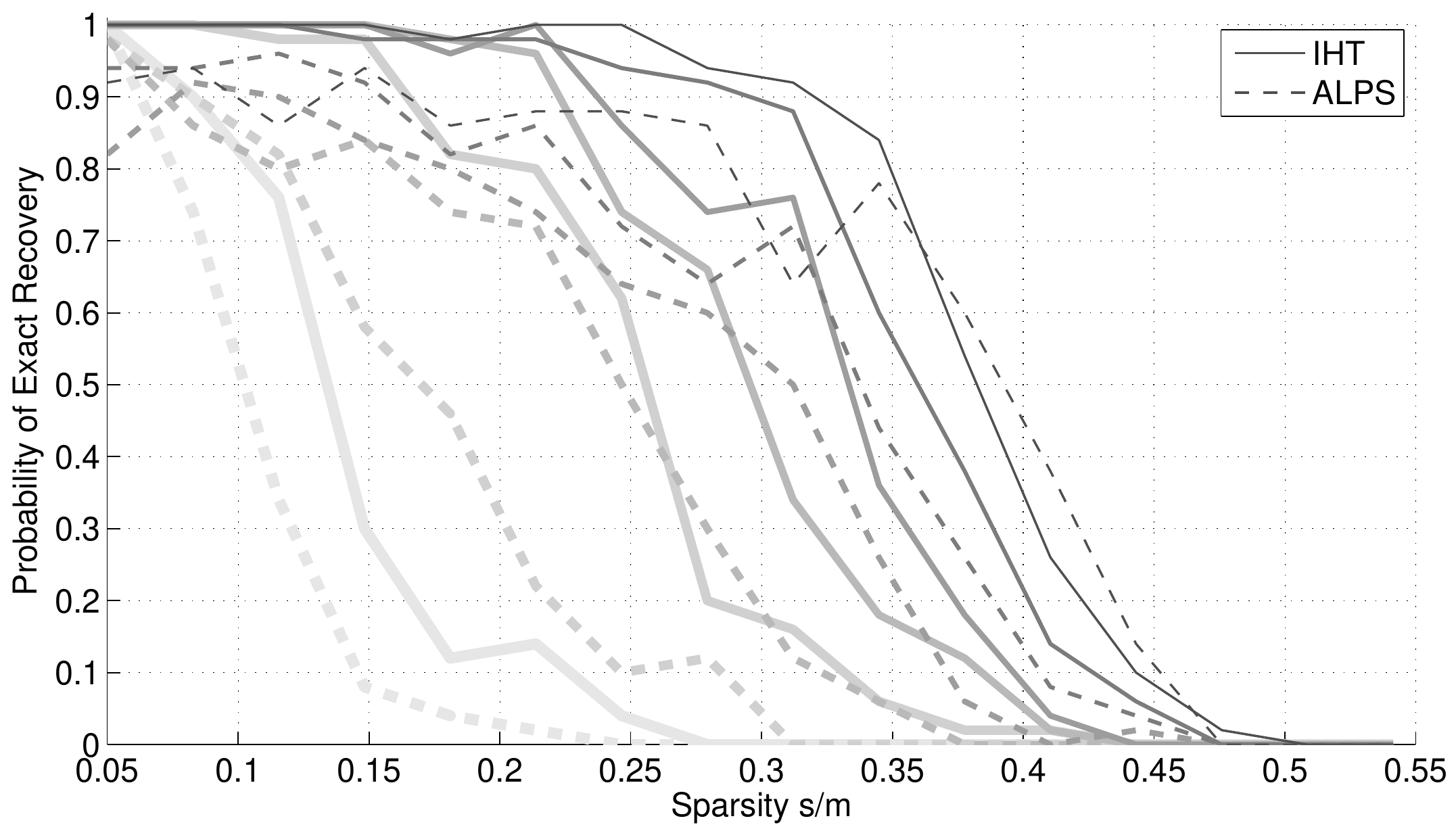}}
\caption{Probability of exact recovery using criterion (\ref{eq:successcriterion2}) 
for CoSaMP (solid) and TST (dashed) as a function of problem sparsity
and six problem indeterminacies from thickest to thinest lines: 
\(\delta = m/N = \{0.05, 0.15, 0.25, 0.34, 0.44, 0.54\}\).}
\label{fig:RecoveryProbabilitiesIHTALPS}
\end{figure}

\begin{figure}[htb]
\centering
\subfigure[Bernoulli]{
\includegraphics[width=0.49\textwidth]{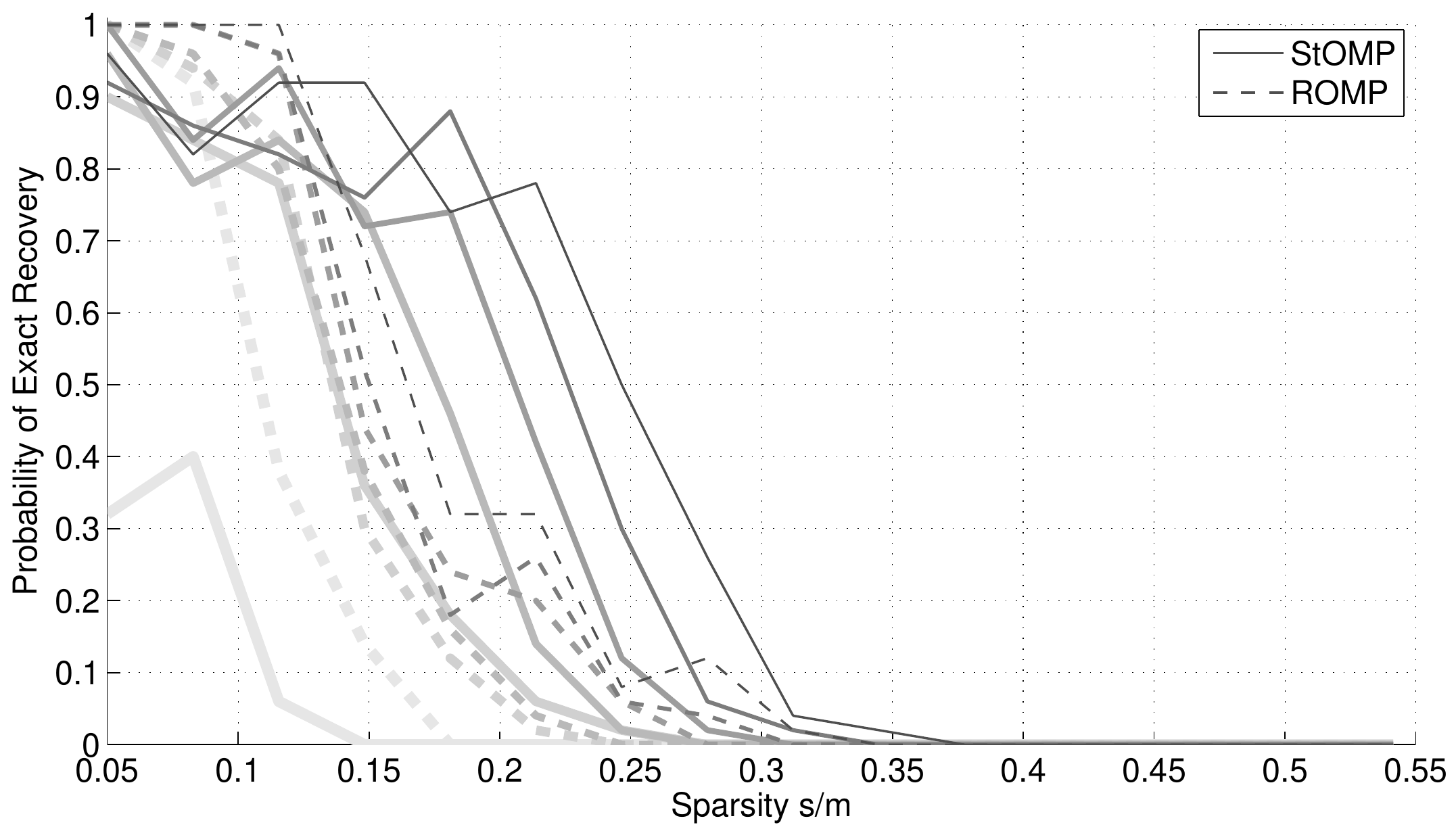}}\hspace{-0.1in}
\subfigure[Bimodal Uniform]{
\includegraphics[width=0.49\textwidth]{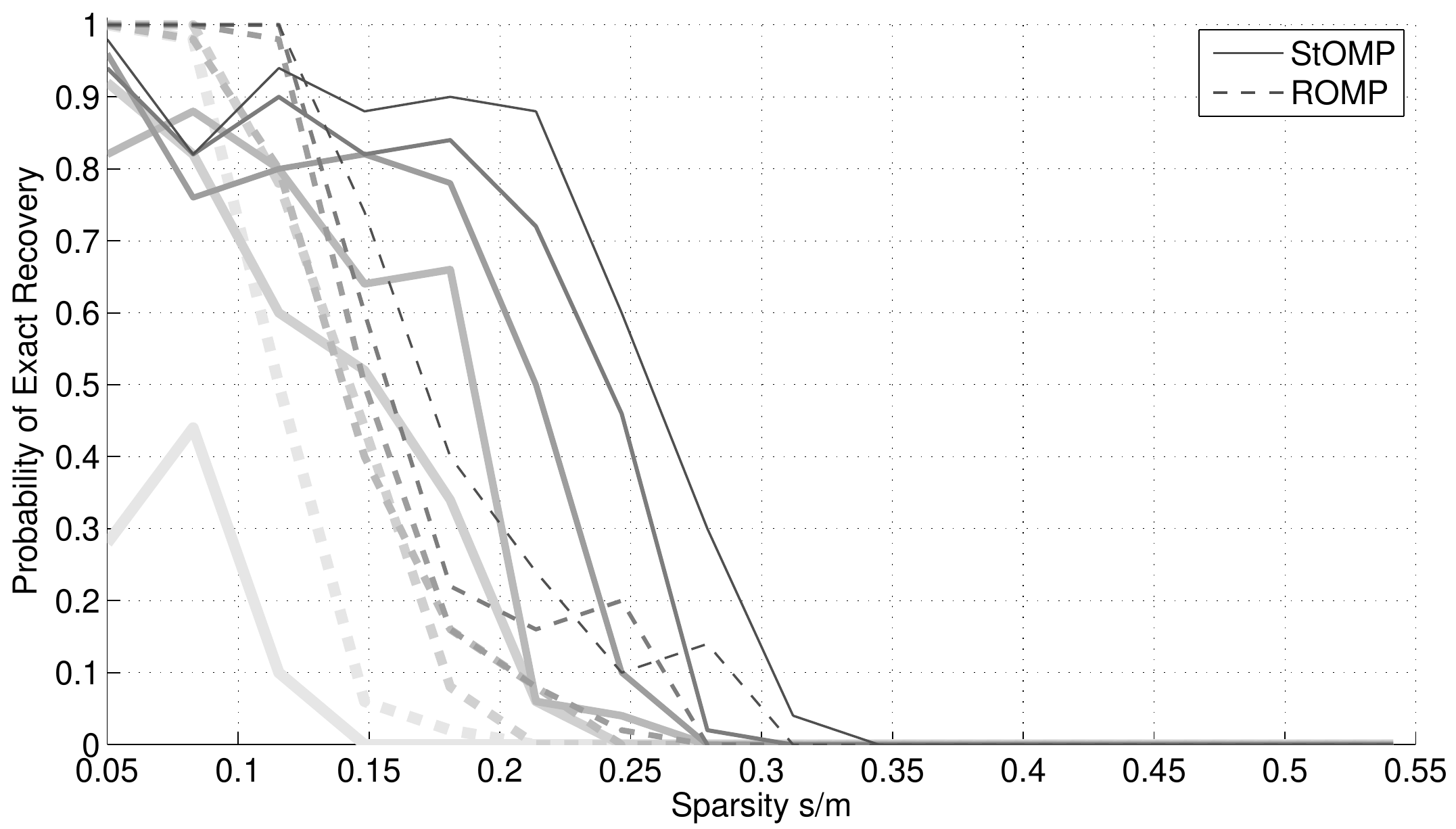}}\\ \vspace{-0.1in}

\subfigure[Bimodal Gaussian]{
\includegraphics[width=0.49\textwidth]{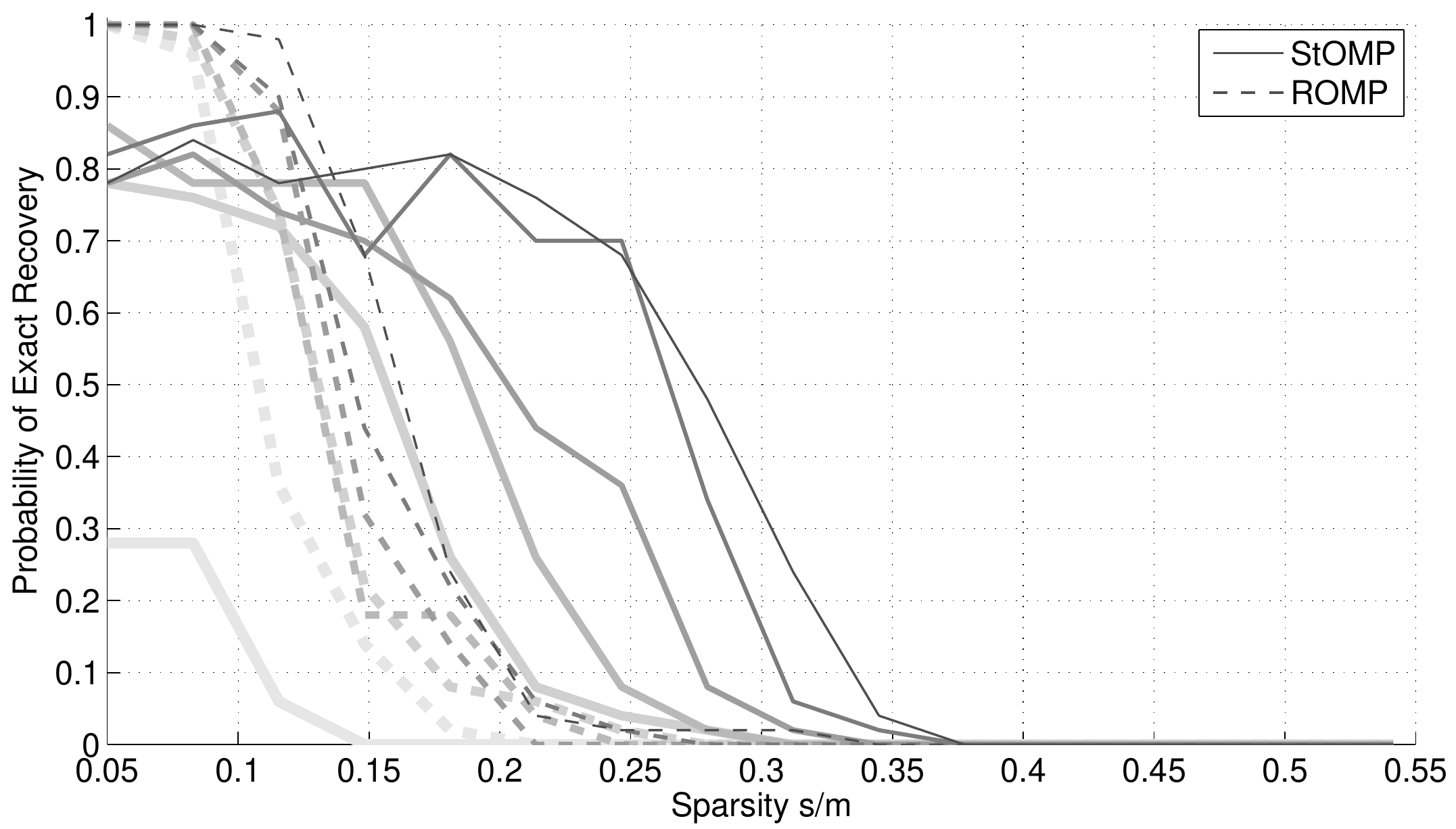}}\hspace{-0.1in}
\subfigure[Uniform]{
\includegraphics[width=0.49\textwidth]{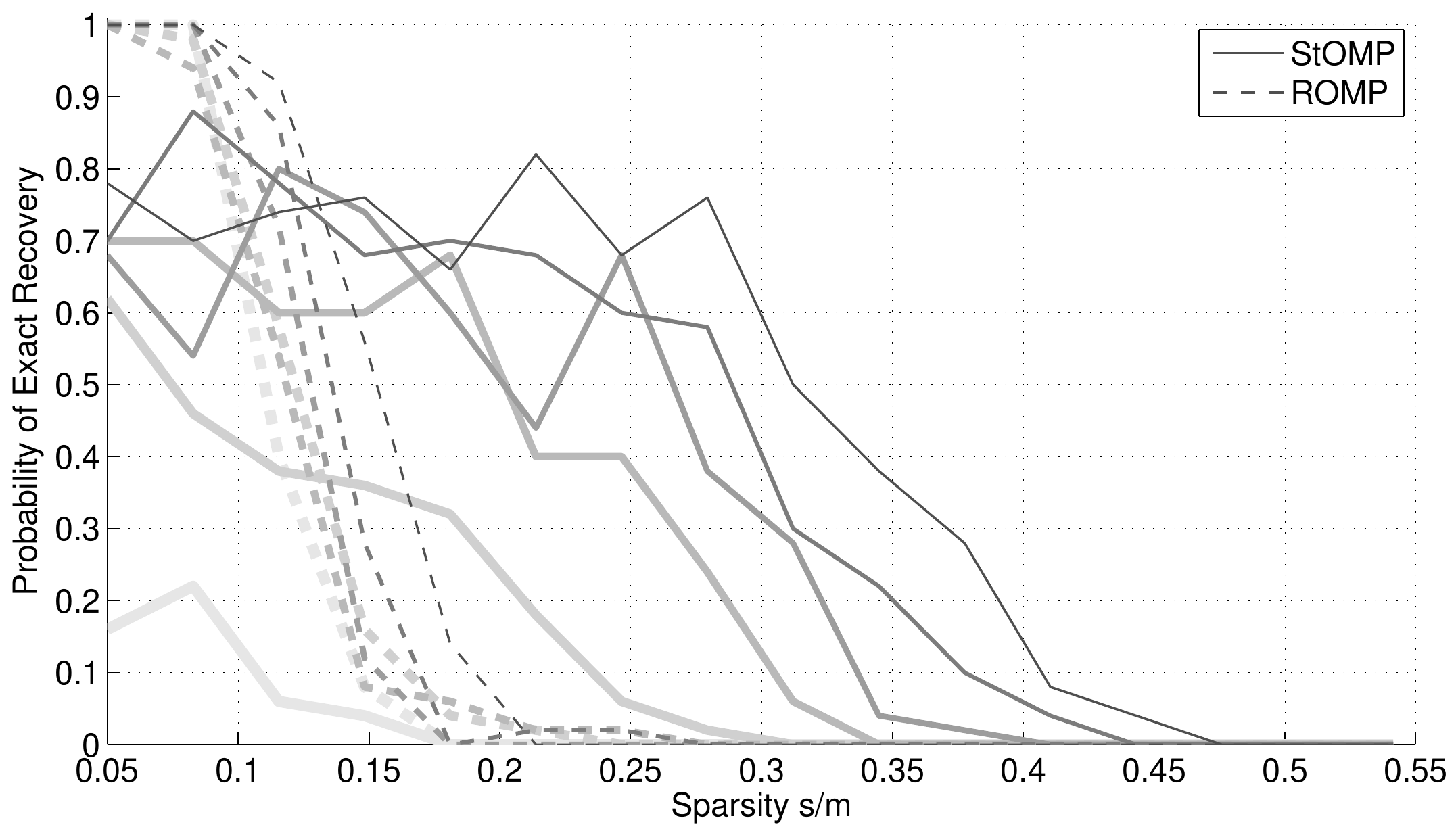}}\\ \vspace{-0.1in}

\subfigure[Bimodal Rayleigh]{
\includegraphics[width=0.49\textwidth]{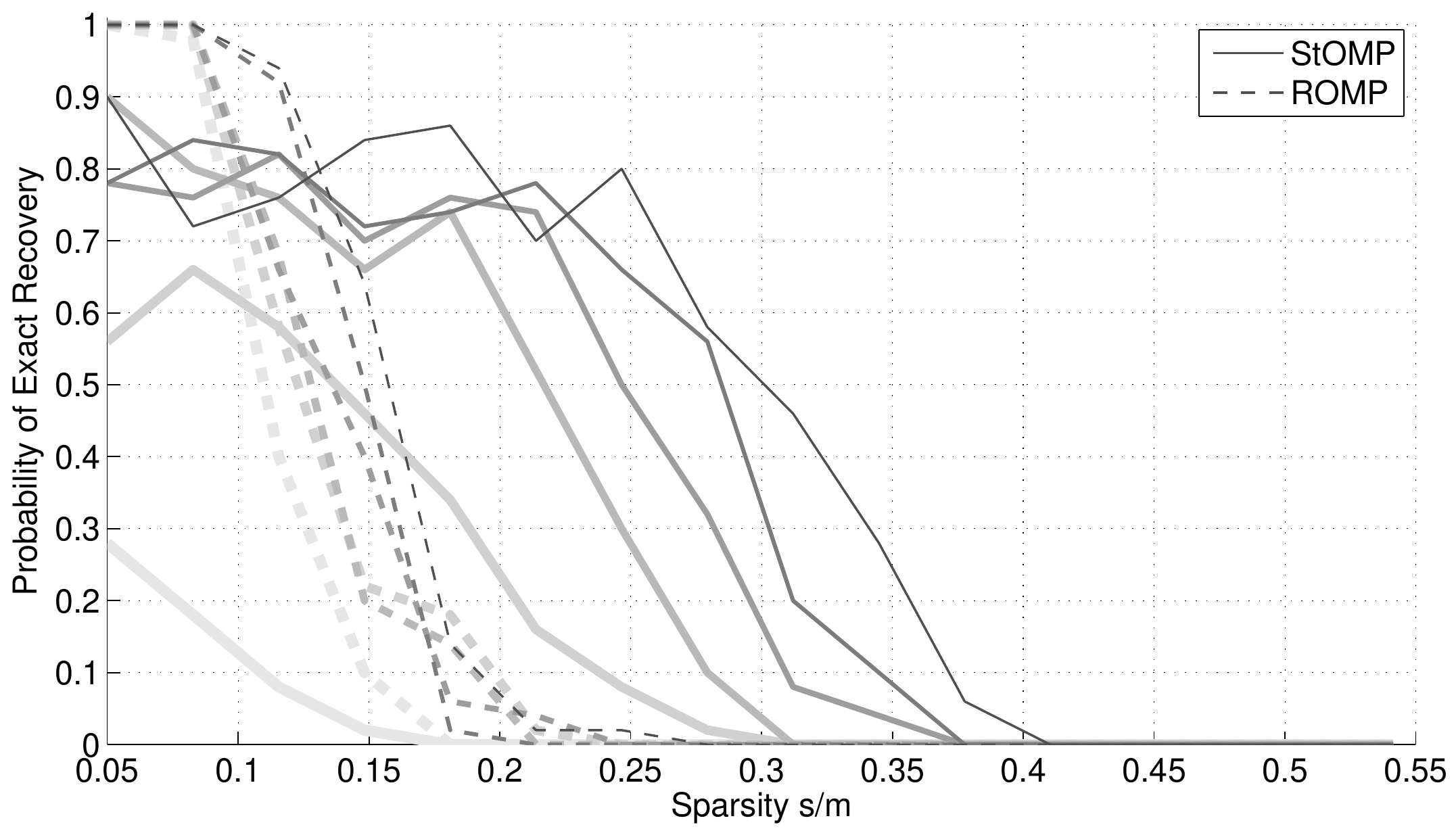}}\hspace{-0.1in}
\subfigure[Normal]{
\includegraphics[width=0.49\textwidth]{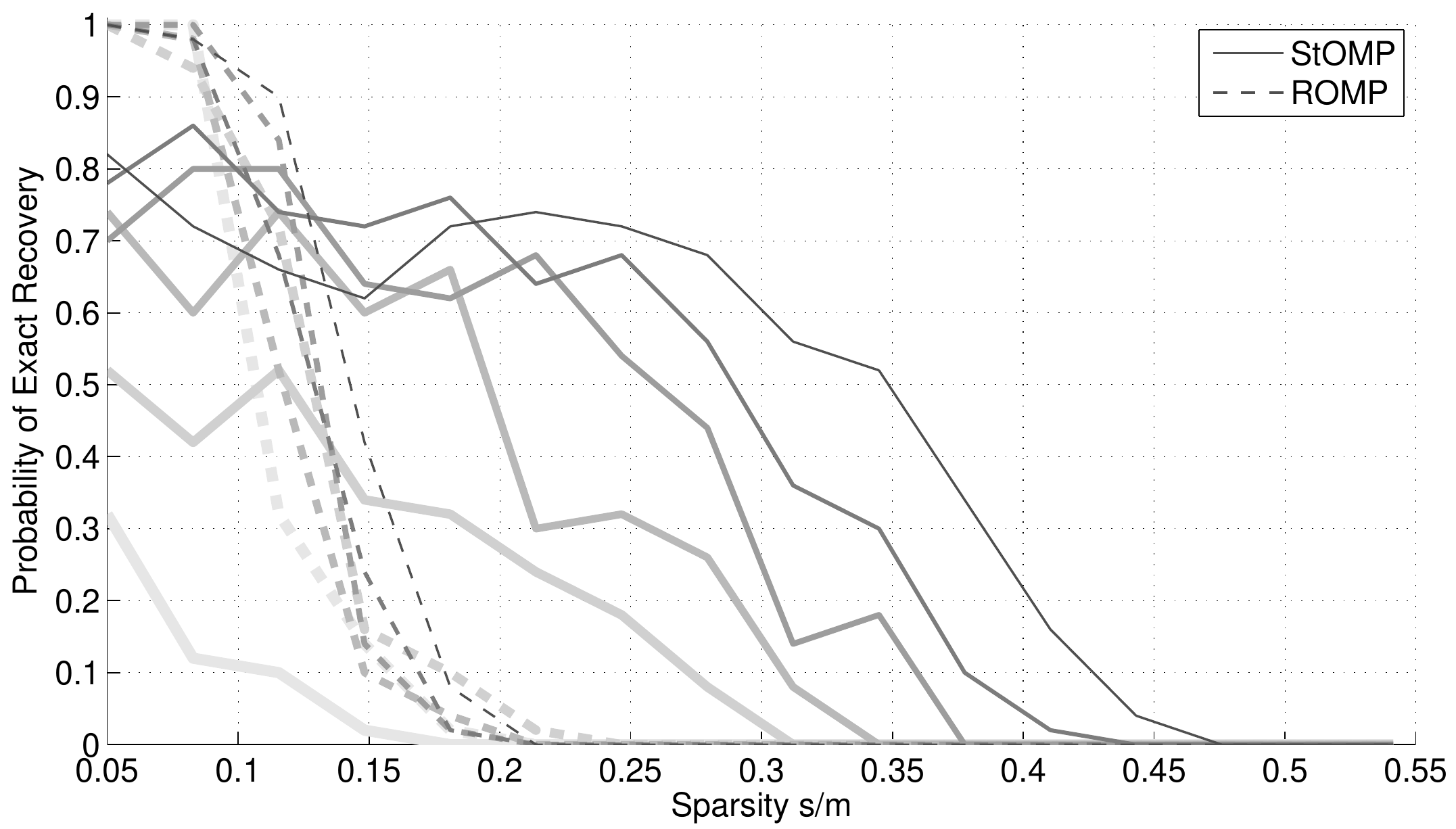}}\\ \vspace{-0.1in}

\subfigure[Laplacian]{
\includegraphics[width=0.49\textwidth]{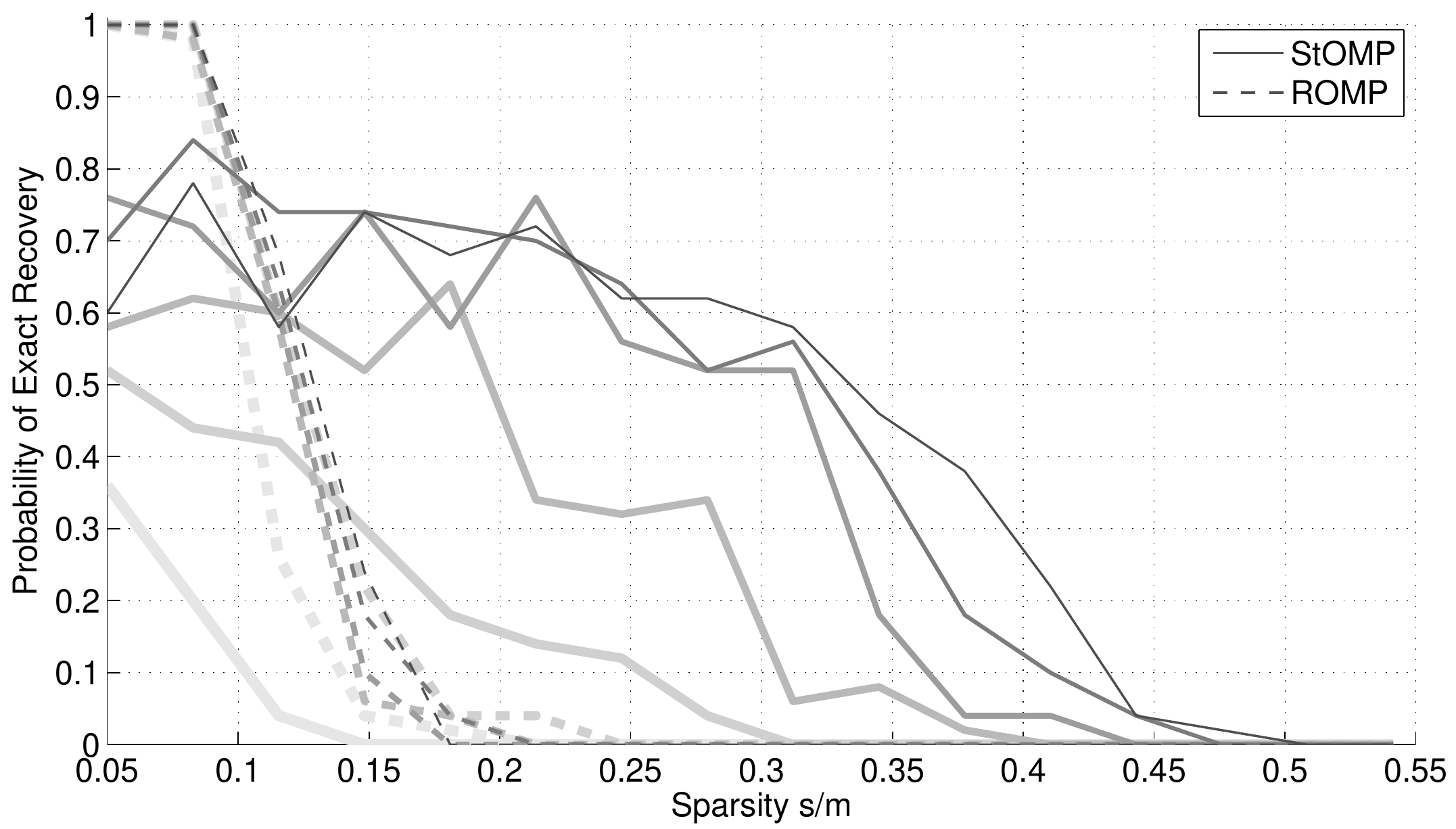}}
\caption{Probability of exact recovery using criterion (\ref{eq:successcriterion2}) 
for StOMP (solid) and ROMP (dashed) as a function of problem sparsity
and six problem indeterminacies from thickest to thinest lines: 
\(\delta = m/N = \{0.05, 0.15, 0.25, 0.34, 0.44, 0.54\}\).}
\label{fig:RecoveryProbabilitiesStOMPROMP}
\end{figure}

Empirical phase transitions only show the regions in $(\delta,\rho)$-space 
where majority recovery does and does not hold.
Another interesting aspect of these algorithms is how fast this transition occurs,
and to what extents we can expect perfect recovery for all vectors.
Figures \ref{fig:RecoveryProbabilitiesBPPrOMP} -- \ref{fig:RecoveryProbabilitiesStOMPROMP}
compare for pairs of algorithms the probability of exact recovery
as a function of sparsity for several problem indeterminacies
for all distributions I test.

In Fig. \ref{fig:RecoveryProbabilitiesBPPrOMP} we see that the transitions of BP and OMP
are quite similar for all distributions except Normal and Laplacian, 
where OMP takes on a smaller transition slope than BP.
At one extreme, for Bernoulli distributed signals,
BP can perfectly recover signals with $\rho < 0.35$ ($s < 76$ for $N = 400$),
while OMP can only recover all signals with $\rho < 0.22$ ($s < 48$ for $N = 400$).
For Laplacian distributed signals, however,
these are reversed, with BP perfectly recovering signals with $\rho < 0.31$ ($s < 67$ for $N = 400$),
while OMP recovers all signals up to $\rho < 0.44$ ($s < 95$ for $N = 400$).

Figure \ref{fig:RecoveryProbabilitiesSL0SP} compares the transition
of probability for SL0 and SP.
They are quite similar for signals distributed 
Bernoulli, bimodal uniform, and bimodal Gaussian.
For all other distributions I test they are much less similar;
and perfect recovery for SL0 for Laplacian distributed signals
extends well beyond that of all the other algorithms.

In Fig. \ref{fig:RecoveryProbabilitiesCoSaMPTST} we see that 
the transitions of probability for CoSaMP and TST are quite similar for 
more distributions except Bernoulli,
and quite different from those of the other ``two-stage thresholding'' algorithm SP in
Fig. \ref{fig:RecoveryProbabilitiesSL0SP}.
Both CoSaMP and TST have very steep transitions
showing that the boundary between perfect recovery
and majority recovery is quite narrow.
A curious thing is that for most distributions I test, 
CoSaMP fails completely at $\rho \approx 0.35$.
In the case of Laplacian signals,
it is quite clear that for CoSaMP this sparsity 
cannot be exceeded no matter $\delta$.
I do not know at this time what causes this behavior.

For the two hard thresholding approaches,
Fig. \ref{fig:RecoveryProbabilitiesIHTALPS}
compares the transitions for IHT and ALPS.
These more than any other, have quite irregular
transitions for all distributions but Bernoulli and bimodal Uniform.
We see for ALPS that only for these distributions
can we expect perfect recovery for some sparsity.
In all the others, ALPS never reaches 100\% recovery,
which is extremely problematic.
The recommended IHT of Maleki and Donoho \cite{Maleki2010}
suffers no such problem, even though
it must estimate the sparsity of the signal,
while ALPS is given that information.

\clearpage 

Figure \ref{fig:RecoveryProbabilitiesStOMPROMP}
shows that StOMP has transition regions that are
similar to those of ALPS, but over all distributions. 
We also see why in Fig. \ref{fig:phasevsalgorithms} StOMP has zero phase transition
at low indeterminacies for vectors distributed bimodal Gaussian, 
bimodal Rayleigh, uniform, and Normal.
I do not yet know the reason for these behaviors,
but it could be due to the necessity to tune
the false alarm rate of StOMP.

\begin{figure}[htb]
\centering
\subfigure[Bernoulli]{
\includegraphics[width=0.49\textwidth]{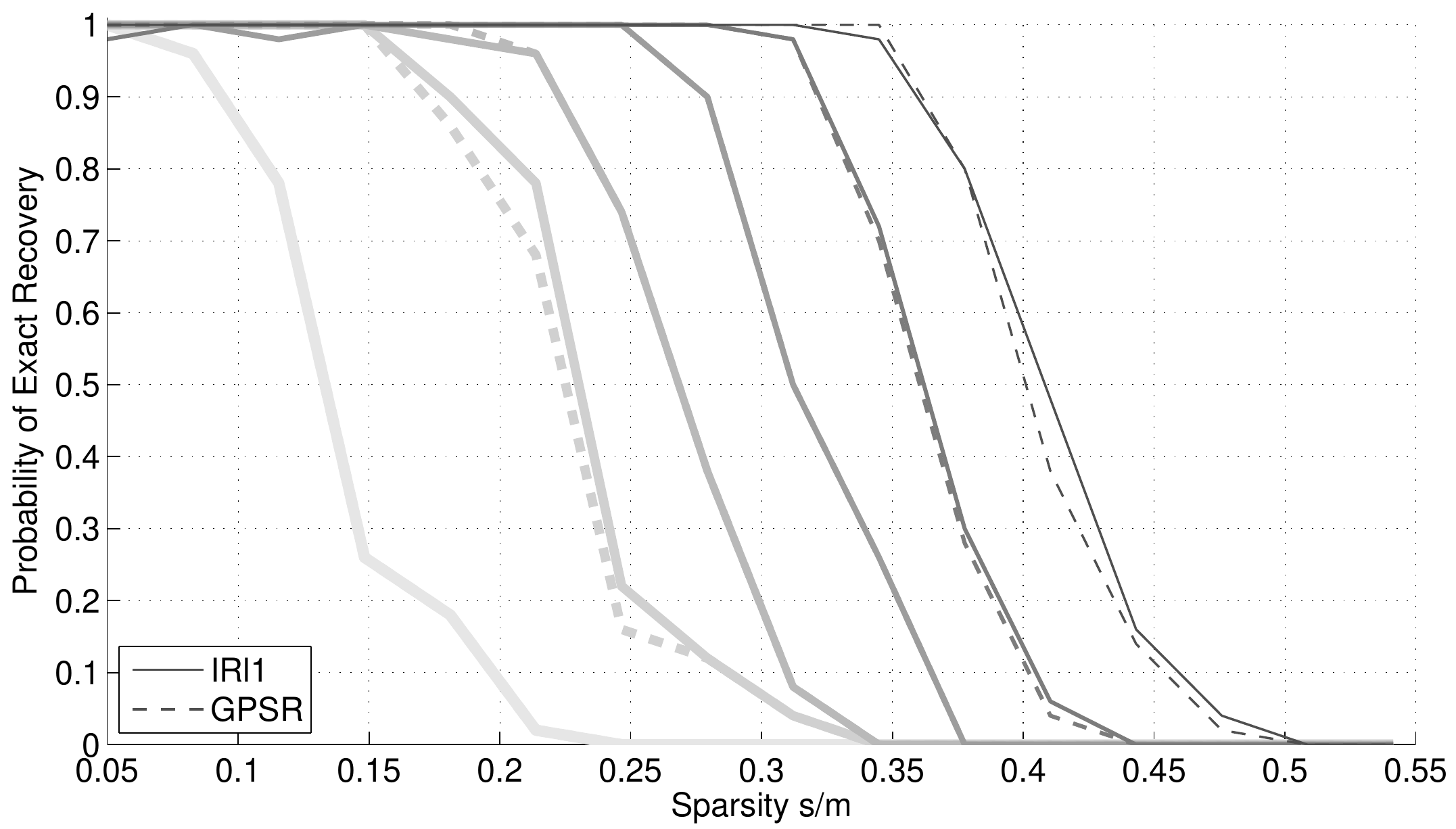}}\hspace{-0.1in}
\subfigure[Bimodal Uniform]{
\includegraphics[width=0.49\textwidth]{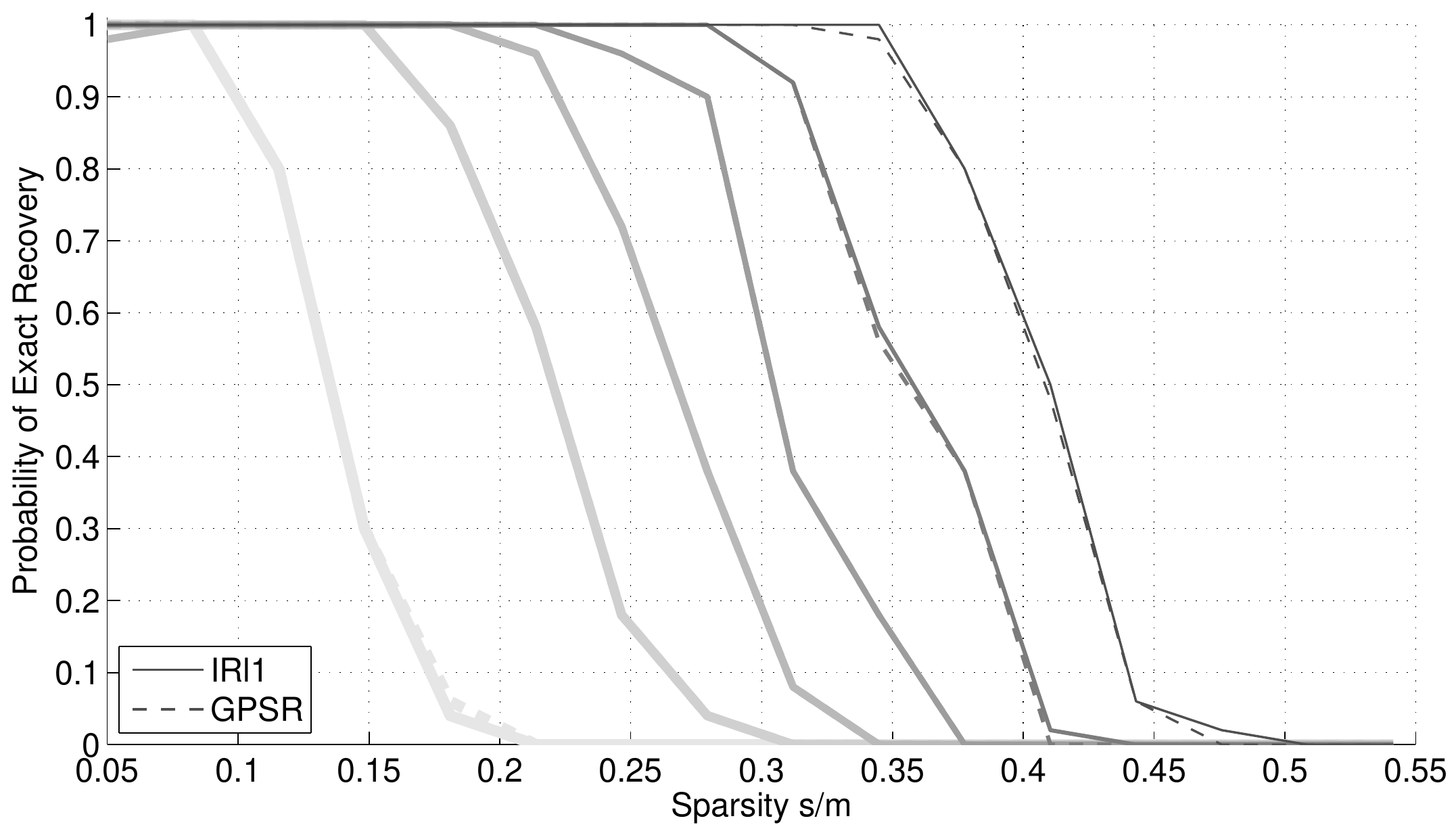}}\\ \vspace{-0.1in}

\subfigure[Bimodal Gaussian]{
\includegraphics[width=0.49\textwidth]{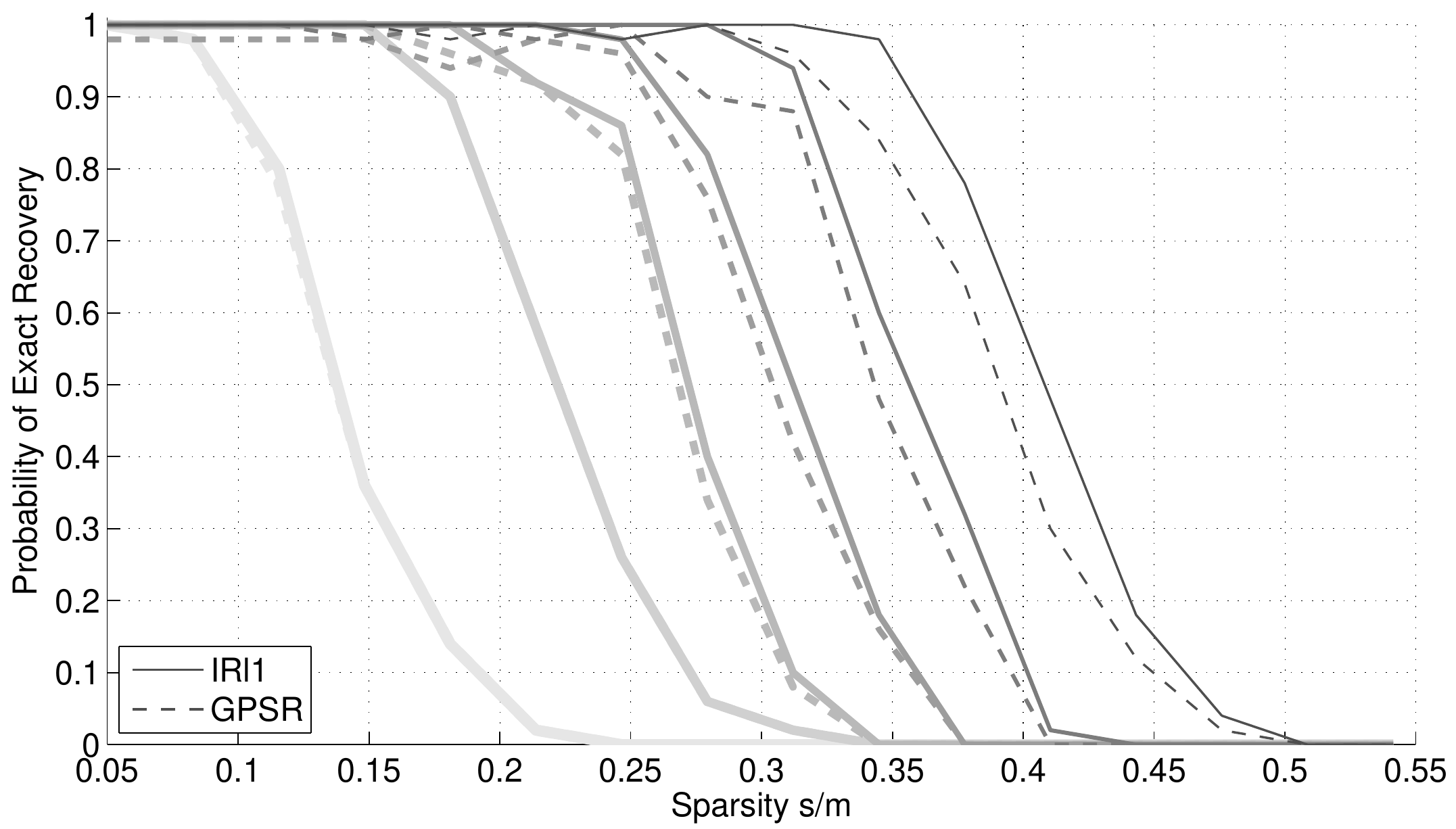}}\hspace{-0.1in}
\subfigure[Uniform]{
\includegraphics[width=0.49\textwidth]{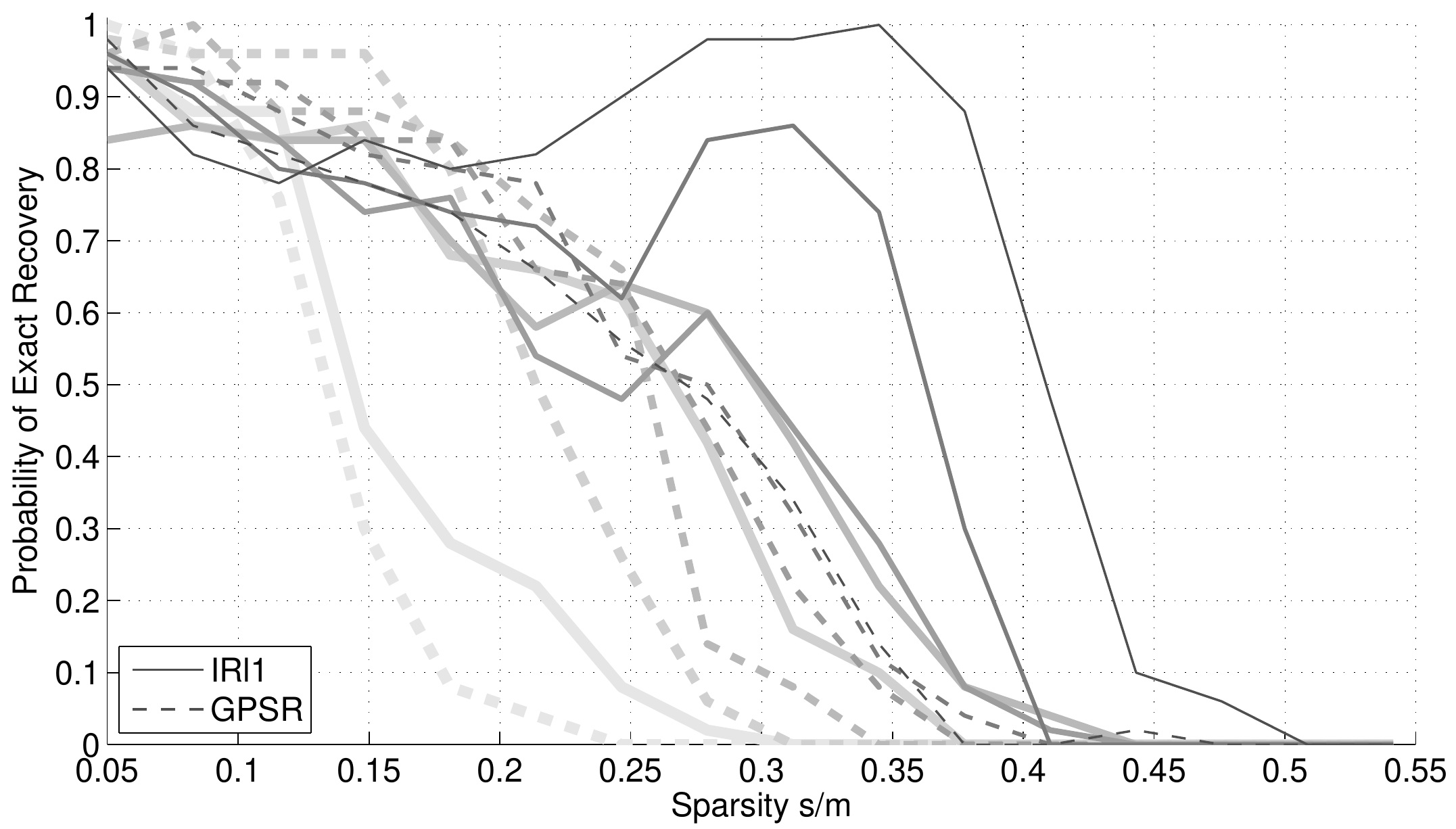}}\\ \vspace{-0.1in}

\subfigure[Bimodal Rayleigh]{
\includegraphics[width=0.49\textwidth]{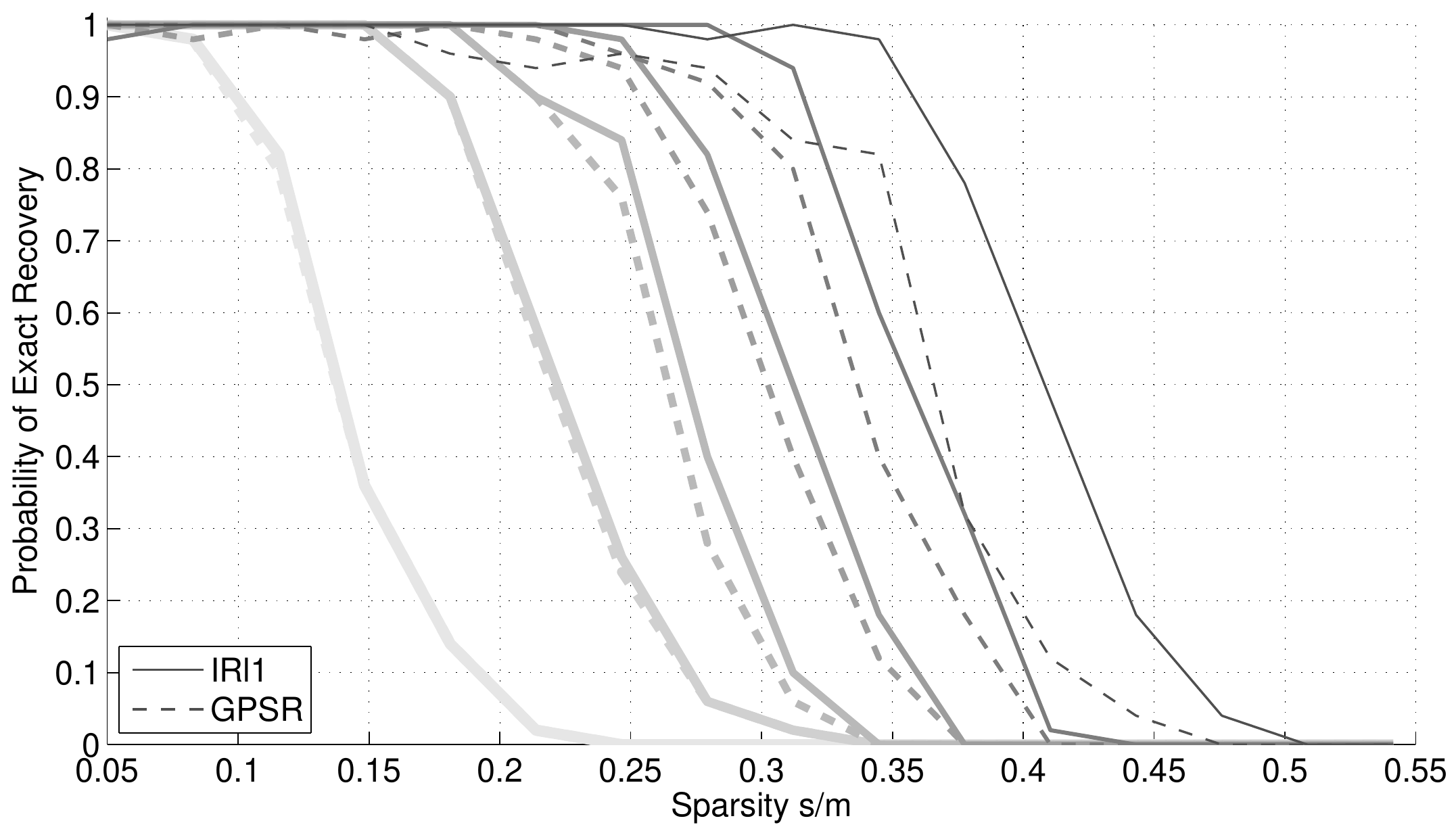}}\hspace{-0.1in}
\subfigure[Normal]{
\includegraphics[width=0.49\textwidth]{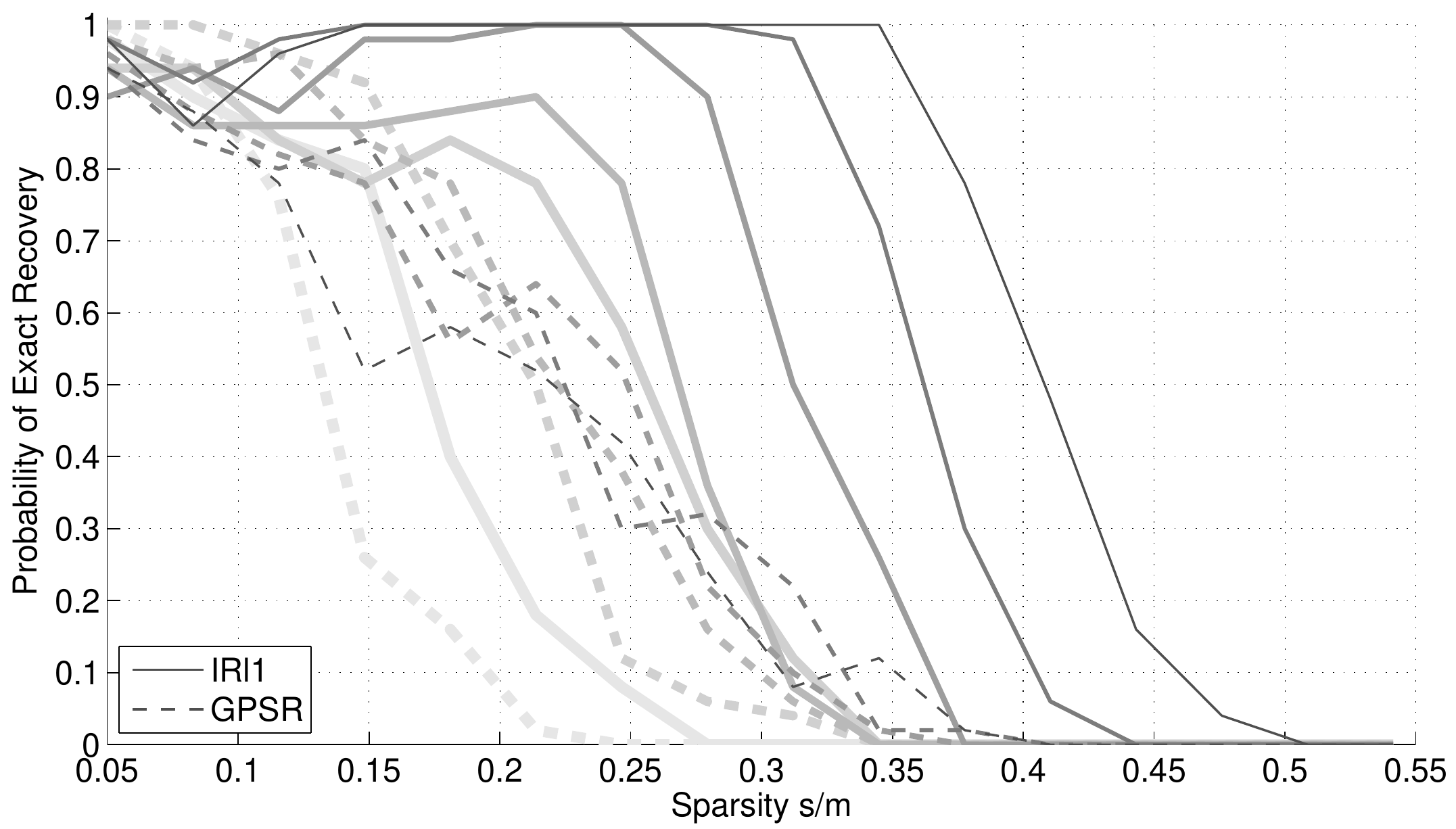}}\\ \vspace{-0.1in}

\subfigure[Laplacian]{
\includegraphics[width=0.49\textwidth]{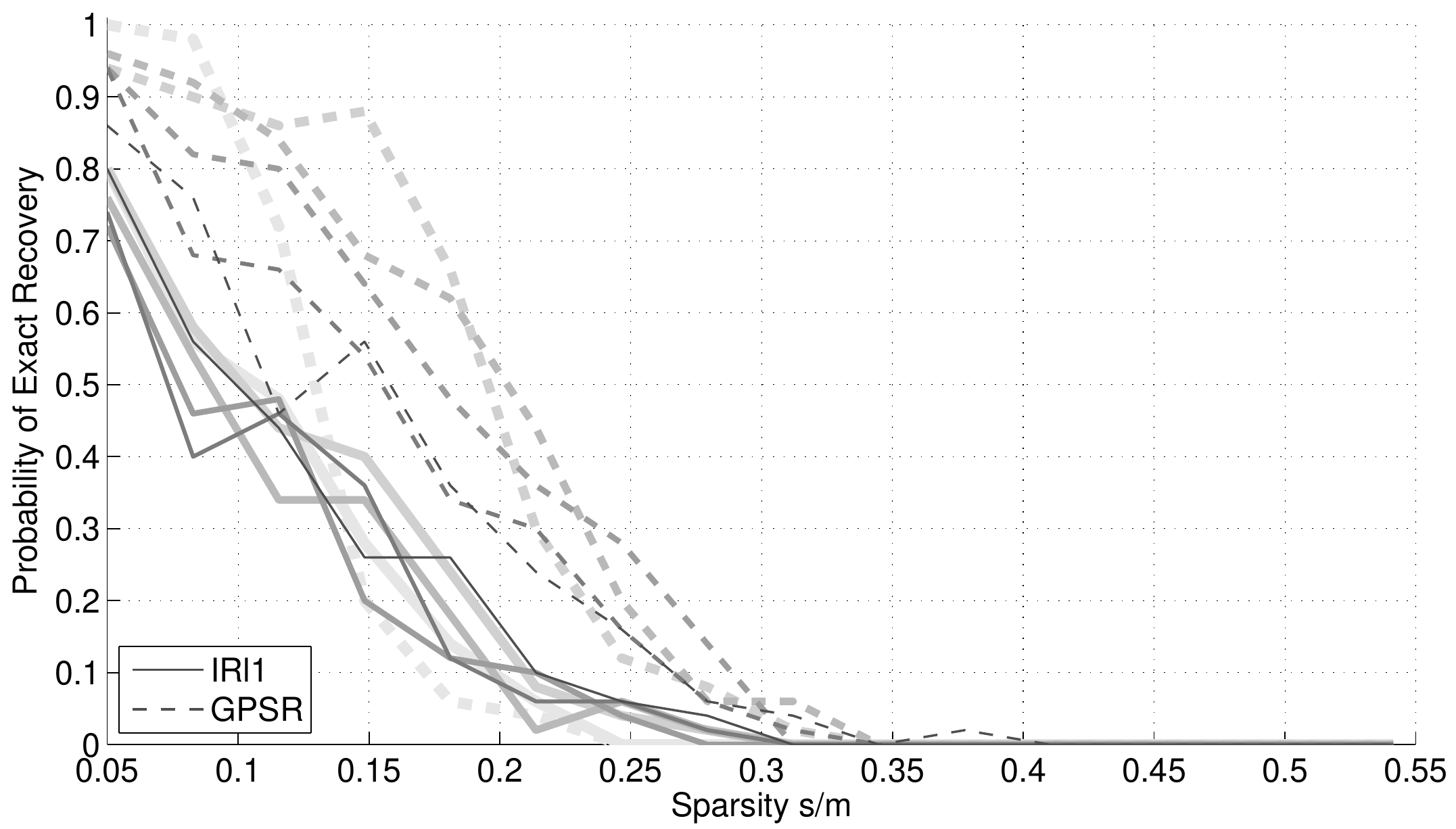}}
\caption{Probability of exact recovery using criterion (\ref{eq:successcriterion2}) 
for IRl1 (solid) and GPSR (dashed) as a function of problem sparsity
and six problem indeterminacies from thickest to thinest lines: 
\(\delta = m/N = \{0.05, 0.15, 0.25, 0.34, 0.44, 0.54\}\).}
\label{fig:RecoveryProbabilitiesIRl1GPSR}
\end{figure}

Finally, Fig. \ref{fig:RecoveryProbabilitiesIRl1GPSR} compares 
the transitions of IRl1 and GPSR.
We see that the two are quite similar for vectors distributed
Bernoulli, bimodal uniform, and bimodal Gaussian;
but GPSR begins to break down as the probability density
becomes more concentrated around zero.
IRl1 has behavior that is inconsistent with what I would predict.
For the most part, it acts fine for Normally distributed vectors,
but its performance is very poor
for vectors distributed uniform and especially Laplacian.
At this time I do not know what troubles IRl1
for these distributions,
or why it cannot perfectly recover Normal vectors with low sparsity,
but it can recovery vectors with higher sparsity.

\clearpage

\section{Conclusion}
In this work I show that any judgement of which algorithm among several
is the best for recovering compressively sensed sparse vectors
must be prefaced by the conditions for which this is observed to be true.
It is clear from my computer simulations
that the performance of a recovery algorithm 
within the context of compressed sensing can be
greatly affected by the distribution underlying the sensed sparse vector,
and that the summary of performance is highly dependent on
the criterion of successful recovery.
These ``findings'' are certainly nothing novel,
and clearly not controversial.
It has already been stated in numerous pieces of research,
and is somewhat codified as ``folk knowledge'' in the 
compressed sensing research community,
that recovery algorithms are sensitive to the nature of a sparse vector,
and that sparse vectors distributed Bernoulli
appear to be the hardest to recover, e.g., \cite{Jin2008, Tropp2004,Dai2009,Donoho2009,Maleki2010,Qui2010}.
The extents to which this is true and meaningful,
and the variability of performance to the criterion
of successful recovery, however,
had yet to be formally and empirically studied and presented.

In light of this work, 
the important thing to ask moves from
what recovery algorithm is the best,
to what distribution underlies a compressively sensed vector,
{\em and} what is the measure of success.
In my experiments, we see that SL0 performs extremely well 
in the sense of recovering the support (and thus the sparse vector exactly)
for five distributions I test,
except for low indeterminacies of Laplacian, Normal, bimodal Rayleigh, 
uniform, and bimodal Gaussian, where the greedy approach PrOMP performs slightly better.
I do not yet know the reason for this,
but it could be that the parameters for SL0 must be adjusted.
SL0 does not perform as well as BP/AMP for 
sparse vectors distributed Bernoulli or bimodal uniform.
With their performance, and because of their speed, 
SL0 and AMP are together extremely attractive algorithms 
for recovering sparse vectors distributed in any of these seven ways.

A critical question to answer is how these results 
change when the sensed vector is corrupted by noise,
or when the sensing matrix is something other than
from the uniform spherical ensemble.
One potential problem with using the algorithms
tuned by Maleki and Donoho \cite{Maleki2010}
is that they are tuned to sparse vectors distributed Bernoulli.
It could be possible that they perform better for vectors distributed Laplacian
if they are tuned to such a distribution; however,
Maleki and Donoho argue that tuning on Bernoulli sparse vectors
is essentially maximizing the best performance for the worst case,
and that this translates to situations that are more forgiving.
In my experiments, I do see that recommended IHT can perform better
than recommended TST for sparse vectors distributed uniform, 
Normal, and Laplacian,
which subverts their ordering of their algorithms in terms of performance.
It stands to be reasoned then that we can better tune
these algorithms for those situations such that
recommended TST does outperform recommended IHT.
Finally, it is important to measure the performance of these algorithms
for real-world signals that are sparsely described only in 
coherent dictionaries \cite{Candes2010}.

\bibliographystyle{IEEEtran}
\bibliography{../bibliographies/BibAnnon}

\end{document}